\documentclass[12pt]{article}
\usepackage[toc,page]{appendix}
\usepackage{xcolor}
\usepackage[compat=1.1.0]{tikz-feynman} \tikzfeynmanset{warn luatex=false}
\usepackage{mathtools}% Loads amsmath
\usepackage{amsthm,amsmath,latexsym,amssymb,amsfonts,amscd}
\usepackage{graphics,lscape,fancyhdr,array,stmaryrd,euscript,wrapfig}
\usepackage{empheq,slashed}
\usepackage{verbatim,slashed}
\numberwithin{equation}{section}
\usepackage{hyperref,setspace}
\usepackage{dsfont}
\usepackage{mathrsfs}

\usepackage[numbers,sort&compress]{natbib}
\usepackage[nottoc]{tocbibind}

\newcommand{\pl}{\partial}

\newcommand{\be}{\begin{equation}}
\newcommand{\ee}{\end{equation}}

\newcommand{\besubeqs}{\begin{subequations}}
\newcommand{\esubeqs}{\end{subequations}}

\newcommand{\PP}{{\mathbb{P}}}

\newcommand{\PPb}{{\bar{\mathbb{P}}}}

\begin{document}
%%%%%%%%%%%%%%%%%%%%%%%%%%%%%%%%%%%%%%%%%%%%%%%%%%%%%%%%%%%%%
\pagenumbering{gobble}
\hfill
\vskip 0.01\textheight
\begin{center}
{\Large\bfseries 
On classification of (self-dual) higher-spin gravities \\
[5mm]
in flat space}

\vspace{0.4cm}

\vskip 0.03\textheight
\renewcommand{\thefootnote}{\fnsymbol{footnote}} Mattia Serrani${}^{\pi}$
\renewcommand{\thefootnote}{\arabic{footnote}}
\vskip 0.03\textheight
\centering
\href{mailto:mattia.serrani@umons.ac.be}{\texttt{mattia.serrani@umons.ac.be}}
\vskip 0.03\textheight

{\em ${}^{\pi}$ Service de Physique de l'Univers, Champs et Gravitation, \\ Universit\'e de Mons, 20 place du Parc, 7000 Mons, 
Belgium}
\end{center}

\vskip 0.02\textheight

\begin{abstract}
There is a great number of higher-spin gravities in $3d$ that can have both finite and infinite spectra of fields and can be formulated as Chern-Simons theories. It was believed that this is impossible in higher dimensions, where higher-spin fields do have propagating degrees of freedom. We show that there are infinitely many higher-spin theories in the $4d$ flat space featuring nontrivial local interactions that can have either a finite or infinite number of fields. We classify all one- and two-derivative (i.e. with gauge and gravitational interactions) higher-spin theories by solving the holomorphic constraint in the light-cone gauge obtained by Metsaev. Therefore, these theories are consistent subsectors of the higher-spin extensions of self-dual Yang-Mills/gravity, which in turn are truncations of the chiral higher-spin gravity.
\end{abstract}

\newpage
\tableofcontents
\newpage
%%%%%%%%%%%%%%%%%%%%%%%%%%%%%%%%%%%%%%%%%%%%%%%%%%%%%%%%%%%%%
\section{Introduction}\label{section1}

%%%%%%%%%%%%%%%%%%%%%%%%%%%%%%%%%%%%%%%%%%%%%%%%%%%%%%%%%%%%%
\pagenumbering{arabic}
\setcounter{page}{2}
%%%%%%%%%%%%%%%%%%%%%%%%%%%%%%%%%%%%%%%%%%%%%%%%%%%%%%%%%%%%%
The basic idea of looking for extensions of gravity by massless fields with higher spins dates back to at least as early as \cite{Fronsdal:1978rb}. The first candidate spectrum of a higher-spin theory was proposed in \cite{Flato:1978qz} and contains massless fields of all spins from $0$ to $\infty$. It was obtained via the representation theory of the $AdS_4$ symmetry group, which, in modern terms, corresponds to working out the spectrum of single-trace operators in the free vector model and using the AdS/CFT dictionary to read off the spectrum of the dual theory. It also came with the idea \cite{Flato:1980zk} that interactions of higher-spin fields have something to do with the scattering of dipoles --- bilinears in the free fields --- on the boundary of $AdS_4$, which was a precursor to the modern AdS/CFT duality with vector models on the boundary, see e.g. \cite{Sezgin:2002rt,Klebanov:2002ja,Sezgin:2003pt,Leigh:2003gk,Giombi:2011kc}. 

Despite the above mentioned initial input, where it is $AdS_4$ and not the flat space that played a role, the systematic study of possible interactions had first been started in the light-front approach\footnote{Introduced in a seminal paper by Dirac \cite{Dirac:1949cp}.} by Bengtsson, Bengtsson and Brink \cite{Bengtsson:1983pg,Bengtsson:1983pd,Bengtsson:1986kh} in the flat space with the complete classification of cubic vertices obtained in \cite{Bengtsson:1986kh}. Around the same time, the first covariant cubic vertices had been found in \cite{Berends:1984wp,Berends:1984rq}, where it also became evident that the smallest spectrum of fields to admit consistent interactions has to be infinite and unbounded in spin. One cannot help mentioning numerous no-go theorems, most notably the Weinberg low energy theorem \cite{Weinberg:1964ew} and the Coleman-Mandula theorem \cite{Coleman:1967ad} that seem to shut down the whole higher-spin idea. Analogues of both Weinberg's and Coleman-Mandula theorems can be extended to $AdS_d$ \cite{Maldacena:2011jn,Fitzpatrick:2012cg,Boulanger:2013zza,Alba:2013yda,Alba:2015upa,Sleight:2021iix}.\footnote{The initial impulse towards anti-de Sitter space \cite{Fradkin:1986qy} was triggered by the fact that these analogues had not yet been available and by the encouraging result \cite{Fradkin:1986qy} on the construction of ``gravitational interactions'' of higher-spin fields that avoid the no-go of \cite{Aragone:1979hx}. However, the cubic gravitational interactions in the flat space were also constructed with the help of the light-front approach, the same year, as a part of the complete classification \cite{Bengtsson:1986kh}. As discussed in \cite{Krasnov:2021nsq}, the ``non-existence'' of the gravitational interactions for higher-spin fields in flat space \cite{Aragone:1979hx} is related to a somewhat degenerate nature of the description based on symmetric (Fronsdal) fields and is not an invariant statement about higher-spin fields. In particular, the ``gravitational'' interaction of \cite{Fradkin:1986qy} is a linear combination of several independent (the minimal and nonminimal) vertices in the light-cone gauge. } However, these results mostly indicate that higher-spin theories should be integrable in some sense, in particular, the S-matrix must be trivial or simple enough, rather than telling anything about the (non)-existence of these theories as field theories.

The present understanding of the problem of higher-spin interactions is that it is difficult for the masslessness of higher-spin fields to coexist with the usual requirements of locality within the field theory approach.\footnote{One of the possible applications of higher-spin gravities, apart from the general search for fundamental interactions, is to provide simple models of quantum gravity where the extended symmetry associated with massless fields should eliminate counterterms. One would not expect a quantum gravity model to be a local field theory, which can partially explain why it is difficult to stick to the usual definitions of locality when higher-spin fields are involved. Therefore, the higher-spin problem can also be rephrased as how to deal with nonlocalities that higher-spin fields usually lead to. } This problem is insensitive to the sign of the cosmological constant. However, it takes a rather technical argument to see how the construction of interactions order by order breaks down in the flat space, see e.g. \cite{Bekaert:2010hp,Ponomarev:2017nrr,Roiban:2017iqg}, while it is much easier to arrive at the same conclusion for the negative cosmological constant by using AdS/CFT \cite{Bekaert:2015tva,Maldacena:2015iua,Sleight:2017pcz,Ponomarev:2017qab}, see, however, \cite{Neiman:2023orj}. 

There are a number of well-defined, i.e. local or perturbatively local, higher-spin gravities known to date. In lower dimensions,  $d<4$, the graviton and massless higher-spin fields do not have propagating degrees of freedom, which neutralizes the locality problem. Also, the famous no-go theorems do not apply. For example, there is a great number of (matter-free) higher-spin gravities in $3d$ (purely massless, conformal and partially-massless) that can be formulated with the help of the Chern-Simons theory with a Lie algebra into which the gravitational $sl_2$ subalgebra is embedded in a sufficiently nontrivial way, \cite{Blencowe:1988gj,Bergshoeff:1989ns,Campoleoni:2010zq,Henneaux:2010xg,Grigoriev:2020lzu,Pope:1989vj,Fradkin:1989xt,Grigoriev:2019xmp}. An example of a matter-coupled theory has recently been constructed in \cite{Sharapov:2024euk}. In $d=2$ one can use the BF-theory in a similar way \cite{Alkalaev:2020kut}.

In $d\geq4$ the situation has been more complicated and, for that reason, less rich in models. The first perturbatively local theory has been the conformal higher-spin gravity of Segal and Tseytlin \cite{Segal:2002gd,Tseytlin:2002gz,Bekaert:2010ky, Basile:2022nou}, which is a higher-spin extension of conformal gravity. The local Weyl symmetry keeps interactions perturbatively local. A recent example is chiral higher-spin gravity \cite{Metsaev:1991mt,Metsaev:1991nb,Ponomarev:2016lrm,Ponomarev:2017nrr}, which can be thought of as a higher-spin extension of self-dual Yang-Mills theory and of self-dual gravity at the same time. The theory admits two simple contractions that are higher-spin extensions either of self-dual Yang-Mills (HS-SDYM) or of self-dual gravity (HS-SDGR) \cite{Ponomarev:2017nrr,Krasnov:2021nsq}. Chiral theory and its contractions can also be double-copied \cite{Ponomarev:2024jyg} to give theories with even bigger spectra. Recently, a quasi-chiral theory has also been found \cite{Adamo:2022lah}. All of these theories have an infinite, unbounded in spin, spectrum of fields.\footnote{One can also mention a higher-spin extension of the IKKT matrix model \cite{Sperling:2017dts}, which is a noncommutative field theory with higher-spin fields in the spectrum. Also, one can try to reverse-engineer the AdS/CFT duals of the vector models, see e.g. \cite{deMelloKoch:2018ivk}. } 

Let us stress that the light-front approach has always played an important role in the study of higher-spin interactions, see e.g. \cite{Metsaev:1993ap,Metsaev:2005ar,Metsaev:2007rn,Metsaev:2018xip}.\footnote{To be fair, one should also mention covariant results, e.g. \cite{Boulanger:2008tg,Manvelyan:2010je,Boulanger:2012dx,Francia:2016weg}, see also a very general review \cite{Bekaert:2022poo}. } Its main appealing feature is that all unphysical degrees of freedom are eliminated and one can get unambiguous results concerning (non)existence of certain interactions.\footnote{Any covariant approach picks a certain embedding of the physical degrees of freedom into a Lorentz-covariant field and, depending on the embedding, may miss important types of interactions. For example, the conclusion of \cite{Aragone:1979hx} that there are no gravitational interactions of higher-spin fields relied on a particular choice of a Lorentz-covariant field, while gravitational two-derivative interactions are present on the list \cite{Bengtsson:1986kh} within the light-front approach. Similarly, while the graviton described by the symmetric tensor allows one to construct the usual gravitational interactions, it is impossible to do so if a different Lorentz tensor is chosen \cite{Bekaert:2002uh}. } In the present paper, we study the consistency condition that appears at the quartic order in the light-cone gauge. It was obtained and analysed by Metsaev in \cite{Metsaev:1991mt,Metsaev:1991nb}. An immediate consequence \cite{Ponomarev:2016lrm} of this result is the existence of chiral higher-spin gravity.\footnote{Note that the light-cone approach is instrumental in getting sharp no-go's \cite{Ponomarev:2017nrr} or existence statements \cite{Ponomarev:2016lrm}. However, once some positive result is established, it is also helpful to have a Lorentz-invariant formulation, e.g. to study solutions, compute amplitudes, extend the results into (anti)-de Sitter space, some of which was achieved for chiral higher-spin gravity in \cite{Sharapov:2022faa,Sharapov:2022wpz,Sharapov:2022awp,Sharapov:2022nps,Sharapov:2023erv,Skvortsov:2024rng,Tran:2025yzd}. Still, the first quantum corrections have been computed in the light-cone gauge \cite{Skvortsov:2018jea,Skvortsov:2020wtf,Skvortsov:2020gpn,Tsulaia:2022csz,Neiman:2024vit}. Another interesting direction is AdS/CFT duality for chiral higher-spin gravity \cite{Skvortsov:2018uru,Sharapov:2022awp,Jain:2024bza,Aharony:2024nqs}. } What we show is that there are more sufficiently interesting solutions hidden in there.  

In a few words, the analysis of the quartic consistency proceeds as follows. One begins with some cubic interaction vertex $V_{\lambda_1,\lambda_2,\lambda_3}$, which also implies the presence of fields with helicities $\lambda_{1,2,3}$ in the spectrum. Every light-cone vertex leads to the standard spinor-helicity amplitude, e.g.
\begin{align}\label{genericV}
   V_{\lambda_1,\lambda_2,\lambda_3}\Big|_{\text{on-shell}} \sim 
        [12]^{\lambda_1+\lambda_2-\lambda_3}[23]^{\lambda_2+\lambda_3-\lambda_1}[13]^{\lambda_1+\lambda_3-\lambda_2}\,, \qquad \sum\lambda_i>0
\end{align}
and there is a similar expression with $\langle | \rangle$ when $\sum\lambda_i<0$. The propagator connects $+\lambda$ to $-\lambda$. The quartic light-cone constraint is equivalent to the Poincaré invariance of the S-matrix, see e.g. \cite{Ponomarev:2016cwi}. To have a nontrivial quartic constraint, one either has $\lambda_i=-\lambda_j$ among the already introduced vertices or has to add more fields/interactions to ``make them talk'' to each other.\footnote{Of course, one can always introduce abelian interactions, which correspond to all $\lambda_{1,2,3}>0$ or $\lambda_{1,2,3}<0$ in a vertex or have a non-abelian interaction that is, nevertheless, consistent on its own because one cannot even form an exchange diagram, e.g. one can take vertex $V_{666,-13,-42}$ and it does not lead to any nontrivial quartic constraint.} However, the quartic constraint may not be satisfied as is and may force one to introduce more exchange diagrams and, hence, more fields/interactions. 

Quite often, the process does not stop till the chiral higher-spin gravity is reached. One can confine oneself to gauge and gravitational interactions (i.e. one- and two-derivative ones) that are present in the contractions --- HS-SDYM and HS-SDGR --- of the chiral higher-spin gravity. As we show, there are still many nontrivial solutions in this subclass of interactions, and we classify them all. Therefore, all the theories we found belong to the class of higher-spin extensions of self-dual Yang-Mills and of self-dual gravity. In a bit more detail, we have found $9$ families of HS-SDGR-like theories with finitely many fields (plus $4$ particular cases) and $3$ families with infinitely many fields. As for HS-SDYM-like theories, we have found $8$ families with finitely many fields (plus $2$ particular cases) and $2$ families with infinitely many fields.

It is tempting to make a parallel with the matter-free higher-spin gravities in $3d$, which can always be formulated as Chern-Simons theories \cite{Grigoriev:2020lzu}. They provide a huge variety of options to have a finite number of higher-spin fields. Our results reveal that the case of $4d$ is somewhat similar if one restricts oneself to holomorphic/anti-holomorphic interactions/amplitudes.  

Among other highlights, one can mention that there are particular low-spin couplings, e.g. $V_{-1,0,2}$, that induce higher-spin amplitudes from the Poincaré invariance of the S-matrix, which is surprising. There are also consistent solutions with both a finite spectrum and higher-derivative nonabelian interactions. We also find that it is possible for the graviton to have colour (multi-graviton theories) once we restrict to the self-dual subsector. Another curiosity is that there are solutions with fractional helicities, e.g. $\lambda=2/3,4/3,...$, which, at least at present, are more of an oddity.  

The outline of the paper is as follows:
In Section \ref{section2}, we briefly summarise the derivation of the holomorphic quartic constraint, along with some basics of the light-front approach to higher-spin interactions.
In Section \ref{section3}, we review and extend the solution of the holomorphic constraint \cite{Metsaev:1991mt,Metsaev:1991nb,Ponomarev:2016lrm,Ponomarev:2017nrr,Monteiro:2022xwq} to all integer spins, allowing for arbitrary couplings involving both even- and odd-derivative interactions. We also analyse the constraint in the presence of a $U(N)$ gauge group.
In Section \ref{section4}, we present our main results: the complete classification of two-derivative chiral higher-spin theories, all of which are contained within HS-SDGR, and the classification of all one-derivative chiral higher-spin theories in the presence of a gauge group, contained within HS-SDYM. We also discuss the higher-derivative case and highlight some interesting features. Additionally, we provide explicit solutions for the couplings in certain chiral higher-spin theories.
In Section \ref{section5}, we compute the $4$-pt amplitudes of a generic chiral higher-spin theory.
Finally, in Section \ref{section6}, we conclude with some comments and outline possible directions for future work.

We also include five appendices. In Appendix \ref{AppendixA}, we settle our notation and review standard results of the light-front approach to massless higher-spin fields. In Appendix \ref{AppendixB}, we provide some simple yet useful formulas used in the main text. In Appendix \ref{AppendixC}, we solve the holomorphic constraint for the low-derivative cases. In Appendix \ref{AppendixD}, we consider the holomorphic constraint for the gauge groups $SO(N)$ and $USp(N)$. In Appendix \ref{AppendixE}, we consider the holomorphic constraint for the case where all fields take values in the adjoint representation of some Lie algebra of internal symmetry.

%%%%%%%%%%%%%%%%%%%%%%%%%%%%%%%%%%%%%%%%%%%%%%%%%%%%%%%%%%%%%
\section{Quartic higher-spin Lorentz constraints}\label{section2}
%%%%%%%%%%%%%%%%%%%%%%%%%%%%%%%%%%%%%%%%%%%%%%%%%%%%%%%%%%%%
We begin by reviewing the derivation of the quartic holomorphic constraint.
To construct higher-spin vertices in $4d$ flat space using the light-cone gauge, one begins by expressing the Poincaré algebra generators in terms of free fields. These are initially quadratic in the fields. The idea is to deform them by adding higher-order local terms and require that the modified generators continue to satisfy the Poincaré algebra, order by order in the fields.

An extensive study of this approach to construct consistent interactions for massless higher-spin fields, along with various results, can be found in \cite{Ponomarev:2016lrm}. We review some relevant notation and results in Appendix \ref{AppendixA}.
Here, we begin by recalling the results for the cubic vertices and boost generators, which are fixed up to field redefinitions to be
\begin{align}\label{CubicVertices}
    h_{\lambda_1,\lambda_2,\lambda_3}&=C^{\lambda_1,\lambda_2,\lambda_3}\frac{\PPb^{\lambda_{123}}}{\beta_1^{\lambda_1}\beta_2^{\lambda_2}\beta_3^{\lambda_3}}+\bar{C}^{-\lambda_1,-\lambda_2,-\lambda_3}\frac{\PP^{-\lambda_{123}}}{\beta_1^{-\lambda_1}\beta_2^{-\lambda_2}\beta_3^{-\lambda_3}}\,,\\
    j_{\lambda_1,\lambda_2,\lambda_3}&=\frac{2}{3}\,C^{\lambda_1,\lambda_2,\lambda_3}\frac{\PPb^{\lambda_{123}-1}}{\beta_1^{\lambda_1}\beta_2^{\lambda_2}\beta_3^{\lambda_3}}\Lambda^{\lambda_1,\lambda_2,\lambda_3}\,,\\
    \bar{j}_{\lambda_1,\lambda_2,\lambda_3}&=-\frac{2}{3}\,\bar{C}^{-\lambda_1,-\lambda_2,-\lambda_3}\frac{\PP^{-\lambda_{123}-1}}{\beta_1^{-\lambda_1}\beta_2^{-\lambda_2}\beta_3^{-\lambda_3}}\Lambda^{\lambda_1,\lambda_2,\lambda_3}\,,
\end{align}
where $\lambda_{123}=\lambda_1+\lambda_2+\lambda_3$. Here, $C^{\lambda_1,\lambda_2,\lambda_3}$ are coupling constants and the main task is to determine which ones lead to consistent theories, i.e. which couplings may not be zero and what the relations between various couplings are. Written in this form, the cubic vertices exhibit a clear separation between holomorphic and anti-holomorphic components, corresponding respectively to the terms involving $\PPb$ and $\PP$ in the equations above.
These expressions can be used to construct the Hamiltonian $H$ and the dynamical boost generators $J^{z-}$ and $J^{\bar{z}-}$ via Eqs.~\eqref{hamiltonian} and \eqref{boostz}.

By examining the expressions for $H$ and the dynamical boosts $J^{z-}, J^{\bar{z}-}$, we can assume certain symmetry properties of the couplings $C^{\lambda_1,\lambda_2,\lambda_3}$ to simplify our computations.\footnote{This assumption is harmless, as it can be made without loss of generality.}
Indeed, one can observe that the parity under the exchange $(\lambda_i, q_i) \leftrightarrow (\lambda_j, q_j)$ of both $h^{q_1,q_2,q_3}_{\lambda_1,\lambda_2,\lambda_3}$ and $j^{q_1,q_2,q_3}_{\lambda_1,\lambda_2,\lambda_3}$ is given by $(-)^{\lambda_{123}}$.
Indeed, we have\footnote{For simplicity, we focus only on the holomorphic part, but the same reasoning applies more generally.}
\begin{alignat}{2}
&H_3^{\lambda_1,\lambda_2,\lambda_3}&&=C^{\lambda_1,\lambda_2,\lambda_3}\int d^9q\;\delta\Big(\sum_i q_i\Big)\frac{\PPb^{\lambda_{123}}}{\beta_1^{\lambda_1}\beta_2^{\lambda_2}\beta_3^{\lambda_3}}\phi^{\lambda_1}_{q_1}\phi^{\lambda_2}_{q_2}\phi^{\lambda_3}_{q_3}\,,\\
\nonumber
&H_3^{\lambda_2,\lambda_1,\lambda_3}&&=C^{\lambda_2,\lambda_1,\lambda_3}\int d^9q\;\delta\Big(\sum_i q_i\Big)\frac{\PPb^{\lambda_{123}}}{\beta_1^{\lambda_2}\beta_2^{\lambda_1}\beta_3^{\lambda_3}}\phi^{\lambda_2}_{q_1}\phi^{\lambda_1}_{q_2}\phi^{\lambda_3}_{q_3}\\
    & &&\overset{q_1\leftrightarrow q_2}{=}(-)^{\lambda_{123}}C^{\lambda_2,\lambda_1,\lambda_3}\int d^9 q\;\delta\Big(\sum_i q_i\Big)\frac{\PPb^{\lambda_{123}}}{\beta_1^{\lambda_1}\beta_2^{\lambda_2}\beta_3^{\lambda_3}}\phi^{\lambda_1}_{q_1}\phi^{\lambda_2}_{q_2}\phi^{\lambda_3}_{q_3}\,,
\end{alignat}
and the same symmetry considerations apply to the boost generators. Therefore, we can assume that the coupling constants are symmetric for even-derivative interactions and antisymmetric for odd-derivative ones:
\begin{equation}\label{nocolour_coupling_sym}
    C^{\lambda_1,\lambda_2,\lambda_3}=(-)^{\lambda_{123}}C^{\lambda_{\sigma_1},\lambda_{\sigma_2},\lambda_{\sigma_3}}\,,
\end{equation}
where $\sigma\in\Sigma_3$ represents an odd permutation. Then we can use the following form for the generators:
\begin{align}\label{generatorssym}
    H_3&=\sum_{\lambda_1,\lambda_2,\lambda_3}\int d^9 q\;\delta\Big(\sum_i q_i\Big)h^{q_1,q_2,q_3}_{\lambda_1,\lambda_2,\lambda_3}\,\phi^{\lambda_1}_{q_1}\phi^{\lambda_2}_{q_2}\phi^{\lambda_3}_{q_3}\,,\\
    J^{z-}_3&=\sum_{\lambda_1,\lambda_2,\lambda_3}\int d^9q\;\delta\Big(\sum_i q_i\Big)\Big[j^{q_1,q_2,q_3}_{\lambda_1,\lambda_2,\lambda_3}-\frac{1}{3}\,h^{q_1,q_2,q_3}_{\lambda_1,\lambda_2,\lambda_3}\Big(\sum_j\frac{\partial}{\partial \bar{q}_j}\Big)\Big]\phi^{\lambda_1}_{q_1}\phi^{\lambda_2}_{q_2}\phi^{\lambda_3}_{q_3}\,,
\end{align}
i.e. to sum over all triplets of helicities instead of all distinct (up to permutation) triplets. 
This observation also implies that odd-derivative couplings involving at least two identical fields\footnote{In this context, ``identical'' refers to fields with the same helicity, since no additional quantum numbers are yet considered.} vanish by symmetry  $C^{\lambda,\lambda,\lambda'}\equiv 0$.

The quartic dynamical constraint takes the simple form
\begin{equation}\label{quartic_constarint}
    H_2j_4^{z-}=J_2^{z-}h_4+[H_3,J_3^{z-}]\,.
\end{equation}
We can observe, by examining the degree of homogeneity of $q$, that these two conditions must be satisfied independently
\begin{align}
    &[H_3(\PPb),J_3]=0,&
    &[H_3(\PP),\bar{J}_3]=0\,.
\end{align}
We refer to them, respectively, as holomorphic and anti-holomorphic quartic constraints.

In particular, it is worth noting that if we assume the theory to be purely holomorphic (or anti-holomorphic) and to satisfy the holomorphic quartic constraints, it will be a well-defined theory at any order and will contain only cubic interactions. These are the theories we will study. We can explicitly write down the holomorphic constraint as
\begin{align}\label{HoloexpressionGR}
\begin{split}
[H_3,J_3^{z-}]=&\sum_{\lambda_i,\alpha_j}\int d^9p\;d^9q\; \delta\left(\sum_i q_i\right) \left[j_3^{\lambda_i}(q_i)-\frac{h_3^{\lambda_i}(q_i)}{3}\left(\sum_{k}\frac{\partial}{\partial \bar{q}_k}\right)\right] \times \\
&\delta\left(\sum_j p_j\right)h_3^{\alpha_j}(p_j)\left[\prod_{i=1}^3\phi_{q_i}^{\lambda_i},\prod_{j=1}^3\phi_{p_j}^{\alpha_j}\right]\,.
\end{split}
\end{align}
In computing the Poisson bracket, we get
\begin{equation}
    \sum_{\lambda_1,\lambda_2,\lambda_3}\sum_{\alpha_1,\alpha_2,\alpha_3}[\phi^{\lambda_1}_{q_1}\phi^{\lambda_2}_{q_2}\phi^{\lambda_3}_{q_3},\phi^{\alpha_1}_{p_1}\phi^{\alpha_2}_{p_2}\phi^{\alpha_3}_{p_3}]=9\sum_{\lambda_1,\lambda_2}\sum_{\alpha_1,\alpha_2}\sum_{\omega}\phi^{\lambda_1}_{q_1}\phi^{\lambda_2}_{q_2}\phi^{\alpha_1}_{p_1}\phi^{\alpha_2}_{p_2}[\phi^{\omega}_{q_3},\phi^{-\omega}_{p_3}]\,,
\end{equation}
and the constraint can be brought to the form
\begin{align}
\begin{split}
[H_3,J_3^{z-}]= &\sum_{\lambda_i,\alpha_j}\int d^9p\;d^9q\; \delta\left(\sum_i q_i\right) \delta\left(\sum_j p_j\right)9\,\delta^{\lambda_3,-\alpha_3}\frac{\delta(q_3+p_3)}{2q_3^+}\,\phi^{\lambda_1}_{q_1}\phi^{\lambda_2}_{q_2}\phi^{\alpha_1}_{p_1}\phi^{\alpha_2}_{p_2}\,\times\\
&\left(j_3^{\lambda_i}(q_i)+\sum_{k\neq 3}\frac{\partial}{\partial \bar{q}_{k}}\frac{h_3^{\lambda_i}(q_i)}{3}\right)h_3^{\alpha_j}(p_j)\,.
\end{split}
\end{align}
Once we substitute the explicit form of the generators in \eqref{generatorssym}, we get
\begin{align}\label{holoGR}
    \begin{split}
    [H_3,J_3^{z-}]=&\sum_{\lambda_i,\omega}\int d^{12}q\;\delta \left(\sum_i q_i\right)\frac{9}{2}\Big[(-)^{\omega}\frac{(\lambda_1+\omega-\lambda_2)\beta_1-(\lambda_2+\omega-\lambda_1)\beta_2}{(\beta_1+\beta_2)\beta_1^{\lambda_1}\beta_2^{\lambda_2}\beta_3^{\lambda_3}\beta_4^{\lambda_4}}\,\times\\
    &C^{\lambda_1,\lambda_2,\omega}C^{-\omega,\lambda_3,\lambda_4}\PPb_{12}^{\lambda_{12}+\omega-1}\PPb_{34}^{\lambda_{34}-\omega}\,\phi^{\lambda_1}_{q_1}\phi^{\lambda_2}_{q_2}\phi^{\lambda_3}_{q_3}\phi^{\lambda_4}_{q_4}\Big]\,,
    \end{split}
\end{align}
where $\lambda_{ij}\equiv\lambda_i+\lambda_j$ and we have to remember that the couplings $C^{\lambda_1,\lambda_2,\lambda_3}$ have the symmetry \eqref{nocolour_coupling_sym}.

In general, we can also assume that the fields live in some representation of a gauge group $G$. The simplest option is to take the fields to be $N\times N$ matrices, i.e. $\phi\in\text{Mat}_N$ and make the replacement:
\begin{align}
    &\phi^{\lambda}_{q}\quad\rightarrow\quad (\phi^{\lambda}_q)_a T^a\equiv (\phi^{\lambda}_{q})^A_{\;B}\,,&
    &H_3\sim \phi^{\lambda_1}_{q_1}\phi^{\lambda_2}_{q_2}\phi^{\lambda_3}_{q_3}\quad\rightarrow\quad H_3\sim\mathrm{Tr}(\phi^{\lambda_1}_{q_1}\phi^{\lambda_2}_{q_2}\phi^{\lambda_3}_{q_3})\,.
\end{align}
In the presence of a gauge group the holomorphic constraint takes the form
\begin{equation}
\begin{split}
[H_3,J_3^{z-}]= & \sum_{\lambda_i,\alpha_j}\int d^9p\;d^9q\;\delta\left(\sum_i q_i\right) \left[j_3^{\lambda_i}(q_i)-\frac{h_3^{\lambda_i}(q_i)}{3}\left(\sum_{k}\frac{\partial}{\partial \bar{q}_k}\right)\right] \times \\
 & \delta\left(\sum_j p_j\right)h_3^{\alpha_j}(p_j)\left[\mathrm{Tr}\prod_{i=1}^3\phi_{q_i}^{\lambda_i},\mathrm{Tr}\prod_{j=1}^3\phi_{p_j}^{\alpha_j}\right]\,,
 \end{split}
\end{equation}
and once we compute the Poisson bracket and substitute the explicit form of the generators, we get
\begin{align}\label{holocolour}
    \begin{split}
    [H_3,J_3^{z-}]=&\sum_{\lambda_i,\omega}\int d^{12}q\,\delta\left(\sum_i q_i\right)\frac{9}{2}\Big[(-)^{\omega}\frac{(\lambda_1+\omega-\lambda_2)\beta_1-(\lambda_2+\omega-\lambda_1)\beta_2}{(\beta_1+\beta_2)\beta_1^{\lambda_1}\beta_2^{\lambda_2}\beta_3^{\lambda_3}\beta_4^{\lambda_4}}\times\\
    &C^{\lambda_1,\lambda_2,\omega}C^{-\omega,\lambda_3,\lambda_4}\PPb_{12}^{\lambda_{12}+\omega-1}\PPb_{34}^{\lambda_{34}-\omega}\,\mathrm{Tr}(\phi^{\lambda_1}_{q_1}\phi^{\lambda_2}_{q_2}\phi^{\lambda_3}_{q_3}\phi^{\lambda_4}_{q_4})\Big]\,.
    \end{split}
\end{align}
In this case, we have used the following form for the generators:
\begin{align}\label{generatorscyc}
    H_3=&\;\sum_{\lambda_1,\lambda_2,\lambda_3}\int d^9 q\;\delta\Big(\sum_i q_i\Big)h^{q_1,q_2,q_3}_{\lambda_1,\lambda_2,\lambda_3}\mathrm{Tr}(\phi^{\lambda_1}_{q_1}\phi^{\lambda_2}_{q_2}\phi^{\lambda_3}_{q_3})\,,\\
    J^{z-}_3=&\;\sum_{\lambda_1,\lambda_2,\lambda_3}\int d^9q\;\delta\Big(\sum_i q_i\Big)\Big[j^{q_1,q_2,q_3}_{\lambda_1,\lambda_2,\lambda_3}-\frac{1}{3}h^{q_1,q_2,q_3}_{\lambda_1,\lambda_2,\lambda_3}\Big(\sum_j\frac{\partial}{\partial \bar{q}_j}\Big)\Big]\mathrm{Tr}(\phi^{\lambda_1}_{q_1}\phi^{\lambda_2}_{q_2}\phi^{\lambda_3}_{q_3})\,.
\end{align}
Note that both  $h^{q_1,q_2,q_3}_{\lambda_1,\lambda_2,\lambda_3}$ and $j^{q_1,q_2,q_3}_{\lambda_1,\lambda_2,\lambda_3}$ are symmetric under the cyclic permutation of $(\lambda_i,q_i)$. By applying the same argument as before, we can assume $C^{\lambda_1,\lambda_2,\lambda_3}$ to be cyclic symmetric.

We will now proceed to solve the holomorphic constraint explicitly, both with and without a gauge group.

%%%%%%%%%%%%%%%%%%%%%%%%%%%%%%%%%%%%%%%%%%%%%%%%%%%%%%%%%%%%%
\section{Quartic holomorphic constraint}\label{section3}
%%%%%%%%%%%%%%%%%%%%%%%%%%%%%%%%%%%%%%%%%%%%%%%%%%%%%%%%%%%%%
In this section, we solve the light-cone holomorphic constraint for any integer helicity. We begin by studying the case without a gauge group, and then extend the analysis to several cases where the fields take values in certain representations of  $G=U(N), SO(N),$ and $USp(N)$. Our primary focus will be on the $U(N)$ case, as the others follow a similar pattern. Additional comments on the other cases are provided in Appendix \ref{AppendixD}.
%%%%%%%%%%%%%%%%%%%%%%%%%%%%%%%%%%%%%%%%%%%%%%%%%%%%%%%%%%%%%
\subsection{Light-cone holomorphic constraint}
%%%%%%%%%%%%%%%%%%%%%%%%%%%%%%%%%%%%%%%%%%%%%%%%%%%%%%%%%%%%%
We solve the holomorphic constraint \eqref{holoGR} for fixed integer helicities $\lambda_{1,2,3,4}$ on the external legs, considering both even- and odd-derivative interactions. As we will see, however, Eq.~\eqref{holoGR} may require additional couplings to be nonzero on top of the initial input. These, in turn, can trigger further instances of Eq.~\eqref{holoGR} involving different helicity configurations, leading to an iterative process where new constraints arise. Consequently, finding a consistent solution to the cubic deformation involves more than solving the constraint for fixed external helicities. Complete solutions to \eqref{holoGR} will be presented in the next section.

The analysis begins by following Appendix A of \cite{Ponomarev:2016lrm}. We rewrite the light-cone holomorphic constraint as\footnote{Note that we use a different notation compared to \cite{Ponomarev:2016lrm}, writing the pair of couplings with the exchanged helicity in the middle, as in $C^{\lambda_1,\lambda_2,\omega}C^{-\omega,\lambda_3,\lambda_4}$. This is equivalent, since, as we have seen above, the couplings are always cyclic symmetric.}
\begin{equation}\label{holo2}
    \sum_{\omega}\text{Sym}\left[(-)^{\omega}\frac{(\lambda_1+\omega-\lambda_2)\beta_1-(\lambda_2+\omega-\lambda_1)\beta_2}{\beta_1+\beta_2}\,C^{\lambda_1,\lambda_2,\omega}C^{-\omega,\lambda_3,\lambda_4}\PPb_{12}^{\lambda_{12}+\omega-1}\PPb_{34}^{\lambda_{34}-\omega}\right]=0\,.
\end{equation}
Here, Sym denotes the sum over $6$ distinct contributions
\begin{equation}
    (1234)=\{1234\}+\{1324\}+\{1423\}+\{3412\}+\{2413\}+\{2314\}\,.
\end{equation}
Starting from the variables $\PPb_{ij}$ and $\beta_i$, using momentum conservation, only $5$ of them are independent. For instance, we can choose $\PPb_{12}$, $\PPb_{34}$ and three of the $\beta_i$'s. The various relations can be found in Appendix \ref{AppendixB}.

Now we can sum the two contributions $(1234)\leftrightarrow (3412)$ and $\omega\leftrightarrow-\omega$, and after defining new independent variables
\begin{equation}
    2A=\PPb_{12}+\PPb_{34}=\PPb_{23}-\PPb_{14}\,,\quad
    2B=\PPb_{13}-\PPb_{24}=\PPb_{34}-\PPb_{12}\,,\quad
    2C=\PPb_{14}+\PPb_{23}=-\PPb_{13}-\PPb_{24}\,,
\end{equation}
and using the relations in Appendix \ref{AppendixB}, the $\beta$ dependence disappears
\begin{align}
    \begin{split}
    &\frac{(\lambda_1+\omega-\lambda_2)\beta_1-(\lambda_2+\omega-\lambda_1)\beta_2}{\beta_1+\beta_2}\,\PPb_{34}+\frac{(\lambda_3-\omega-\lambda_4)\beta_3-(\lambda_4-\omega-\lambda_3)\beta_4}{\beta_3+\beta_4}\,\PPb_{12}\\
    &=(\lambda_1-\lambda_2)\PPb_{34}+(\lambda_3-\lambda_4)\PPb_{12}+\frac{\omega}{2}(\PPb_{13}-\PPb_{23}+\PPb_{24}-\PPb_{14})\\
    &=(\lambda_1-\lambda_2+\lambda_3-\lambda_4)A+(\lambda_1-\lambda_2-\lambda_3+\lambda_4)B-2\omega C\,.
    \end{split}
\end{align}
The constraint \eqref{holo2} can then be rewritten as a polynomial in just $3$ independent variables
\begin{align}\label{LCholo}
\nonumber
    &\sum_{\omega}(-)^{\omega}\big[((\lambda_{13}-\lambda_{24})A+(\lambda_{14}-\lambda_{23})B-2\omega C)\,
    \mathcal{C}^{1234\omega}(A-B)^{\lambda_{12}+\omega-1}(A+B)^{\lambda_{34}-\omega-1}\\
    \nonumber
    &+(-)^{\lambda_{24}+\omega}((\lambda_{14}-\lambda_{23})B+(\lambda_{12}-\lambda_{34})C-2\omega A)\,
    \mathcal{C}^{1324\omega}(B-C)^{\lambda_{13}+\omega-1}(B+C)^{\lambda_{24}-\omega-1}\\
    &+((\lambda_{12}-\lambda_{34})C+(\lambda_{13}-\lambda_{24})A-2\omega B)\,
     \mathcal{C}^{1423\omega}(C-A)^{\lambda_{14}+\omega-1}(C+A)^{\lambda_{23}-\omega-1}\big]=0\,,
\end{align}
where $\lambda_{ij}\equiv\lambda_i+\lambda_j$ and $\mathcal{C}^{1234\omega}\equiv C^{\lambda_1,\lambda_2,\omega}C^{-\omega,\lambda_3,\lambda_4}$. We will use the fact that $\mathcal{C}^{1234\omega}$ is a product of two couplings only at the end; for now, we will treat it as a generic function. 

To avoid unnecessary minus signs between products of couplings and extra $i$ factors between couplings, we assign an additional factor of $i$ to odd-helicity fields. This is equivalent to taking even-helicity fields to be Hermitian and odd-helicity fields to be anti-Hermitian matrices (with singlets treated as $1\times 1$ matrices). With this convention, the factor $(-)^{\omega}$ in \eqref{LCholo} is removed.

Before we begin solving the constraint, we set the following definitions:
\begin{align}\label{definitions1}
\nonumber
&k^{1234}_+\equiv(-)^{\lambda_{12}}\sum_{\omega}(-)^{\omega}\mathcal{C}^{1234\omega}\,,\quad
    k^{1234}_-\equiv\sum_{\omega}\mathcal{C}^{1234\omega}\,,\;\;\;\;\;\;\;\quad
    k^{1234}_{\omega+}\equiv(-)^{\lambda_{12}}\sum_{\omega}(-)^{\omega}\omega\,\mathcal{C}^{1234\omega}\,,\\
    &k^{1234}_{\omega-}\equiv\sum_{\omega}\omega\, \mathcal{C}^{1234\omega}\,,\quad
    k^{1234}_{\omega^2+}\equiv(-)^{\lambda_{12}}\sum_{\omega}(-)^{\omega}\omega^2\, \mathcal{C}^{1234\omega}\,,\quad
    k^{1234}_{\omega^2-}\equiv\sum_{\omega}\omega^2\, \mathcal{C}^{1234\omega}\,,\\ \label{definitions2}
    \begin{split}
    &f_+^{1234}(A,B)\equiv(-)^{\lambda_{34}}\sum_{\omega}(-)^{\omega}\mathcal{C}^{1234\omega}(A-B)^{\lambda_{12}+\omega-1}(A+B)^{\lambda_{34}-\omega-1}\,,\\
    &f^{1234}_{\omega+}(A,B)\equiv(-)^{\lambda_{34}}\sum_{\omega}(-)^{\omega}\omega\,\mathcal{C}^{1234\omega}(A-B)^{\lambda_{12}+\omega-1}(A+B)^{\lambda_{34}-\omega-1}\,,\\
    &f_-^{1234}(A,B)\equiv\sum_{\omega}\,\mathcal{C}^{1234\omega}(A-B)^{\lambda_{12}+\omega-1}(A+B)^{\lambda_{34}-\omega-1}\,,\\
    &f^{1234}_{\omega-}(A,B)\equiv\sum_{\omega}\omega\,\mathcal{C}^{1234\omega}(A-B)^{\lambda_{12}+\omega-1}(A+B)^{\lambda_{34}-\omega-1}\,,
     \end{split}
\end{align}
where the upper index $1234$ indicates the order of the external helicities. We can then rewrite the constraint \eqref{LCholo} using the functions \eqref{definitions2} just defined as
\begin{align}\label{short_LCholo}
    \begin{split}
    &\;\;\;\;((\lambda_{13}-\lambda_{24})A+(\lambda_{14}-\lambda_{23})B)f^{1234}_-(A,B)-2\omega Cf^{1234}_{\omega -}(A,B)\\
    &+((\lambda_{14}-\lambda_{23})B+(\lambda_{12}-\lambda_{34})C)f^{1324}_+(B,C)-2\omega Af^{1324}_{\omega+}(B,C)\\
    &+((\lambda_{12}-\lambda_{34})C+(\lambda_{13}-\lambda_{24})A)f^{1423}_-(C,A)-2\omega Bf^{1423}_{\omega-}(C,A)=0\,.
    \end{split}
\end{align}
We divide the solution into four distinct cases, which differ based on the form of the constraint once specific relations among the external helicities are imposed. The general approach proceeds as follows:

\begin{itemize}
\item First, we determine the polynomial form of the functions appearing in \eqref{short_LCholo} --- up to some free coefficients (later identified with those in \eqref{definitions1}) --- by setting to zero all monomials except those that can be generated by at least two different functions $f^{\cdots\cdot}_{\pm}$.

\item Second, we substitute these polynomial expressions back into the constraint \eqref{short_LCholo} to derive relations among the coefficients in \eqref{definitions1}. 

\item Third, once all coefficients in \eqref{definitions1} are fixed, the equation \eqref{LCholo} is solved and we can extract the products of couplings $\mathcal{C}^{1234\omega}$ from the functions $f^{\cdots\cdot}_{\pm}$ via a simple change of variables. 
\end{itemize}
As we will see, we always end up with a system of the general form
\begin{equation}
    \mathcal{C}^{1234\omega}\sim\frac{(\Lambda-2)!}{2^{\Lambda-2}(\lambda_{12}+\omega-1)!(\lambda_{34}-\omega-1)!}\qquad 
    \forall\;\omega\,,\qquad
    \Lambda\equiv \lambda_1+\lambda_2+\lambda_3+\lambda_4\,.
\end{equation}
Assuming this form and using the formulas collected in Appendix \ref{AppendixB}, we obtain the following simple relations, which will be useful later:
\begin{align}\label{usefulrelations}
\nonumber
 &(\lambda_{34}-\lambda_{12})k^{1234}_{\omega+}-2k^{1234}_{\omega^2+}=\frac12\,(2-\Lambda)k^{1234}_+\,,&
   &(\lambda_{34}-\lambda_{12})k^{1234}_{\omega-}-2 k^{1234}_{\omega^2-}=\frac12\,(2-\Lambda)k^{1234}_-\,,\\
    &k^{1234}_{\omega+}=\frac12\,(\lambda_{34}-\lambda_{12})k^{1234}_+\,,&
   &k^{1234}_{\omega-}=\frac12\,(\lambda_{34}-\lambda_{12})k^{1234}_-\,,\\
\nonumber
    &k^{1212}_{\omega^2+}=\frac{\Lambda-2}{4}\,k^{1212}_{+}\,,&
    &k^{1212}_{\omega^2-}=\frac{\Lambda-2}{4}\,k^{1212}_{-}\,.
\end{align}
These relations allow us to solve $k^{\cdots\cdot}_{\omega\pm}$ and $k^{\cdots\cdot}_{\omega^2\pm}$ in terms of $k^{\cdots\cdot}_{\pm}$.
Given the symmetry properties of the couplings \eqref{nocolour_coupling_sym}, we obtain the relations
\begin{align}\label{useful_relations}
    &k^{1234}_{\mp}=k^{2134}_{\pm}=(-)^{\Lambda}k^{1243}_{\pm}\,,&
    &k^{3412}_-=k^{1234}_-\,,&
    &k^{3412}_+=(-)^{\Lambda}k^{1234}_+\,.
\end{align}
In all cases, we will present both the general solution and the solution assuming only even-derivative vertices (i.e. $k^{\cdots\cdot}_+=k^{\cdots\cdot}_-$).  In the following, we focus on $\Lambda\geq 4$; for lower-derivative cases, fewer constraints arise due to the somewhat degenerate structure of the monomials. The cases $\Lambda=2,3$ are analysed separately in Appendix \ref{AppendixC}.

\paragraph{Case 1.} Here, we assume that all helicities are equal, i.e.  $\lambda_1=\lambda_2=\lambda_3=\lambda_4$. The constraint \eqref{LCholo} becomes
\begin{align}\label{case1}
    \begin{split}
    \sum_{\omega}\omega\, \mathcal{C}^{1111\omega}&\big[ C
    (A-B)^{2\lambda+\omega-1}(A+B)^{2\lambda-\omega-1}+(-)^{\omega}A
    (B-C)^{2\lambda+\omega-1}(B+C)^{2\lambda-\omega-1}\\
    &+B (C-A)^{2\lambda+\omega-1}(C+A)^{2\lambda-\omega-1}\big]=0\,.
    \end{split}
\end{align}
First, we determine the polynomial form of the functions \eqref{definitions2} as
\begin{subequations}\label{case11}
\begin{align}
    f^{1111}_{\omega-}(A,B)&=2(k^{1111}_{\omega^2+} AB^{\Lambda-3}-k^{1111}_{\omega^2-}A^{\Lambda-3}B)\,,\\
    f^{1111}_{\omega+}(B,C)&=2(k^{1111}_{\omega^2-} BC^{\Lambda-3}-k^{1111}_{\omega^2+}B^{\Lambda-3}C)\,,\\
    f^{1111}_{\omega-}(C,A)&=2(k^{1111}_{\omega^2+} CA^{\Lambda-3}-k^{1111}_{\omega^2-}C^{\Lambda-3}A)\,.
\end{align}
\end{subequations}
By changing variables and using $S=A+B$, $D=A-B$, we can derive a system of equations for the couplings by solving \eqref{case11} for fixed $\omega$. The resulting system is
\begin{equation}
    \mathcal{C}^{1111\omega}=\frac{k^{1111}_{\omega^2-}+(-)^{\omega}k^{1111}_{\omega^2+}}{2^{\Lambda-4}}  \begin{pmatrix}
        \Lambda-3\\
        2\lambda-\omega-1
    \end{pmatrix}\qquad
    \forall\,\omega\neq 0\,.
\end{equation}
Now, substituting \eqref{case11} into \eqref{case1}, we obtain an additional constraint
\begin{align}
    \begin{split}
   &(k^{1111}_{\omega^2-}-k^{1111}_{\omega^2+})A^{\Lambda-3}BC=0\quad
    \Rightarrow\quad
    k^{1111}_{\omega^2-}=k^{1111}_{\omega^2+}\quad
    \Rightarrow\quad 
    k^{1111}_{\omega^2E}\neq 0,\; k^{1111}_{\omega^2O}=0\,,\\
    &k^{1111}_{\omega^2E}\equiv\sum_{\omega\in\text{even}}\omega^2\, \mathcal{C}^{1111\omega}\,,\qquad
    k^{1111}_{\omega^2O}\equiv\sum_{\omega\in\text{odd}}\omega^2\, \mathcal{C}^{1111\omega}\,.
    \end{split}
\end{align}
Only the solutions with even $\omega$ survive. This was expected, since by symmetry, odd-derivative vertices with at least two equal helicities vanish. The final solution is
\begin{equation}
    \mathcal{C}^{1111\omega}=\frac{k^{1111}_{\omega^2E}(\Lambda-3)!}{2^{\Lambda-5}(2\lambda+\omega-1)!(2\lambda-\omega-1)!}\qquad
    \forall\,\omega\neq 0\,\text{even}\,.
\end{equation}
Consistency is guaranteed by the relation
\begin{equation}
    \sum_{\omega\in\text{even}}\frac{\omega^2(\Lambda-3)!}{2^{\Lambda-5}(2\lambda+\omega-1)!(2\lambda-\omega-1)!}=1\,.
\end{equation}
Note that the product $C^{\lambda,\lambda,0}C^{0,\lambda,\lambda}$ decouples from the system; indeed, it does not contribute to the constant $k^{1111}_{\omega^2E}$ and does not appear in the constraint \eqref{case1}; it is then unconstrained. As we will see, this has important consequences, such as allowing for non-vanishing amplitudes and ensuring the correct relations among the couplings.

\paragraph{Case 2.}
Here, we assume $\lambda_{12}=\lambda_{34}$ and $\lambda_{14}=\lambda_{23}$, which implies $\lambda_1=\lambda_3$ and $\lambda_2=\lambda_4$. The constraint \eqref{LCholo} becomes
\begin{align}\label{case2}
    \begin{split}
    \sum_{\omega}&\big[((\lambda_1-\lambda_2)A-\omega C)\,
    \mathcal{C}^{1212\omega}(A-B)^{\lambda_{12}+\omega-1}(A+B)^{\lambda_{12}-\omega-1}\\
    &-(-)^{\omega}\omega A\,
    \mathcal{C}^{1122\omega}(B-C)^{2\lambda_1+\omega-1}(B+C)^{2\lambda_2-\omega-1}\\
    &+((\lambda_1-\lambda_2)A-\omega B)
     \,\mathcal{C}^{1221\omega}(C-A)^{\lambda_{12}+\omega-1}(C+A)^{\lambda_{12}-\omega-1}\big]=0\,.
     \end{split}
\end{align}
First, we determine the polynomial form of the functions \eqref{definitions2} as
\begin{subequations}\label{case2_funct}
\begin{align}
    \nonumber
    f_-^{1212}(A,B)=&k^{1212}_-A^{\Lambda-2}-2k^{1212}_{\omega-}A^{\Lambda-3}B-k^{1212}_+ B^{\Lambda-2}\\ \label{case2f1}
    =&k^{1212}_-A^{\Lambda-2}-k^{1212}_+ B^{\Lambda-2}\,,\\
    \nonumber
    f^{1212}_{\omega-}(A,B)=&k^{1212}_{\omega-}A^{\Lambda-2}-2k^{1212}_{\omega^2-}A^{\Lambda-3}B+2k^{1212}_{\omega^2+}AB^{\Lambda-3}\\\label{case2f2}
    =&2(k^{1212}_{\omega^2+}AB^{\Lambda-3}-k^{1212}_{\omega^2-}A^{\Lambda-3}B)\,,\\ 
    \nonumber
    f^{1122}_{\omega+}(B,C)=&k^{1122}_{\omega+}B^{\Lambda-2}+2((\lambda_2-\lambda_1)k^{1122}_{\omega+}-k^{1122}_{\omega^2+})B^{\Lambda-3}C\\ \label{case2f3}
    &-2((\lambda_2-\lambda_1)k^{1122}_{\omega-}-k^{1122}_{\omega^2-})BC^{\Lambda-3}-k^{1122}_{\omega-}C^{\Lambda-2}\,.
\end{align}
\end{subequations}
In this case $k^{1212}_{\omega+}$ and $k^{1212}_{\omega-}$ are zero because the sum runs over both $\omega$ and $-\omega$, then
\begin{align}
    &k^{1212}_{\omega+}=(-)^{\lambda_{12}}\sum_{\omega}\omega\, \mathcal{C}^{1212\omega}=0\,,&
    k^{1212}_{\omega-}=\sum_{\omega}(-)^{\omega}\omega\, \mathcal{C}^{1212\omega}=0\,.
\end{align}
The same happens to the coefficients $k^{1221}_{\omega+}$ and $k^{1221}_{\omega-}$.
The functions $f^{1221}_-(C,A)$ and $f^{1221}_{\omega-}(C,A)$ have the same form as $f^{1212}_-(A,B)$ and $f^{1212}_{\omega-}(A,B)$.

Now, by using \eqref{useful_relations}, we have $k_{\pm}^{1212}=k_{\mp}^{1221}$ and by substituting the functions back into \eqref{case2}, we obtain the constraints
\begin{equation}\label{case2conditions}
    k^{1122}_{\omega-}=(\lambda_2-\lambda_1)k^{1212}_-=k^{1122}_{\omega+}\,,
\end{equation}
and
\begin{align}\label{case2further}
    &k^{1122}_{\omega^2-}+(\lambda_1-\lambda_2)k^{1122}_{\omega-}=k^{1212}_{\omega^2-}\,,&
    &k^{1122}_{\omega^2+}+(\lambda_1-\lambda_2)k^{1122}_{\omega+}=k^{1212}_{\omega^2+}\,.
\end{align}
Notice that the only nontrivial constraints are those in \eqref{case2conditions}, while \eqref{case2further} are automatically satisfied, as can be seen using \eqref{usefulrelations}.
From the relations above, we can rewrite \eqref{case2_funct} as
\begin{subequations}
\begin{align}\label{case2f1new2}
    f_-^{1212}(A,B)&=k^{1212}_-A^{\Lambda-2}-k^{1212}_+ B^{\Lambda-2}\,,\\\label{case2f2new2}
    f^{1212}_{\omega-}(A,B)&=2(k^{1212}_{\omega^2+}AB^{\Lambda-3}-k^{1212}_{\omega^2-}A^{\Lambda-3}B)\,,\\\label{case2f3new2}
    f^{1122}_{\omega+}(B,C)&=(\lambda_2-\lambda_1)k^{1212}_-(B^{\Lambda-2}-C^{\Lambda-2})+2(k^{1212}_{\omega^2-}BC^{\Lambda-3}-k^{1212}_{\omega^2+}B^{\Lambda-3}C)\,.
\end{align}
\end{subequations}
From the functions \eqref{case2f1new2} and \eqref{case2f2new2}, we obtain
\begin{align}\label{case2_1212}
     \mathcal{C}^{1212\omega}&=\frac{k^{1212}_-+(-)^{\lambda_{12}+\omega}k^{1212}_+}{2^{\Lambda-2}}\begin{pmatrix}
         \Lambda-2\\
         \lambda_{12}-\omega-1
     \end{pmatrix}\qquad
     \forall\,\omega\,,\\
    \mathcal{C}^{1212\omega}&=\frac{k^{1212}_{\omega^2-}+(-)^{\lambda_{12}+\omega}k^{1212}_{\omega^2+}}{2^{\Lambda-4}(\Lambda-2)}
   \begin{pmatrix}
        \Lambda-2\\
        \lambda_{12}-\omega-1
    \end{pmatrix}
    \qquad
    \forall\,\omega\,.
\end{align}
These systems are, in fact, equivalent due to the relation given in \eqref{usefulrelations}. From the function \eqref{case2f3new2}, we obtain
\begin{equation}\label{case2_1122}
    \mathcal{C}^{1122\omega}=\frac{k^{1212}_-+(-)^{\omega}k^{1212}_+}{2^{\Lambda-2}}
   \begin{pmatrix}
        \Lambda-2\\
        2\lambda_2-\omega-1
    \end{pmatrix}
    \qquad
    \forall\,\omega\neq 0\,.
\end{equation}
Also in this case, the product $C^{\lambda_1,\lambda_1,0}C^{0,\lambda_2,\lambda_2}$ remains unconstrained. Now, assuming we only have even-derivative vertices, we can rewrite \eqref{case2_1212} and \eqref{case2_1122} as
\begin{align}
     \mathcal{C}^{1212\omega}&=\frac{k^{1212}_-(\Lambda-2)!}{2^{\Lambda-3}(\lambda_{12}+\omega-1)!(\lambda_{12}-\omega-1)!}\qquad
     \forall\,\omega\,,\\
     \mathcal{C}^{1122\omega}&=\frac{k^{1212}_-(\Lambda-2)!}{2^{\Lambda-3}(2\lambda_1+\omega-1)!(2\lambda_2-\omega-1)!}\qquad
     \forall\,\omega\neq 0\,.
\end{align}
Here, the sum runs over odd or even values of $\omega$, depending on the parity of the external helicities. Everything discussed above holds for $\Lambda \geq 5$. For $\Lambda = 4$, although some monomial powers become degenerate, the conditions in \eqref{case2conditions} are still satisfied, leading to the same solution.

\paragraph{Case 3.}
Here, we assume $\lambda_{12}=\lambda_{34}$, in which case the constraint becomes
\begin{align}\label{case3}
\begin{split}
    \sum_{\omega}&\big[((\lambda_{13}-\lambda_{24})A+(\lambda_{14}-\lambda_{23})B-2\omega C)\,
    \mathcal{C}^{1234\omega}(A-B)^{\lambda_{12}+\omega-1}(A+B)^{\lambda_{12}-\omega-1}\\
    &+(-)^{\lambda_{24}+\omega}((\lambda_{14}-\lambda_{23})B-2\omega A)\,
    \mathcal{C}^{1324\omega}(B-C)^{\lambda_{13}+\omega-1}(B+C)^{\lambda_{24}-\omega-1}\\
    &+((\lambda_{13}-\lambda_{24})A-2\omega B)\,
     \mathcal{C}^{1423\omega}(C-A)^{\lambda_{14}+\omega-1}(C+A)^{\lambda_{23}-\omega-1}\big]=0\,.
     \end{split}
\end{align}
First, we determine the polynomial form of the functions \eqref{definitions2} as
\begin{subequations}\label{case_3func}
\begin{align}\label{case3f1}
   f_-^{1234}(A,B)&=k^{1234}_-A^{\Lambda-2}-k^{1234}_+B^{\Lambda-2}\,,\\ \label{case3f2}
   f_+^{1324}(B,C)&=(-)^{\Lambda}(k^{1324}_+B^{\Lambda-2}+((\lambda_{24}-\lambda_{13})k^{1324}_+-2k^{1324}_{\omega+})CB^{\Lambda-3}-k^{1324}_-C^{\Lambda-2})\,,\\ \label{case3f3}
   f_-^{1423}(C,A)&=k^{1423}_-C^{\Lambda-2}-((\lambda_{23}-\lambda_{14})k^{1423}_+-2k^{1423}_{\omega+})CA^{\Lambda-3}-k^{1423}_+A^{\Lambda-2}\,,\\ \label{case3f4}
   \nonumber
    f^{1234}_{\omega-}(A,B)&=k^{1234}_{\omega-}A^{\Lambda-2}+((\lambda_{34}-\lambda_{12})k^{1234}_{\omega-}-2k^{1234}_{\omega^2-})A^{\Lambda-3}B\\
    &-((\lambda_{34}-\lambda_{12})k^{1234}_{\omega+}-2k^{1234}_{\omega^2+})AB^{\Lambda-3}-k^{1234}_{\omega+}B^{\Lambda-2}\,.
\end{align}
\end{subequations}
The functions $f_{\omega+}^{1324}$ and $f_{\omega-}^{1423}$ have the same form as $f_{\omega-}^{1234}$. We can now substitute these functions into \eqref{case3}, leading to the constraints
\begin{align}\label{case3conditions}
    \begin{split}
    &k^{1234}_-=k^{1423}_+,\qquad
    k^{1324}_+=(-)^{\Lambda}k^{1234}_+,\qquad
    k^{1234}_{\omega+}=k^{1234}_{\omega-}=0\,,\\
    &k^{1324}_{\omega-}=\frac{1}{2}(-)^{\Lambda}(\lambda_{24}-\lambda_{13})k^{1423}_-\,,\qquad
    k^{1423}_{\omega-}=\frac{1}{2}(-)^{\Lambda}(\lambda_{23}-\lambda_{14})k^{1324}_-\,,\\
    &((\lambda_{23}-\lambda_{14})k^{1423}_{\omega-}-2k^{1423}_{\omega^2-})=(-)^{\Lambda}((\lambda_{24}-\lambda_{13})k^{1324}_{\omega-}-2k^{1324}_{\omega^2-})\,,\\
    \end{split}
\end{align}
and
\begin{equation}
\begin{aligned}\label{case3further}
    &k^{1324}_{\omega+}=\frac{1}{2}(\lambda_{24}-\lambda_{13})k^{1324}_+\,,\qquad
    &&k^{1423}_{\omega+}=\frac{1}{2}(\lambda_{23}-\lambda_{14})k^{1423}_+\,,\\
    &k^{1234}_{\omega^2-}=\frac{1}{2}((\lambda_{14}-\lambda_{23})k^{1423}_{\omega+}+2k^{1423}_{\omega^2+})\,,
    &&k^{1234}_{\omega^2+}=\frac{1}{2}(-)^{\Lambda}((\lambda_{23}-\lambda_{14})k^{1324}_{\omega+}+2k^{1324}_{\omega^2+})\,.
\end{aligned}
\end{equation}
Using the relations above, we can rewrite \eqref{case_3func} as
\begin{subequations}
\begin{align}\label{case3f1new}
    f_-^{1234}(A,B)&=k^{1234}_-A^{\Lambda-2}-k^{1234}_+B^{\Lambda-2}\,,\\ \label{case3f2new}
    f^{1324}_+(B,C)&=(-)^{\Lambda}(k^{1324}_+B^{\Lambda-2}-k^{1324}_-C^{\Lambda-2})\,,\\ \label{case3f3new}
    f^{1423}_-(C,A)&=k^{1423}_-C^{\Lambda-2}-k^{1423}_+A^{\Lambda-2}\,,\\ \label{case3f4new}
    f_{\omega-}^{1234}(A,B)&=2(k_{\omega^2+}^{1234}AB^{\Lambda-3}-k_{\omega^2-}^{1234}A^{\Lambda-3}B)\,. 
\end{align}
\end{subequations}
Notice that the only nontrivial constraints are those in \eqref{case3conditions}, while \eqref{case3further} are automatically
satisfied, as can be seen using \eqref{usefulrelations}.
From the function \eqref{case3f1new}, we find
\begin{equation}
    \mathcal{C}^{1234\omega}=\frac{(k^{1234}_- +(-)^{\omega+\lambda_{12}}k^{1234}_+)(\Lambda-2)!}{2^{\Lambda-2}(\lambda_{12}+\omega-1)!(\lambda_{34}-\omega-1)!}\qquad 
    \forall\;\omega\,,
\end{equation}
and an identical expression holds for the external helicity configurations $(1324)$ and $(1423)$.

Now, assuming only even-derivative vertices, from the constraints \eqref{case3conditions} we obtain
\begin{equation}
    k^{1234}_+=k^{1324}_+=k^{1423}_+\,,
\end{equation}
that leads to the solution
\begin{equation}
    \mathcal{C}^{1234\omega}=\frac{k^{1234}_+(\Lambda-2)!}{2^{\Lambda-3}(\lambda_{12}+\omega-1)!(\lambda_{34}-\omega-1)!}\qquad 
    \forall\;\omega\,.
\end{equation}
The same solution applies to the other two orderings of external helicities $(1324)$ and $(1423)$. Everything discussed above holds for $\Lambda \geq 5$. For $\Lambda = 4$, although some monomial powers become degenerate, the conditions in \eqref{case3conditions} are still satisfied, leading to the same solution.

\paragraph{Case 4.}
Here, we solve \eqref{LCholo} assuming generic helicities. First, we determine the polynomial form of the functions \eqref{definitions2} as
\begin{subequations}\label{case4_func}
\begin{align}\label{case4f1}
   f_-^{1234}(A,B)=&k^{1234}_-A^{\Lambda-2}-k^{1234}_+B^{\Lambda-2}\,,\\ \label{case4f2}
   \nonumber
    f^{1234}_{\omega-}(A,B)=&k^{1234}_{\omega-}A^{\Lambda-2}+((\lambda_{34}-\lambda_{12})k^{1234}_{\omega-}-2k^{1234}_{\omega^2-})A^{\Lambda-3}B\\
    &+((\lambda_{12}-\lambda_{34})k^{1234}_{\omega+}+2k^{1234}_{\omega^2+})AB^{\Lambda-3}-k^{1234}_{\omega+}B^{\Lambda-2}\,.
\end{align}
\end{subequations}
The functions $f_+^{1324}(B,C)$ and $f_-^{1423}(C,A)$ have the same form as $f_-^{1234}(A,B)$, and the functions $f_{\omega+}^{1324}(B,C)$ and $f_{\omega-}^{1423}(C,A)$ have the same form as $f^{1234}_{\omega-}(A,B)$.

From equation \eqref{case4f1}, we can extract the system for the couplings and by substituting the functions \eqref{case4_func} into \eqref{LCholo}, also the relations among them:
\begin{align}\label{case4system1}
    \begin{split}
&\mathcal{C}^{1234\omega}=\frac{(k^{1234}_- +(-)^{\omega+\lambda_{12}}k^{1234}_+)(\Lambda-2)!}{2^{\Lambda-2}(\lambda_{12}+\omega-1)!(\lambda_{34}-\omega-1)!}\qquad 
    \forall\;\omega\,,\quad \text{same for $(1324)$ and $(1423)$}\,,\\
     &k^{1234}_-=k^{1423}_+\,,\qquad
    k^{1324}_+=(-)^{\Lambda}k^{1234}_+\,,\qquad
   k^{1423}_-=(-)^{\Lambda}k^{1324}_-\,.
   \end{split}
\end{align}
This gives the most general solution to the constraint \eqref{LCholo} in the presence of both even- and odd-derivative couplings. We also find these further constraints
\begin{align}\label{case4further}
    \begin{split}
   &k^{1234}_{\omega-}=\frac{1}{2}(\lambda_{34}-\lambda_{12})k^{1234}_-\,,\quad
   k^{1234}_{\omega+}=\frac{1}{2}(\lambda_{34}-\lambda_{12})k^{1234}_+\,,\quad
   k^{1324}_{\omega+}=\frac{1}{2}(\lambda_{24}-\lambda_{13})k^{1324}_+\,,\\
   &k^{1324}_{\omega-}=\frac{1}{2}(\lambda_{24}-\lambda_{13})k^{1324}_-\,,\quad
   k^{1423}_{\omega-}=\frac{1}{2}(\lambda_{23}-\lambda_{14})k^{1423}_-\,,\quad
   k^{1423}_{\omega+}=\frac{1}{2}(\lambda_{23}-\lambda_{14})k^{1423}_+\,,\\
   &(\lambda_{34}-\lambda_{12})k^{1234}_{\omega-}-2 k^{1234}_{\omega^2-}=((\lambda_{23}-\lambda_{14})k^{1423}_{\omega+}-2 k^{1423}_{\omega^2+})\,,\\
   &(\lambda_{24}-\lambda_{13})k^{1324}_{\omega+}-2 k^{1324}_{\omega^2+}=(-)^{\Lambda}((\lambda_{34}-\lambda_{12})k^{1234}_{\omega+}-2 k^{1234}_{\omega^2+})\,,\\
   &(\lambda_{23}-\lambda_{14})k^{1423}_{\omega-}-2 k^{1423}_{\omega^2-}=(-)^{\Lambda}((\lambda_{24}-\lambda_{13})k^{1324}_{\omega-}-2 k^{1324}_{\omega^2-})\,.
   \end{split}
\end{align}
However, these are automatically satisfied, as can be seen using \eqref{usefulrelations}. Everything discussed above holds for $\Lambda \geq 5$. For $\Lambda = 4$, although some monomial powers become degenerate, the conditions in \eqref{case4system1} are still satisfied, leading to the same solution.

The next problem is to find a meaningful solution\footnote{By this, we mean a solution that involves several couplings, perhaps, including the gravitational ones.} to \eqref{case4system1}. This can be done by examining each case individually; however, finding a general procedure to classify all solutions is not straightforward.

For instance, we observe that a solution involving only odd-derivative vertices (i.e. $k^{\cdots\cdot}_+=-k^{\cdots\cdot}_-$) is inconsistent with \eqref{case4system1}. Consequently, any consistent solution must involve a nontrivial combination of both even- and odd-derivative couplings (or only even).

Let us rewrite the coefficients in a more convenient form, which will facilitate the identification of additional constraints:
\begin{align}\label{new_variablesE}
    k^{1234}_E=\frac{1}{2}(k^{1234}_-+k^{1234}_+)&=\sum_{(\lambda_{12}+\omega)\in\text{even}}\mathcal{C}^{1234\omega}\,,\\ \label{new_variablesO}
    k^{1234}_O=\frac{1}{2}(k^{1234}_--k^{1234}_+)&=\sum_{(\lambda_{12}+\omega)\in\text{odd}}\mathcal{C}^{1234\omega}\,.
\end{align}
We can then rewrite the system governing the couplings in the form
\begin{align}\label{even_first}
    \mathcal{C}^{1234\omega}&=\frac{k^{1234}_E (\Lambda-2)!}{2^{\Lambda-3}(\lambda_{12}+\omega-1)!(\lambda_{34}-\omega-1)!}\qquad 
    \forall\;(\lambda_{12}+\omega)\in\text{even}\,,\\ \label{odd_first}
    \mathcal{C}^{1234\omega}&=\frac{k^{1234}_O (\Lambda-2)!}{2^{\Lambda-3}(\lambda_{12}+\omega-1)!(\lambda_{34}-\omega-1)!}\qquad 
    \forall\;(\lambda_{12}+\omega)\in\text{odd}\,.
\end{align}
Recall that $\mathcal{C}^{1234\omega}$ is a product of two terms. Therefore, following the procedure outlined in Appendix A of \cite{Ponomarev:2016lrm}, assuming that all even couplings are non-vanishing, and using \eqref{even_first} we find that the unique solution (Metsaev solution) for the couplings is
\begin{equation}\label{alleven}
    C^{\lambda_1,\lambda_2,\lambda_3}\sim \frac{1}{(\lambda_1+\lambda_2+\lambda_3-1)!}\,,\qquad
    \sum_{i=1}^3\lambda_i\in\text{even}\,.
\end{equation}
Similarly, assuming that all odd-derivative vertices are activated and using \eqref{odd_first}, we obtain
\begin{equation}\label{allodd}
    C^{\lambda_1,\lambda_2,\lambda_3}\sim \frac{\epsilon^{\lambda_1\lambda_2\lambda_3}}{(\lambda_1+\lambda_2+\lambda_3-1)!}\,,\qquad
    \sum_{i=1}^3\lambda_i\in\text{odd}\,,
\end{equation}
where $\epsilon^{\lambda_1\lambda_2\lambda_3}$ is a totally antisymmetric tensor. Using \eqref{new_variablesE} and \eqref{new_variablesO}, the constraint \eqref{case4system1} becomes
\begin{align}
    \begin{split}
    k^{1234}_E+k^{1234}_O&=k^{1423}_E-k^{1423}_O\,,\qquad
   k^{1324}_E-k^{1324}_O=(-)^{\Lambda}(k^{1234}_E-k^{1234}_O)\,,\\
   k^{1423}_E+k^{1423}_O&=(-)^{\Lambda}(k^{1324}_E+k^{1324}_O)\,.
   \end{split}
\end{align}
Now, assuming \eqref{alleven}, which implies $k^{1234}_E=k^{1324}_E=k^{1423}_E$, we arrive at
\begin{align}
    &k^{1234}_O=-k^{1423}_O\,,&
    &k^{1234}_O=(-)^{\Lambda}k^{1324}_O=k^{1423}_O\,,
\end{align}
that implies $k^{\cdots\cdot}_O=0$. Therefore, in the presence of all even-derivative couplings, no consistent solution can accommodate the inclusion of any odd-derivative interactions.

On the other hand, assuming \eqref{allodd}, we have $k^{1234}_O=k^{1324}_O=k^{1423}_O=k_O$, which leads to
\begin{align}
\begin{split}
&k^{1234}_E-k^{1423}_E+2k_O=0,\qquad
k^{1324}_E+(-)^{\Lambda+1}k^{1234}_E+(-)^{\Lambda}k_O-k_O=0\,,\\
&k^{1423}_E+(-)^{\Lambda+1}k^{1324}_E+(-)^{\Lambda+1}k_O+k_O=0\,.
\end{split}
\end{align}
Attempting to find a solution for even or odd $\Lambda$ leads to $k_O=0$. Therefore, this case is also ruled out.

We conclude that any viable solution must involve a truncated spectrum --- i.e. one that cannot contain all even- and/or odd-derivative couplings --- and must feature nontrivial relations among the couplings. A detailed investigation of this possibility is left for future work. From this point onward, we restrict our analysis to the case in which only even-derivative vertices are present.

When only even-derivative vertices are present (i.e. $k^{\cdots\cdot}_+=k^{\cdots\cdot}_-)$, the solution to \eqref{case4system1} yields
\begin{equation}\label{unique_even_solution}
    k^{1234}_+=k^{1324}_+=k^{1423}_+=k^{1234}_-=k^{1324}_-=k^{1423}_-\,,
\end{equation}
and these are the solutions we will study in detail.

\paragraph{Summary.} The general solution to the holomorphic constraint \eqref{LCholo} is given by
\begin{equation}
\boxed{
\begin{aligned}\label{case4system}
&\mathcal{C}^{1234\omega}=\frac{(k^{1234}_- +(-)^{\omega+\lambda_{12}}k^{1234}_+)(\Lambda-2)!}{2^{\Lambda-2}(\lambda_{12}+\omega-1)!(\lambda_{34}-\omega-1)!}\quad 
    \forall\;\omega\,,\quad \text{same for $(1324)$ and $(1423)$}\,,\\ 
    &k^{1234}_-=k^{1423}_+\,,\;\;
    k^{1324}_+=(-)^{\Lambda}k^{1234}_+\,,\;\;
   k^{1423}_-=(-)^{\Lambda}k^{1324}_-\,,\;\;C^{\lambda_1,\lambda_1,0}C^{0,\lambda_2,\lambda_2}=\;\text{generic}\,.
\end{aligned}
}
\end{equation}
Assuming only even-derivative vertices, the solution to the holomorphic constraint is given by
\begin{equation}
\boxed{
\begin{aligned}\label{symfinalsystem}
    &\mathcal{C}^{1234\omega}=\frac{k^{1234}_+(\Lambda-2)!}{2^{\Lambda-3}(\lambda_{12}+\omega-1)!(\lambda_{34}-\omega-1)!}\quad 
    \forall\;\omega\,,\quad \text{same for $(1324)$ and $(1423)$}\,,\\
    &k_+^{1234}= k_+^{1324}= k_+^{1423}\,,\qquad
    C^{\lambda_1,\lambda_1,0}C^{0,\lambda_2,\lambda_2}=\;\text{generic}\,.
\end{aligned}
}
\end{equation}
Let us stress that $\mathcal{C}^{1234\omega}$ is not assumed to factorize into $C^{\lambda_1,\lambda_2,\lambda_3}$.

\paragraph{Lower-spin analysis.}
We begin by examining solutions for lower helicities, specifically $0,1,2$. The corresponding even-derivative couplings are given by
\begin{equation}
    \{C^{-2,2,2},C^{-1,1,2},C^{0,0,2},C^{0,1,1},C^{0,2,2},C^{1,1,2},C^{2,2,2}\}\,.
\end{equation}
In total, we have $7$ couplings: $5$ abelian and $2$ non-abelian.\footnote{The non-abelian vertices are the most interesting ones, as they deform the gauge algebra in a covariant formulation. The abelian ones do not. Nevertheless, it is worth noting that, in general, abelian vertices can also be constrained and may play an important role in ensuring the consistency of the theory. Since we work in the light-cone gauge and with holomorphic theories, we define abelian couplings as those with $\lambda_i\geq0$ and $\sum_i\lambda_i>0$. Note that $(0,s,s)$ is still considered abelian this way, which is justified since the spin-zero exchange disappears from the constraint. Such couplings are, obviously, consistent on their own unless there are non-abelian couplings. } We can now explore all possible chiral theories that can be constructed from these couplings.

We begin with two-derivative theories, involving the couplings $\{C^{-2,2,2},C^{-1,1,2},C^{0,0,2},C^{0,1,1}\}$.
All possible two-derivative chiral theories with lower-spin fields that solve the constraints are
\begin{subequations}\label{lower_der_lower_spin}
\begin{align}
        &\{C^{-2,2,2}\}\,,&&\text{graviton coupling}\,,\\
        &\{C^{1,1,0}\}\,,&&\text{photons coupled to scalars}\,,\\
        &\{C^{-2,2,2}=C^{0,0,2}\}\,,&&\text{graviton coupled to scalars}\,,\\\label{case4lowerspin}
        &\{C^{-2,2,2}=C^{-1,1,2}\}\,,&&\text{graviton coupled to photons}\,,\\\label{case5lowerspin}
        &\{C^{-2,2,2}=C^{-1,1,2}=C^{0,0,2},C^{0,1,1}\}\,,&& \text{graviton, photons and scalars}\,.
\end{align}
\end{subequations}
We use the following notation: within $\{\cdots\}$, we list the active couplings. When a coupling appears alone, it is considered unconstrained --- that is, it can appear in the theory’s action with a free coefficient. However, the constraints may sometimes impose nontrivial relations among these couplings. It is important to specify the active couplings before solving the constraints. Notably, the resulting theories do not always correspond to truncations of the larger one. For instance, the above theories do not always correspond to a truncation of the larger one \eqref{case5lowerspin} by setting to zero certain couplings, but rather to distinct theories. For example, to obtain \eqref{case4lowerspin} from \eqref{case5lowerspin}, we remove $C^{0,0,2}$, which is not equivalent to setting it to zero in \eqref{case5lowerspin}.
Note that the relations above follow from the well-known universality of gravitational interactions.

If we allow for the higher-derivative terms, the most general theory we obtain is
\begin{equation}\label{HDlowerspin}
    \{C^{-2,2,2}=C^{-1,1,2}=C^{0,0,2},C^{0,1,1},C^{0,2,2},C^{1,1,2},C^{2,2,2}\}\,.
\end{equation}
The fact that the couplings satisfy the constraints implies that a consistent action can be written in the light-cone gauge. In principle,\footnote{In some cases, a covariant formulation of the theory may not exist or may require introducing non-localities. However, a covariant description can often be found, as is the case here.} it may be possible to find a covariant formulation of the theory with an appropriate gauge symmetry.

Let us consider, for example, the largest two-derivative theory \eqref{case5lowerspin}. In this case, the covariant (parity completed) action corresponds to the truncation to the cubic order of the Einstein-Maxwell-scalar theory
\begin{equation}
    S=\int d^4x\sqrt{-g}\left(\frac{1}{2k}R-\frac{1}{2}\nabla^{\mu}\varphi\nabla_{\mu}\varphi-\frac{1}{4}F^{\mu\nu}F_{\mu\nu}+\frac{1}{4}a\,\varphi F^{\mu\nu}F_{\mu\nu}\right),
\end{equation}
where $\nabla$ is the gravitational covariant derivative, $g$ is the determinant of the metric, $k = 8\pi G$ is the gravitational coupling, $R$ is the Ricci scalar, $\varphi$ is the scalar  field, and $F_{\mu\nu} = \pl_{\mu} A_{\nu} - \pl_{\nu} A_{\mu}$ is the Maxwell–Faraday $2$-form. The parameter $a$ is the coupling constant between the scalar and the vector field. The values of the coefficients in the action are directly related to those appearing in light-cone gauge: we have $k \sim C^{-2,2,2}$, while the relations among the other couplings are determined by minimal couplings imposed by diffeomorphism invariance, and $a \sim C^{0,1,1}$.
Note that we can always include a scalar cubic coupling $C^{0,0,0}$, corresponding to a potential term of the form $V(\varphi) \sim \varphi^3$.
As expected from the covariant perspective, most of the abelian couplings remain unconstrained.\footnote{This is a typical feature of lower-spin theories. As we introduce higher-spin interactions, we will see that many abelian couplings become constrained as well.}

The covariant (parity completed) higher-derivative action of \eqref{HDlowerspin} corresponds to the truncation to the cubic order of 
\begin{align}
\nonumber
    S&=\int d^4x\sqrt{-g}\left(\frac{1}{2k}R-\frac{1}{2}\nabla^{\mu}\varphi\nabla_{\mu}\varphi-\frac{1}{4}F^{\mu\nu}F_{\mu\nu}+\frac{1}{4}a\,\varphi F^{\mu\nu}F_{\mu\nu}+b R^3 +c RF^2+dR^2\varphi\right)\,,\\
    \nonumber
    &R^3=R_{\mu\nu\rho\sigma}R^{\mu\nu}_{\phantom{\mu\nu}\alpha\beta}R^{\rho\sigma\alpha\beta}\,,\qquad
    RF^2=R_{\mu\nu\rho\sigma}F^{\mu\nu}F^{\rho\sigma}\,,\qquad
    R^2=R_{\mu\nu\rho\sigma}R^{\mu\nu\rho\sigma}\,,\\
    & k\sim C^{-2,2,2}\,,\qquad 
    a\sim C^{0,1,1}\,,\qquad 
    b\sim C^{2,2,2}\,,\qquad 
    c\sim C^{1,1,2}\,,\qquad 
    d\sim C^{0,2,2}\,.
\end{align}
It is worth emphasising the importance of the freedom in the product $C^{\lambda_1,\lambda_1,0}C^{0,\lambda_2,\lambda_2}$. Without this freedom, we would obtain the constraint (as in the case of chiral higher-spin gravity)
\begin{align}\label{wrong_condition}
    &C^{2, 2, 2}=\frac{3}{10}\frac{(C^{0, 2, 2})^2}{C^{-2, 2, 2}}\,,&
    &C^{1, 1, 2}=\frac{C^{0, 1, 1} C^{0, 2, 2}}{C^{-2, 2, 2}}\,.
\end{align}
This would be in contrast with the covariant framework, where it is perfectly possible to introduce both $R^3$ and $FR^2$ terms with arbitrary coefficients in a Lorentz-invariant way. This is a further consistency check for our solutions.

We also point out that a method for constructing covariant chiral Lagrangians was recently developed in \cite{Krasnov:2021nsq}, using the spinorial representation of the Lorentz group in four dimensions, $SO(3,1)$. This formalism appears to be the most suitable for formulating Lagrangian descriptions of all consistent chiral higher-spin theories identified in the light-cone approach.

\paragraph{Higher-spin analysis.}
We now begin to search for complete solutions involving higher-spin fields. In particular, we need to require that $\mathcal{C}^{1234\omega}$ factorizes into the product of two couplings. While it is commonly believed that the presence of higher-spin fields imposes strong constraints on the theory, requiring an infinite spectrum of fields, we will show that this is not always the case. In fact, we can find many interesting families of solutions, some with finite spectra and others with infinite ones. Well-known examples such as HS-SDGR, HS-SDYM, and chiral higher-spin gravity will appear as special cases.

To better understand the underlying logic, particularly for higher-derivative interactions, we will sometimes label the couplings by their number of derivatives, rather than by the specific helicities. We denote a generic $n$-derivative coupling as $C^n$, allowing us to schematically rewrite the system of couplings \eqref{symfinalsystem} as
\begin{equation}\label{LCcouplingder}
C^nC^m=C^mC^n=\frac{k_{\Lambda}(\Lambda-2)!}{2^{\Lambda-3}(n-1)!(m-1)!}\,,\qquad
k_{\Lambda}=\sum_{\omega}\mathcal{C}^{1234\omega}\,,\qquad
\Lambda=n+m\,.
\end{equation}
Remember, we consider $n,m$ even and $n,m\geq 2$.
The key observation is that \eqref{LCholo} decomposes into independent constraints at fixed total derivative order $\Lambda$; the same can be seen from the system above. Let us assume a chiral theory that includes $2$- and $4$-derivative interactions. Using \eqref{LCcouplingder}, we can write the system (very schematically)\footnote{We emphasize that this is just a sketch to understand the main logic behind the existence of a truncated spectrum, especially for higher derivative theories. To identify consistent theories, one must solve all the constraints explicitly. This will be carried out in the next section for lower-derivative theories.} as
\begin{equation}
    \begin{cases}
    C^2C^2=C^2C^2\,,\\
    C^2C^4=C^2C^4\,,\\
    C^4C^4=\frac{5}{8}(C^4C^4+2\,C^2C^6)\,.
    \end{cases}
\end{equation}
The first line corresponds to the $2$-derivative sector, which is consistent on its own and includes HS-SDGR and other theories all contained in the former. However, problems arise when we try to go beyond $2$-derivative interactions.

If we include $4$-derivative interactions, the second constraint above is consistent and admits solutions, but the third equation becomes inconsistent unless $6$-derivative couplings are introduced.
Proceeding to the next order, we can resolve the previous inconsistency, but another one will arise:
\begin{equation}
    \begin{cases}
    C^4C^4=\frac{5}{8}(C^4C^4+2\,C^2C^6)\,,\\
    C^2C^6=\frac{3}{16}(C^4C^4+2\,C^2C^6)\,,\\
    C^4C^6=\frac{7}{16}(2\,C^4C^6+2\,C^2C^8)\,,\\
    C^6C^6=\frac{63}{128}(C^6C^6+2\,C^4C^8+2\,C^2C^{10})\,.\\
    \end{cases}
\end{equation}
The first two equations are consistent and fix the previous problem. However, truncating the theory at $6$ derivatives makes the remaining equations inconsistent, thereby requiring the inclusion of terms up to  $10$-derivative couplings. This structure arises from the identity
\begin{equation}
    \sum_{\omega}\frac{(\Lambda-2)!}{2^{\Lambda-3}(\lambda_{12}+\omega-1)!(\lambda_{34}-\omega-1)!}=1\qquad 
    \forall\;\omega\,,
\end{equation}
which underlies the consistency of the system \eqref{symfinalsystem}.

This pattern persists at all orders: truncating the theory at any finite number of derivatives generally introduces inconsistencies at higher orders, pointing toward a unique solution with all couplings turned on, which is the full chiral higher-spin theory.

However, there is an alternative. One can impose that lower-derivative couplings do not generate higher-order constraints. For instance, a theory involving only $C^2$ and $C^4$  couplings must be arranged so that no $C^4C^4$ term ever arises. Surprisingly, this condition can be satisfied, and it provides a pathway for classifying all consistent chiral higher-spin theories in $4d$ flat space.

%%%%%%%%%%%%%%%%%%%%%%%%%%%%%%%%%%%%%%%%%%%%%%%%%%%%%%%%%%%%%
\subsection{Light-cone holomorphic constraint with U(N) gauge group}
%%%%%%%%%%%%%%%%%%%%%%%%%%%%%%%%%%%%%%%%%%%%%%%%%%%%%%%%%%%%%
Here, we solve the holomorphic constraint in the presence of a $U(N)$ gauge group in the most general setting, allowing for both even- and odd-derivative vertices. The generators of the $u(N)$ algebra satisfy the Poisson bracket
\begin{equation}\label{U(N)poisson_bracket}
  [(\phi^{\lambda}_p)^A_{\;B},(\phi^{\mu}_q)^C_{\;D}]=\frac{\delta^{\lambda,-\mu}\delta^3(p+q)}{2p^+}\,\theta_{\lambda}\delta^C_{\;B}\delta^A_{\;D}\,,
\end{equation}
where $\theta_{\lambda}=e^{ix\lambda}$ is a phase factor reflecting a possible ambiguity in the commutator.

The constraint is given in \eqref{holocolour}. Due to the trace, we can decompose the contributions into several colour-ordered terms, as
\begin{align}\label{colour-ordered_terms}
    \begin{split}
    (1234)=&\;[1234]+[1342]+[1423]+[1324]+[1243]+[1432]\,,\\
    [1234]=&\;\{1234\}+\{2341\}+\{3412\}+\{4123\}\,.
    \end{split}
\end{align}
We will focus on a single colour-ordered term, $[1234]$, as the same will follow for the other orders. The constraint for $[1234]$ takes the form
\begin{equation}\label{Unconstraint}
\sum_{\omega}\text{Cycl}\Big[(-)^{\omega}\theta_{\omega}\frac{(\lambda_1+\omega-\lambda_2)\beta_1-(\lambda_2+\omega-\lambda_1)\beta_2}{\beta_1+\beta_2}\,
\mathcal{C}^{1234\omega}\PPb_{12}^{\lambda_{12}+\omega-1}\PPb_{34}^{\lambda_{34}-\omega}\Big]=0\,.
\end{equation}
As before, we can apply the trick of summing the terms $(1234)\leftrightarrow (3412)$ and $\omega\leftrightarrow -\omega$, followed by a change of variables to $A,B,C$. However, we must now account for the phase $\theta_{\omega}$, and the simplification works only if we have $\theta_{\omega}=\theta_{-\omega}$. This is the case due to the antisymmetry of the Poisson bracket.\footnote{Using the antisymmetric property of the Poisson bracket, we have
\begin{equation}
    [(\phi^{\lambda}_p)^A_{\;B},(\phi^{\mu}_q)^C_{\;D}]=\frac{\delta^{\lambda,-\mu}\delta^3(p+q)}{2p^+}\,\theta_{\lambda}\delta^C_{\;B}\delta^A_{\;D}=-[(\phi^{\mu}_q)^C_{\;D},(\phi^{\lambda}_p)^A_{\;B}]=\frac{\delta^{\mu,-\lambda}\delta^3(p+q)}{2p^+}\,\theta_{-\lambda}\delta^A_{\;D}\delta^C_{\;B}\,,
\end{equation}
that implies $\theta_{\lambda}=\theta_{-\lambda}$.}
The constraint then becomes
\begin{align}\label{LCholocolour}
\nonumber
    &\sum_{\omega} \big[(-)^{\omega}\theta_{\omega}((\lambda_{13}-\lambda_{24})A+(\lambda_{14}-\lambda_{23})B-2\omega C)\,
    \mathcal{C}^{1234\omega}(A-B)^{\lambda_{12}+\omega-1}(A+B)^{\lambda_{34}-\omega-1}\\
    &+(-)^{\lambda_{14}}\theta_{\omega}((\lambda_{12}-\lambda_{34})C+(\lambda_{13}-\lambda_{24})A-2\omega B)\,
    \mathcal{C}^{4123\omega}(C-A)^{\lambda_{14}+\omega-1}(C+A)^{\lambda_{23}-\omega-1}\big]=0\,.
\end{align}
Following the same conventions used in \cite{Skvortsov:2020wtf} and in analogy with the previous case, we fix $\theta_{\omega}=(-)^{\omega}$. This choice is again equivalent to taking even-helicity fields to be Hermitian and odd-helicity fields to be anti-Hermitian matrices.

As in the previous case, we divide the solution into several subcases. However, we note that the constraints \eqref{LCholo} and \eqref{LCholocolour} are structurally very similar. In fact, the only difference lies in the form of the functions, which, in this case, include only half of the terms present in the previous one.
Therefore, we focus on two representative cases, as the others follow straightforwardly.

In the following, we assume $\Lambda \geq 3$. Indeed, as in the previous case, for lower derivatives, fewer constraints arise due to the partially degenerate structure of the monomials. Therefore, the case $\Lambda = 2$ is treated separately in Appendix \ref{AppendixC}.

\paragraph{Case 1.}
Here, we assume all helicities are equal, i.e. $\lambda_1=\lambda_2=\lambda_3=\lambda_4$. In this case, the constraint \eqref{LCholocolour} becomes
\begin{align}
\begin{split}
    \sum_{\omega}\omega\,\mathcal{C}^{1111\omega}\big(&C
    (A-B)^{2\lambda+\omega-1}(A+B)^{2\lambda-\omega-1}\\
    &+(-)^{\omega}B
(C-A)^{2\lambda+\omega-1}(C+A)^{2\lambda-\omega-1}\big)=0\,.
\end{split}
\end{align}
First, we determine the polynomial form of the functions \eqref{definitions2} as
\begin{align}\label{U(1)_case1}
    &f^{1111}_{\omega-}(A,B)=-2k^{1111}_{\omega^2_-}A^{\Lambda-3}B\,,&
    &f^{1111}_{\omega+}(C,A)=2k^{1111}_{\omega^2_-}CA^{\Lambda-3}\,.
\end{align}
Upon substituting them back into the constraint, we find it is trivially satisfied. From the functions \eqref{U(1)_case1}, we find the system
\begin{align}
\begin{split}
    \mathcal{C}^{1111\omega}=&\;\frac{k^{1111}_{\omega^2-}(\Lambda-3)!}{2^{\Lambda-4}(2\lambda+\omega-1)!(2\lambda-\omega-1)!}\qquad
    \forall\;\omega\neq 0\,,\\
    k^{1111}_{\omega^2-}=&\;\sum_{\omega}\,\omega^2\,\mathcal{C}^{1111\omega}\,.
    \end{split}
\end{align}
As in the previous analysis, here too, just by looking at \eqref{LCholocolour}, one can see that the product $C^{\lambda_1,\lambda_1,0}C^{0,\lambda_2,\lambda_2}$ decouples from the system; therefore, it is unconstrained. 

\paragraph{Case 2.} Here, we solve \eqref{LCholocolour} assuming generic helicities. First, we determine the polynomial form of the functions \eqref{definitions2} as
\begin{subequations}
\begin{align}\label{case4colourf1}
   f_-^{1234}(A,B)=&\;k^{1234}_-A^{\Lambda-2}\,,\\ \label{case4colour2}
   f^{1234}_{\omega-}(A,B)=&\;k^{1234}_{\omega-}A^{\Lambda-2}+((\lambda_{34}-\lambda_{12})k^{1234}_{\omega-}-2k^{1234}_{\omega^2-})A^{\Lambda-3}B\,.
\end{align}
\end{subequations}
The functions $f^{4123}_+$ and $f^{4123}_{\omega+}$ have the same form as $f_-^{1234}(A,B)$ and $f^{1234}_{\omega-}(A,B)$, respectively. Upon substituting these into \eqref{LCholocolour} gives
\begin{equation}
    k^{1234}_-=k^{4123}_-\,.
\end{equation}
From the function \eqref{case4colourf1}, we find the system
\begin{align}
    \begin{split}
    \mathcal{C}^{1234\omega}=&\;\frac{k^{1234}_-(\Lambda-2)!}{2^{\Lambda-2}(\lambda_{12}+\omega-1)!(\lambda_{34}-\omega-1)!}\qquad 
    \forall\;\omega\,,\\
    k^{1234}_-=&\;\sum_{\omega}\,\mathcal{C}^{1234\omega}\,.
    \end{split}
\end{align}
and an identical expression holds for the external helicity configurations $(4123)$. In all other cases, when particular relations among the couplings are imposed, the same pattern follows.

The same procedure used to find a solution to the constraint for $U(N)$ can be applied to other gauge groups, such as $G = SO(N),\,USp(N)$, where the case of $SO(N)$ was first studied in \cite{Metsaev:1991nb}. We comment on them in Appendix \ref{AppendixD}. 

Another interesting case is when all fields are taken to live in the adjoint representation of a Lie algebra. This scenario is studied in Appendix \ref{AppendixE}.

\paragraph{Summary.} A solution to the holomorphic constraint in the presence of $U(N)$ gauge group is given by even- and odd-derivative vertices that satisfy 
\begin{equation}
\boxed{
\begin{aligned}\label{symfinalsystem_U(N)}
    &\mathcal{C}^{1234\omega}=\frac{k^{1234}_-(\Lambda-2)!}{2^{\Lambda-2}(\lambda_{12}+\omega-1)!(\lambda_{34}-\omega-1)!}\quad 
    \forall\;\omega\,,\quad\text{same for $(4123)$}\,,\\
    &k_-^{1234}= k_-^{4123}\,,\qquad
     C^{\lambda_1,\lambda_1,0}C^{0,\lambda_2,\lambda_2}=\;\text{generic}\,.
\end{aligned}
}
\end{equation}

\paragraph{Lower-spin analysis.}
We now examine all solutions in the presence of a gauge group, focusing on lower helicities $0,1,2$. The complete set of even- and odd-derivative couplings that can be constructed is given by
\begin{align}\label{all_colour_couplings}
\begin{split}
    \{&C^{-2,1,2},\textcolor{blue}{C^{-2,2,2}},\textcolor{blue}{C^{-1,0,2}},C^{-1,1,1},\textcolor{blue}{C^{-1,1,2}},\textcolor{red}{C^{-1,2,2}},C^{0,0,1},\textcolor{blue}{C^{0,0,2}},\\
&C^{0,1,1},\textcolor{blue}{C^{0,1,2}},\textcolor{blue}{C^{0,2,2}},C^{1,1,1},\textcolor{red}{C^{1,1,2}},\textcolor{red}{C^{1,2,2}},\textcolor{red}{C^{2,2,2}}\}\,.
\end{split}
\end{align}
In total, we have $15$ couplings: $9$ abelian and $6$ non-abelian. We now search for possible consistent chiral theories.

We highlight the lower-spin couplings as follows: blue indicates those which, on their own, require higher-spin couplings for consistency; red denotes those which, when combined with any other coupling, also require higher-spin ones; uncoloured couplings are, as we will show in more detail below, the only consistent lower-spin couplings.

We begin by considering one-derivative interactions, involving the subset of couplings
\begin{equation}\label{one_derivative_colour}
    \{C^{-2,1,2},C^{-1,0,2},C^{-1,1,1},C^{0,0,1}\}\,.
\end{equation}
Consistent theories are given by
\begin{subequations}
\begin{align}
    &\{C^{-1,1,1}\}\,, && \text{Yang-Mills coupling}\,,\\
    &\{C^{-1,1,1}=C^{0,0,1}\}\,,&& \text{Yang-Mills coupled to scalars}\,,\\
    &\{C^{-2,1,2}=C^{-1,1,1}\}\,,&& \text{coloured graviton}\,.
\end{align}
\end{subequations}
Note that including $C^{-1,0,2}$ and at least one other of the couplings in \eqref{one_derivative_colour} leads, by consistency, to an infinite tower of higher-spin fields. 

If instead we look for the minimal consistent theory that includes $C^{-1,0,2}$, we still necessarily generate at least one higher-spin field. Indeed, $C^{-1,0,2}$ gives rise to an exchange via the scalar field, which is proportional to $C^{-1,0,2}C^{-1,0,2}$. Next, the minimal consistent set of couplings is given by\footnote{We are making use of the concept of the small crystal \eqref{small_crystal_1d}, which will be introduced and justified in the next section.}  
\begin{equation}
    \begin{aligned}
     &C^{2,-1,0}C^{0,2,-1},&&\;\;&& C^{2,2,-3}C^{3,-1,-1},&&\;\;&&C^{2,-1,0}C^{0,-1,2}\,.
     \end{aligned}
\end{equation}
In particular, the solutions for the couplings give
\begin{equation}
    \{C^{-1,0,2},C^{-3, 2, 2},C^{-1, -1, 3}=\frac{(C^{-1, 0, 2})^2}{C^{-3, 2, 2}}\}\,.
\end{equation}
This is a surprising result on its own: a one-derivative lower-spin coupling induces higher-spin couplings!

Another interesting option is the coloured graviton. It is well known that, under the usual assumptions, gravitons cannot carry colour \cite{Boulanger:2000rq}. However, we find that it is possible to have multi-graviton theories provided we restrict to the self-dual case, which is thanks to $C^{-2,1,2}$. However, we expect that once the unitary completion is considered --- via the inclusion of its parity-related anti-self-dual counterpart $C^{-2,-1,2}$ --- the general quartic constraint will lead to non-localities.

Now we can start to include higher-derivative couplings, then the most general consistent theory is
\begin{equation}\label{HDcolourtheory}
\{C^{-1,1,1}=C^{-2,1,2}=C^{0,0,1},C^{0,1,1},C^{1,1,1}\}\,.
\end{equation}
Indeed, any attempt to include one of the couplings highlighted in blue or red in \eqref{all_colour_couplings} leads to inconsistencies, forcing us to introduce higher-spin fields.
This is consistent with expectations from the covariant formalism. For example, the coupling $C^{-2,2,2}$ would lead to ``colour gravity'', which is known not to be consistent \cite{Boulanger:2000rq}, at least including only lower spins.

Save for the $C^{-2,1,2}$ coupling, the covariant (parity completed) action, corresponding to \eqref{HDcolourtheory}, is the truncation to the cubic order of
\begin{align}
    \begin{split}
    S=&\;\int d^4x\left(-\frac{1}{2\alpha^2}\mathrm{Tr}(F_{\mu\nu}F^{\mu\nu})-\frac{1}{2}D^{\mu}\varphi D_{\mu}\varphi+aF^3+b\,\varphi F^2\right)\,,\\
    F^3=&\;\mathrm{Tr}(F_{\mu\nu}F^{\mu}_{\phantom{\mu}\alpha}F^{\nu\alpha})\,,\qquad
    \varphi F^2=\mathrm{Tr}(\varphi F_{\mu\nu}F^{\mu\nu})\,,\\
    \alpha\sim&\;C^{-1,1,1}\,,\qquad 
    a\sim C^{1,1,1}\,,\qquad 
    b\sim C^{0,1,1}\,.
    \end{split}
\end{align}
Here, $D$ denotes the covariant derivative, $F^a_{\mu\nu}=\pl_\mu A^{a}_\nu-\pl_\nu A^{a}_\mu+f^{abc}A_{\mu}^bA_{\nu}^c$ is the YM field strength and $\alpha\sim C^{-1,1,1}$ indicates the YM coupling. The relations among the couplings are determined by the minimal coupling, required by the gauge invariance.

\paragraph{Higher-spin analysis.}The study of solutions involving higher-spin fields follows the same logic as in the case without a gauge group.

%%%%%%%%%%%%%%%%%%%%%%%%%%%%%%%%%%%%%%%%%%%%%%%%%%%%%%%%%%%%%
\section{Low-derivative chiral higher-spin theories}\label{section4}
%%%%%%%%%%%%%%%%%%%%%%%%%%%%%%%%%%%%%%%%%%%%%%%%%%%%%%%%%%%%%
In this section, we provide a complete classification of chiral higher-spin theories with one- and two-derivatives. In particular, we identify new families of theories beyond the well-known HS-SDYM, HS-SDGR, and full chiral higher-spin gravity. These novel higher-spin theories feature a reduced spectrum, which in turn allows more freedom to the couplings.\footnote{Let us note that the problem of solving the light-cone constraints should be closely related to the problem of classifying higher-spin algebras, as in \cite{Fradkin:1986ka,Boulanger:2013zza}, see also \cite{Ponomarev:2017nrr} for the steps towards establishing such a relation.}

In the following, we always consider that at least two cubic couplings ``talk'' with each other (i.e. they share at least one pair of opposite helicities.). In fact, it is always possible to construct well-defined theories with isolated couplings that do not ``talk'', but such cases are rather trivial.
%%%%%%%%%%%%%%%%%%%%%%%%%%%%%%%%%%%%%%%%%%%%%%%%%%%%%%%%%%%%%
\subsection{Two-derivative chiral higher-spin theories}
%%%%%%%%%%%%%%%%%%%%%%%%%%%%%%%%%%%%%%%%%%%%%%%%%%%%%%%%%%%%%
Here, we classify all inequivalent solutions of chiral higher-spin theories involving two-derivative interactions.
As previously established, satisfying the holomorphic constraint \eqref{holo2} requires respecting two basic rules:
\begin{itemize}
    \item Starting from a single pair of two-derivative couplings $C^2C^2$, four additional ones are required. In particular, the following set of couplings must be present together:
    \begin{equation}\label{small_crystal_2d}
        C^{\lambda_1,\lambda_2,2-\lambda_{12}}C^{\lambda_{12}-2,\lambda_3,\lambda_4},\;\;C^{\lambda_1,\lambda_3,2-\lambda_{13}}C^{\lambda_{13}-2,\lambda_2,\lambda_4},\;\;C^{\lambda_1,\lambda_4,2-\lambda_{14}}C^{\lambda_{14}-2,\lambda_2,\lambda_3}\,.
    \end{equation}
    We also impose  $\sum_{i=1}^4\lambda_i=4$ to ensure that only two-derivative interactions are present. We refer to the collection of the $6$ couplings above as a small crystal. An exception is the pair $C^{\lambda_1,\lambda_1,0}C^{0,\lambda_2,\lambda_2}$, which is excluded from the small crystal, as it does not participate in the constraint \eqref{LCholo}.
    \item These couplings have to satisfy the following system of equations:
    \begin{align}\label{coupling2d}
    \begin{split}
    &C^{\lambda_1,\lambda_2,2-\lambda_{12}}C^{\lambda_{12}-2,\lambda_3,\lambda_4}= C^{\lambda_1,\lambda_3,2-\lambda_{13}}C^{\lambda_{13}-2,\lambda_2,\lambda_4}=C^{\lambda_1,\lambda_4,2-\lambda_{14}}C^{\lambda_{14}-2,\lambda_2,\lambda_3}\,,\\
    &C^{\lambda_1,\lambda_1,0}C^{0,\lambda_2,\lambda_2}=\text{generic}\,.
     \end{split}
    \end{align}
    While this system does not affect the classification below, it plays a role in determining relations among the couplings in any given theory. Notably, a solution is always guaranteed to exist --- namely the one corresponding to HS-SDGR --- in which all two-derivative couplings in the theory take the same value. However, in general, for smaller theories than HS-SDGR, some couplings remain unconstrained and correspond to free parameters. In what follows, we assume that all such free parameters are kept different from zero.
\end{itemize}
We begin by outlining the method used to identify solutions and introducing some key terminology. We define a crystal as a self-consistent set of couplings that, taken together, satisfy the condition \eqref{small_crystal_2d}. Each crystal is uniquely determined by the $C^2C^2$ pair of couplings from which it originates.

In particular, starting from a set of external helicities $(\lambda_1, \lambda_2, \lambda_3, \lambda_4)$, they uniquely identify a crystal. We refer to this configuration as the seed.

For instance, the seed $(2, 2 - 2\lambda, \lambda, \lambda)$ generates the crystal
\begin{equation}
\begin{aligned}\label{example}
     &C^{2,2-2 \lambda,-2+2 \lambda}C^{2-2 \lambda,\lambda,\lambda},&&\;\;&& C^{2,\lambda,-\lambda}C^{\lambda,2-2 \lambda,\lambda},&&\;\;&&C^{2,\lambda,-\lambda}C^{\lambda,2-2 \lambda,\lambda}\\
     &C^{2,2,-2}C^{2,-2,2},&&\;\;&&C^{2,-2,2}C^{-2,2,2},&&\;\;&&C^{2,2,-2}C^{2,2,-2}\\
     &C^{2,2,-2}C^{2,2-2 \lambda,-2+2 \lambda},&&\;\;&&C^{2,2-2 \lambda,-2+2 \lambda}C^{2-2 \lambda,2,-2+2 \lambda},&&\;\;&&C^{2,-2+2 \lambda,2-2 \lambda}C^{-2+2 \lambda,2,2-2 \lambda}\\
     &C^{2,\lambda,-\lambda}C^{\lambda,2,-\lambda},&&\;\;&&C^{2,2,-2}C^{2,\lambda,-\lambda},&&\;\;&&C^{2,-\lambda,\lambda}C^{-\lambda,\lambda,2}\,.
\end{aligned}
\end{equation}
To construct the crystal above, we do as follows. The first line represents the small crystal generated by the seed. We then search for all possible pairs that we can connect (via fields with opposite helicity) to generate other small crystals. We repeat this process till it terminates. Note that all couplings involved are two-derivative ones.

The crystal above defines a specific chiral higher-spin theory with two-derivative interactions. It contains a finite amount of couplings, a finite spectrum, and includes the SDGR interaction $C^{2,-2,2}$.
We can observe that starting from a different seed, corresponding to a small crystal other than the initial one, would generate a different crystal. For instance, starting from $(2,2,-2,2)$ leads to SDGR, producing only the second line of \eqref{example}. Starting from $(2,\lambda,2,-\lambda)$ or $(2,2,2-2\lambda,-2+2\lambda)$ would still generate smaller solutions.

We say that two crystals are equivalent if their seeds can be mapped into each other via a combination of the following two operations:
\begin{itemize}
    \item A general permutation of the external legs:
    \begin{align}\label{eq_rel_permutations}
   &(\lambda_1,\lambda_2,\lambda_3,\lambda_4)\rightarrow (\lambda_{\sigma_1},\lambda_{\sigma_2},\lambda_{\sigma_3},\lambda_{\sigma_4})\,,& 
    &\sigma\in S_4\,.
\end{align}
This is a simple consequence of the symmetry of the corresponding amplitude, which is also seen from the three products of couplings in \eqref{small_crystal_2d} that contribute to the same holomorphic constraint.
    \item An affine transformation of the helicities:
    \begin{align}\label{eq_rel_affine}
    &(\lambda_1,\lambda_2,\lambda_3,\lambda_4)\rightarrow
    (\lambda'_1,\lambda'_2,\lambda'_3,\lambda'_4)\,,&
    &\vec{{\lambda'}}_i=\hat{A}\vec{\lambda_i}+\vec{b}\,.
    \end{align}
    What we mean is that a solution can have some of the helicities $\lambda_i$ as free parameters and two solutions that differ by an affine integer-valued transformation of these free parameters should be considered equivalent.\footnote{We just mean that, for example, if the classification contains theories with helicities $\lambda,\lambda+1,\lambda+2$ and $2\lambda'-1,2\lambda',2\lambda'+1$ for any $\lambda$, $\lambda'$, then this is the same solution (parameterised by $\lambda=2\lambda'-1$).}
\end{itemize}

\paragraph{Observations and simple solutions.} Here, we present some general observations that enable us to readily recover known solutions and identify new ones.

First, HS-SDGR is a well-defined theory. It contains all higher-spin fields and all possible chiral two-derivative couplings, and is thus defined by the largest possible crystal.
Moreover, starting from a seed containing only even helicities, we will never generate odd ones. Conversely, it is not possible to construct a theory with only odd helicities. This tells us that HS-SDGR admits a consistent truncation to the even-helicity sector.

Some solutions have only (++$-$) cubic couplings. This possibility was first pointed out in \cite{Ponomarev:2017nrr} while exploring possible subalgebras of the ``gauge algebra'' of HS-SDGR (see also \cite{Monteiro:2022xwq}). A covariant action for the case including all (++$-$) couplings and no scalar fields was constructed in \cite{Krasnov:2021nsq}. 

To see the existence of such theories using our method, we have to start from a small crystal \eqref{small_crystal_2d} and impose that the newly generated couplings never produce any ($--$+) terms. We can achieve this by requiring
\begin{equation}
    \lambda_1,\lambda_2,\lambda_3\geq n\geq 2\,,\qquad n\in\mathbb{Z}\,.
\end{equation}
As a consequence
\begin{align}
    \begin{split}
    \lambda_4=4-\lambda_{123}\leq -n\leq -2\,,&\\
    2-\lambda_{ij}\leq -n\leq -2,&\qquad i,j=\{1,2,3\}\,,\\
    \lambda_{ij}-2\geq n\geq 2,&\qquad i,j=\{1,2,3\}\,. 
    \end{split}
\end{align}
These define consistent truncations of HS-SDGR: we have a truncation for every $n\geq 2$ where all couplings will only contain fields of helicities $|\lambda|\geq n$.
We can also observe that there exist solutions containing all possible cubic couplings constructed from the truncated spectrum.\footnote{This is, in fact, a more general statement. As we will see, all consistent two-derivative chiral higher-spin theories are entirely determined by their spectrum.}
To see it, we need to prove that any coupling with $|\lambda|\geq n$ can be reached by a small crystal generated by two couplings with helicities $|\lambda|\geq n$. To do this, we simply use that pair of couplings as the seed of the small crystal. These are exactly the truncations found in \cite{Ponomarev:2017nrr}.

Note that this does not prevent the existence of consistent ``smaller''\footnote{Here by smaller we mean theories that contain fewer fields and/or fewer cubic couplings.} theories. Indeed, as we will see later, this is the case. We should interpret the solutions we just described as a collection (finite or infinite) of crystals, nested one inside the other. This idea is related to the notion of subalgebras, see \cite{Ponomarev:2017nrr}.

An additional interesting observation arises when we attempt to include a scalar field. A two-derivative cubic coupling involving a scalar takes the form $C^{\lambda,2-\lambda,0}$ and generates (even on its own, but here we assume the presence of two such couplings) the small crystal
\begin{equation}
    C^{\lambda_1,2-\lambda_1,0}C^{0,\lambda_2,2-\lambda_2},\;\;
    C^{\lambda_1,\lambda_2,2-\lambda_{12}}C^{\lambda_{12}-2,2-\lambda_1,2-\lambda_2},\;\;
    C^{\lambda_1,2-\lambda_2,\lambda_2-\lambda_1}C^{\lambda_1-\lambda_2,2-\lambda_1,\lambda_2}\,.
\end{equation}
This crystal contains both (++$-$) and ($--$+) couplings if at least one of the helicities satisfies $\lambda\geq 3$ or $\lambda\leq -1$. Therefore, it is not possible to consistently add a scalar field and retain only (++$-$) couplings while preserving Lorentz invariance; adding a scalar necessarily introduces ($--$+) couplings as well.

The only exceptions occur for $\lambda=1$, which gives a crystal with a single coupling $C^{1,1,0}$, or $\lambda=0,2$, yielding the two couplings $C^{2,0,0}$ and $C^{-2,2,2}$.

We can now ask which is the simplest theory that includes the SDGR coupling $C^{-2,2,2}$ and a generic coupling containing at least one scalar field $C^{\lambda,2-\lambda,0}$. The smallest crystal containing them is generated by the seed $(2,\lambda,0,2-\lambda)$, which gives the spectrum
\begin{equation}\label{crystal_sdgr_scalar}
    \pm\{0,2,\lambda,\lambda-2,2\lambda-2\}\,.
\end{equation}
The cubic couplings are then all possible ones constructed from the spectrum above:
\begin{equation}
\{C^{-2,2,2},C^{2,-\lambda ,\lambda},C^{2,2-\lambda ,\lambda -2},C^{2,2-2 \lambda ,2 \lambda -2},C^{0,0,2},C^{2-2 \lambda ,\lambda ,\lambda},C^{0,2-\lambda ,\lambda},C^{2-\lambda ,2-\lambda ,2 \lambda -2}\}\,.
\end{equation}
One final observation is that we can easily recover the minimal coupling\footnote{For massless fields, the minimal coupling corresponds to the cubic interaction with the lowest number of derivatives, two for gravitational interactions and one for gauge interactions. Note also that the mass dimension of the coupling scales as $1-|\lambda_1+\lambda_2-\lambda_3|$.  Consequently, the gravitational minimal coupling has mass dimension $-1$ and is proportional to $1/M_P$, while the gauge (spin-$1$) coupling is dimensionless and corresponds to the electric charge $e$ for photons or the gauge coupling $g$ for gluons. For a more detailed explanation, including a generalization to massive higher-spin fields, we refer to \cite{Arkani-Hamed:2017jhn}. There, all cubic interactions --- not only the minimal ones --- are classified using the spinor-helicity formalism, which is also adapted to the massive case.} to gravity for higher-spin fields, thanks to the small crystal 
\begin{align}\label{sdgr_universality}
    &C^{2,2,-2}C^{2,\lambda,-\lambda},&
    &C^{2,\lambda,-\lambda}C^{\lambda,2,-\lambda},&
    &C^{2,-\lambda,\lambda}C^{-\lambda,2,\lambda}\,.
\end{align}
This configuration yields the following solution to the system of equations:
\begin{align}
    &C^{2,2,-2}C^{2,\lambda,-\lambda}=C^{2,\lambda,-\lambda}C^{\lambda,2,-\lambda}&
    &\Rightarrow&
    C^{2,2,-2}=C^{2,\lambda,-\lambda}\,.
\end{align}
This result demonstrates the universality of gravitational interactions, even for higher-spin fields. An expression of what is commonly referred to as the equivalence principle.\\
Both the crystal \eqref{crystal_sdgr_scalar} and \eqref{sdgr_universality} will appear again in the classification we present below.

\paragraph{Spectrum $\Rightarrow$ couplings.} There is an interesting ``experimental fact'' that all two- and one-derivative theories we classify below have all possible cubic couplings that can be constructed from the spectrum of the helicities that participate in the interactions, which we can call ``interacting spectrum''. In other words, suppose that the free approximation contains fields with helicities $S'=\{\lambda_i'\}$, while the cubic couplings contain a subset of these helicities $S'\ni S=\{\lambda_i\}$. Then, the theory contains all cubic vertices $V_{\lambda_i,\lambda_j,\lambda_k}$, $\lambda_i+\lambda_j+\lambda_k=1 \text{ or } 2$. Some theories contain free parameters, and for the above to hold, we assume these parameters are generic --- that is, they do not vanish. The fact that the (interacting) spectrum determines the couplings is nontrivial, and as we will show later, this property no longer holds in the higher-derivative case.

\paragraph{All two-derivative solutions.}
To construct inequivalent crystals, we begin with a seed containing generic helicities $(\lambda_1,\lambda_2,\lambda_3,\lambda_4)$, with the condition $\sum_{i=1}^4\lambda_i=4$. We then proceed as follows:
\begin{itemize}
    \item First, we construct the small crystal generated by the seed and identify all possible relations among the helicities that would generate new small crystals. This is done by fixing two generic helicities to be opposite to each other.
    \item For each such relation, once imposed, we compute the newly generated crystals and then look for further constraints they induce.
    \item Since the initial small crystal contains three independent helicities --- taken to be $(\lambda_1,\lambda_2,\lambda_3)$ --- we iterate this process up to three times and then stop.
    \item Finally, we identify inequivalent crystals, modulo the equivalence relations \eqref{eq_rel_permutations} and \eqref{eq_rel_affine}.
\end{itemize}
In general, the solutions may also include non-integer helicities,\footnote{In particular, most of them do not contain half-integer spins, which should be interpreted as (bosonic) fermions (note, that the generators for fermions are slightly different and are not just obtained but letting $\lambda$ be half-integer). Instead, they involve helicities like $\lambda=\frac{2}{3},\frac{4}{3}$, or other generic rational values. These configurations, even though they may be mathematically interesting they are not physically relevant. For this reason, we decided to omit them in the classification. Nevertheless, it is tempting to consider $\lambda \in \mathbb{R}$ since the light-cone generators tolerate that. One can also speculate that the light-cone gauge allows one to define ``fractional spin'' in four dimensions. } that would correspond to non-physical configurations. In the classification below, we only report the physical solutions.

We present the solutions starting from the three-parameter family, then proceed to the two- and one-parameter families, followed by particular cases. However, it is important to interpret the classification in the opposite direction. The solution with the greatest helicity freedom remains valid until an additional relation is imposed, reducing the number of free parameters.

As all solutions are completely determined by the (interacting) spectrum, for each inequivalent crystal, we only write down the interacting spectrum in the following way: $\pm\{ \lambda_1, ...\}\cup\{s_1,...\}$. The spectrum of a theory consists of fields with helicity $\pm \lambda_1$, ..., $\pm s_1$, ...\footnote{The free Hamiltonian always contains pairs of fields with opposite helicities.}. The cubic vertices are given by all possible triplets of fields with helicities $\pm \lambda_1$, ... and $+s_1$, ... (no $-s_1$!) such that the total helicity is $(1)2$ for (one-)two-derivative theories.

Note that a crystal can exhibit two possible behaviours when a new relation among the helicities is imposed: 
\begin{itemize}
    \item The new relation may trigger the formation of new and different small crystals. This leads to an inequivalent crystal compared to the original one, but with one less degree of freedom in the helicities. 
    \item The new relation does not result in any new small crystals. In this case, the resulting crystal is simply a contraction of the original one, in the sense that, although it contains fewer fields and couplings, it is not independent of the original. 
\end{itemize}

We begin by describing inequivalent crystals with a finite spectrum of fields and a finite number of cubic couplings. The first case is the simplest and least interesting, involving only a single small crystal. This case gives a three-parameter family of solutions, with the following seed $(\lambda_1,\lambda_2,\lambda_3,4-\lambda_{123})$ and the crystal contains $10$ fields:
\begin{equation}\label{2dgeneral}
    \pm\{2-\lambda_{12},2-\lambda_{13},2-\lambda_{23}\}\cup \{\lambda_1,\lambda_2,\lambda_3,4-\lambda_{123}\}\,.
\end{equation}
There are $4$ two-parameter families of finite crystals:
\begin{enumerate}
    \item The seed $(\lambda_1,\lambda_2,0,4-\lambda_{12})$ generates a crystal with $22$ fields:
    \begin{align}\label{2pcase1}
        \begin{split}
        \pm\{&0,\lambda_1-2,\lambda_2-2,\lambda_1-\lambda_2,2\lambda_1-2,2\lambda_2-2,6-2 \lambda_{12},\\
        &2-\lambda_{12},4-\lambda_1-2 \lambda_2,4-2 \lambda_1-\lambda_2\}\cup\{\lambda_1,\lambda_2,4-\lambda_{12}\}\,.
         \end{split}
    \end{align}
    It is obtained by imposing $\lambda_3=-\lambda_3\Rightarrow\lambda_3=0$.
    \item The seed $(\lambda_1,\lambda_2,2,2-\lambda_{12})$ generates a crystal with $8$ fields:
    \begin{equation}\label{2pcase2}
        \pm\{2,\lambda_1,\lambda_2,2-\lambda_{12}\}\,.
    \end{equation}
    It is obtained by imposing $\lambda_{13}-2=\lambda_1\Rightarrow\lambda_3=2$.
    \item The seed $(\lambda_1,\lambda_2,2-2 \lambda_1,2+\lambda_1-\lambda_2)$ generates a crystal with $13$ fields:
    \begin{equation}\label{2pcase3}
        \pm\{\lambda_1,2\lambda_2-2,2\lambda_1-\lambda_2,2-\lambda_{12},2+2 \lambda_1-2 \lambda_2\}\cup \{\lambda_2,2-2 \lambda_1,2+\lambda_1-\lambda_2\}\,.
    \end{equation}
    It is obtained by imposing $2-\lambda_{13}=\lambda_1\Rightarrow\lambda_3=2-2\lambda_1$.
    \item The seed $(\lambda_1,\lambda_2,2-\lambda_1,2-\lambda_2)$ generates a crystal with $13$ fields:
    \begin{equation}\label{2pcase4}
        \pm\{0,2\lambda_1-2,2\lambda_2-2,\lambda_1-\lambda_2,2-\lambda_{12}\}\cup\{\lambda_1,\lambda_2,2-\lambda_1,2-\lambda_2\}\,.
    \end{equation}
    It is obtained by imposing $2-\lambda_{13}=\lambda_{13}-2\Rightarrow\lambda_3=2-\lambda_1$.
\end{enumerate}
There are $5$ one-parameter families of finite crystals:
\begin{enumerate}
    \item The seed $(0,6-4 \lambda ,\lambda ,3 \lambda -2)$ generates a crystal with $24$ fields:
   \begin{align}
        \begin{split}
       \pm&\{0,\lambda-2,2\lambda-2,3 \lambda -4,4 \lambda -4,5 \lambda -6,6 \lambda -6,\\
       &7 \lambda -8, 8 \lambda -10,9 \lambda -10,12 \lambda -14\} \cup\{\lambda,3\lambda-2,6-4\lambda\}\,.
        \end{split}
   \end{align}
   It comes from \eqref{2pcase1} by imposing $\lambda_2-\lambda_1=2\lambda_1-2\Rightarrow\lambda_2=3\lambda_1-2$.
   \item The seed $(0,4-3 \lambda ,\lambda ,2 \lambda)$ generates a crystal with $31$ fields:
   \begin{align}
       \begin{split}
       \pm&\{0,\lambda,\lambda-2,2\lambda-2,3 \lambda -2,4 \lambda -4,4 \lambda -2,5 \lambda -4,6 \lambda -6,6 \lambda -4,\\
       &7 \lambda -6,8 \lambda -6,9\lambda -8,10 \lambda -8,12 \lambda -10\}\cup\{2\lambda,4-3 \lambda\}\,.
       \end{split}
   \end{align}
    It comes from \eqref{2pcase1} by imposing $\lambda_2-\lambda_1=\lambda_1\Rightarrow\lambda_2=2\lambda_1$.
     \item The seed $(4-3 \lambda,2-2\lambda,\lambda,4 \lambda-2)$ generates a crystal with $16$ fields:
    \begin{equation}
    \pm\{12 \lambda-10,9 \lambda-8,8 \lambda-6,6 \lambda-6,5\lambda-4,2 \lambda-2, \lambda\}\cup\{4-3 \lambda,4 \lambda-2\}\,.
    \end{equation}
    It comes from \eqref{2pcase3} by imposing $\lambda_1=2-2\lambda_2$.
    \item The seed $(1,2-2 \lambda ,\lambda ,\lambda +1)$ generates a crystal with $16$ fields:
   \begin{equation}
       \pm\{0,\lambda,\lambda-1,2\lambda,2 \lambda -1,3 \lambda -1,4 \lambda -2\}\cup\{1,\lambda+1,2-2\lambda\}\,.
   \end{equation}
   It comes from \eqref{2pcase3} by imposing $2\lambda_2-2=2-2\lambda_2\Rightarrow\lambda_2=1$.
   \item The seed $(6 \lambda -6,\lambda ,4-3 \lambda ,6-4 \lambda)$ generates a crystal with $13$ fields:
   \begin{equation}
       \pm\{2 \lambda -2,3 \lambda -4,7 \lambda -8,8 \lambda -10,12 \lambda -14\}\cup\{\lambda,6 \lambda -6,6-4 \lambda\}\,.
   \end{equation}
   It comes from \eqref{2pcase3} by imposing $\lambda_1+\lambda_2-2=2-2\lambda_2\Rightarrow\lambda_1=4-3\lambda_2$.
\end{enumerate}
There are $4$ particular finite crystals:
\begin{enumerate}
    \item The seed $(0,1,1,2)$ generates a crystal with $5$ fields:
    \begin{equation}\label{2d_lower_spin}
        \pm\{0,1,2\}\,.
    \end{equation}
    This corresponds to SDGR coupled to the photon and a scalar field. 
    \item The seed $(0,0,2,2)$ generates a crystal with $3$ fields:
    \begin{equation}
        \pm\{0,2\}\,.
    \end{equation}
    This corresponds to SDGR with a scalar field.
    \item The seed $(-2,2,2,2)$ generates a crystal with $2$ fields:
    \begin{equation}
        \pm\{2\}\,.
    \end{equation}
    This corresponds to SDGR.
    \item The seed $(1,1,1,1)$ generates a crystal with $2$ fields:
    \begin{equation}
        \{0,1\}\,.
    \end{equation}
    This corresponds to the abelian vertex $\varphi F^2$ alone.
\end{enumerate}
Below, we present the crystals with infinite spectra and infinitely many cubic couplings, which we refer to as infinite crystals. We explicitly indicate whether helicities $0$ and $\pm 2$ are present in the theory, as these fields are of particular relevance.  There is $1$ two-parameter family of infinite crystals:
\begin{enumerate}
    \item The seed $(-\lambda_1,\lambda_1,4-\lambda_2,\lambda_2)$ generates a crystal with the following fields:
    \begin{equation}\label{2inf}
        \pm\{2,(2-\lambda_2)k\pm 2,(2-\lambda_2)k\pm\lambda_1\}\,,\qquad
        k\in \mathbb{Z}_{\geq 0}\,.
    \end{equation}
    It is obtained by imposing $\lambda_3=-\lambda_1$.
\end{enumerate}
There are $2$ one-parameter families of infinite crystals:
\begin{enumerate}
    \item The seed $(-2,6,-\lambda,\lambda)$ generates a crystal with the following fields:
    \begin{subequations}
    \begin{align}
        &\pm\{2,4k\}\,,&&k\in \mathbb{Z}_{\geq 0}\,,\qquad
        \lambda\equiv 2\;(\text{mod}\;4)\,,\\
        &\pm\{0,2,2k\}\,,&&k\in \mathbb{Z}_{\geq 0}\,,\qquad
        \lambda\equiv 0\;(\text{mod}\;4)\,,\\
        &\pm\{0,2,k\}\,,&&k\in \mathbb{Z}_{\geq 0}\,,\qquad
        \lambda\;\text{odd}\,.
    \end{align}
    \end{subequations}
    It comes from \eqref{2pcase3} by imposing $\lambda_2=\lambda_2-\lambda_1-2\Rightarrow\lambda_1=-2$ or from \eqref{2inf} by imposing $\lambda_2=-2$.
    \item The seed $(0,6-2 \lambda,\lambda-2,\lambda)$ generates a crystal with the following fields:
    \begin{equation}
        \pm\{0,2,(2-\lambda)k,(2-\lambda)k\pm 2\}\,,\qquad k\in \mathbb{Z}_{\geq 0}\,.
    \end{equation}
    It comes from \eqref{2pcase1} by imposing $\lambda_1=\lambda_2-2$.
\end{enumerate}
There are some particular infinite solutions:\footnote{
We include these cases because they arise by fixing specific helicities in the finite crystals discussed above.}
\begin{itemize}
    \item The following seeds generate all integer helicities:
    \begin{align}\label{exceptions1}
         \begin{split}
        \{&(0,0,1,3),(0,1,-1,4),(0,3,-3,4),(1,2,-2,3),\\
        &(1,-1,-2,6),(1,4,-4,3),(1,-1,1,3)\}\,.
        \end{split}
    \end{align}
    \item The following seeds generate all even helicities:
    \begin{align}\label{exceptions2}
        \begin{split}
        \{&(0,2,-2,4),(0,0,0,4),(0,4,-4,4),(0,0,-2,6),\\
        &(0,6,-6,4),(0,10,-10,4),(4,-4,-2,6),(0,10,-4,-2)\}\,.
        \end{split}
    \end{align}
    \item The following seeds generate all helicities $\lambda\equiv 2$ (mod $4$):
    \begin{equation}
        \{(2,-2,-2,6),(10,-10,-2,6),(6,-6,-2,6)\}\,.
    \end{equation}
\end{itemize}
In the following, we present additional crystals which, as described above, are merely contractions of larger crystals, even though they exhibit a different spectrum. Nevertheless, they are relevant, and we include them here as well.
We also indicate whether the crystal coincides with its small crystal, and whether it contains only (++$-$) couplings. There is $1$ two-parameter family of finite crystals:
\begin{enumerate}
    \item The seed $(\lambda_1,\lambda_2,\lambda_1,4-2 \lambda_1-\lambda_2)$ generates a crystal with $7$ fields:
    \begin{equation}\label{2pcase5}
        \pm\{2 \lambda_1-2,2-\lambda_{12}\}\cup\{\lambda_1,\lambda_2,4-2 \lambda_1-\lambda_2\}\,.
    \end{equation}
    It is a contraction of \eqref{2dgeneral} by imposing $2-\lambda_{12}=2-\lambda_{23}\Rightarrow\lambda_3=\lambda_1$. It coincides with the small crystal.
\end{enumerate}
There are $13$ one-parameter families of finite crystals:
\begin{enumerate}
   \item The seed $(0,1,3-\lambda ,\lambda)$ generates a crystal with $21$ fields:
   \begin{equation}
       \pm\{0,1,\lambda-1,\lambda-2,2 \lambda -4,2 \lambda
   -3,2 \lambda -2,3 \lambda -5,3 \lambda -4,4 \lambda-6\}\cup\{\lambda,3-\lambda\}\,.
   \end{equation}
   It is a contraction of \eqref{2pcase1} by imposing $\lambda_2=2-\lambda_2\Rightarrow\lambda_2=1$.
   \item The seed $(2,\lambda,0,2-\lambda)$ generates a crystal with $9$ fields:
   \begin{equation}\label{example3}
       \pm\{0,2,\lambda,\lambda -2,2 \lambda -2\}\,.
   \end{equation}
   It is a contraction of \eqref{2pcase1} by imposing $2\lambda_2-2=\lambda_2\Rightarrow\lambda_2=2$.
   \item The seeds $(0,\lambda ,\lambda ,4-2 \lambda)$ and $(2 \lambda -2,\lambda ,2-\lambda ,4-2 \lambda)$ generate a crystal with $11$ fields:
   \begin{equation}
       \pm\{0,\lambda-2,2 \lambda -2,3 \lambda -4,4 \lambda -6\}\cup\{\lambda,4-2 \lambda\}\,.
   \end{equation}
   It is a contraction of \eqref{2pcase1} by imposing $\lambda_1-\lambda_2=\lambda_2-\lambda_1\Rightarrow\lambda_2=\lambda_1$ and of \eqref{2pcase3} by imposing $2-2\lambda_1=2\lambda_2-2\Rightarrow\lambda_1=2-\lambda_2$, respectively.
   \item The seed $(0,6-3 \lambda ,\lambda ,2 \lambda -2)$ generates a crystal with $17$ fields:
   \begin{equation}
       \pm\{0,2-\lambda,2 \lambda -4,2 \lambda -2,3 \lambda -4,4 \lambda
   -6,5 \lambda -8,6 \lambda -10\}\cup\{\lambda,6-3\lambda\}\,.
   \end{equation}
   It is a contraction of \eqref{2pcase1} by imposing $\lambda_2=2\lambda_1-2$.
   \item The seed $(2,2,-\lambda ,\lambda)$ generates a crystal with $4$ fields:
   \begin{equation}\label{simplestcase}
       \pm\{2,\lambda \}\,.
   \end{equation}
   It is a contraction of \eqref{2pcase2} by imposing $\lambda_2=2$. This crystal contains only (++$-$) couplings. It coincides with the small crystal.
   \item The seed $(2,2-2 \lambda ,\lambda ,\lambda)$ generates a crystal with $6$ fields:
   \begin{equation}
       \pm\{2,\lambda ,2 \lambda -2\}\,.
   \end{equation}
   It is a contraction of \eqref{2pcase2} by imposing $\lambda_2=\lambda_1$. This crystal contains only (++$-$) couplings.
   \item The seed $(6-4 \lambda ,\lambda ,\lambda ,2 \lambda -2)$ generates a crystal with $6$ fields:
   \begin{equation}
       \pm\{3 \lambda -4,2 \lambda -2\}\cup\{\lambda,6-4 \lambda\}\,.
   \end{equation}
   It is a contraction of \eqref{2pcase3} by imposing $\lambda_1=2\lambda_2-2$. This crystal contains only (++$-$) couplings. It coincides with the small crystal.
   \item The seed $(2-2 \lambda ,2-2 \lambda ,\lambda ,3 \lambda)$ generates a crystal with $8$ fields:
   \begin{equation}
       \pm\{\lambda,4 \lambda -2,6 \lambda -2\}\cup\{2-2\lambda,3\lambda\}\,.
   \end{equation}
   It is a contraction of \eqref{2pcase3} by imposing  $\lambda_2=2-2\lambda_1$.
   \item The seed $(10-6 \lambda ,\lambda ,2 \lambda -2,3 \lambda -4)$ generates a crystal with $10$ fields:
   \begin{equation}
       \pm\{2 \lambda -2,3 \lambda -4,4 \lambda -6,5 \lambda -8\}\cup\{\lambda,10-6 \lambda\}\,.
   \end{equation}
   It is a contraction of \eqref{2pcase3} by imposing $2+\lambda_1-\lambda_2=2\lambda_2-2\Rightarrow\lambda_1=3\lambda_2-4$. This crystal contains only (++$-$) couplings.
   \item The seed $(1,1,2-\lambda ,\lambda)$ generates a crystal with $8$ fields:
   \begin{equation}
       \pm\{0,\lambda -1,2 \lambda -2\}\cup\{1,\lambda,2-\lambda\}\,.
   \end{equation}
   It is a contraction of \eqref{2pcase4} by imposing $2\lambda_2-2=2-2\lambda_2\Rightarrow\lambda_2=1$.
   \item The seed $(\lambda ,\lambda ,2-\lambda ,2-\lambda)$ generates a crystal with $5$ fields:
   \begin{equation}
       \pm\{0,2 \lambda -2\}\cup\{\lambda,2-\lambda\}\,.
   \end{equation}
    It is a contraction of \eqref{2pcase5} by imposing $\lambda_1+\lambda_2-2=2-\lambda_1-\lambda_2\Rightarrow\lambda_2=2-\lambda_1$. It coincides with the small crystal.
   \item The seed $(4-3\lambda ,2-\lambda ,\lambda ,3\lambda-2)$ generates a crystal with $11$ fields:
   \begin{equation}\label{example4}
       \pm\{0,2 \lambda-2 ,4 \lambda-4 ,6 \lambda-6 \}\cup\{ 4-3\lambda ,2-\lambda ,\lambda ,3\lambda-2\}\,.
   \end{equation}
   It is a contraction of \eqref{2pcase4} by imposing $\lambda_2-\lambda_1=2\lambda_1-2\Rightarrow\lambda_2=3\lambda_1-2$.
   \item The seed $(4-3\lambda ,\lambda ,\lambda ,\lambda)$ generates a crystal with $4$ fields:
   \begin{equation}
       \pm\{2\lambda-2 \}\cup\{\lambda ,4-3\lambda \}\,.
   \end{equation}
   It is a contraction of \eqref{2pcase5} by imposing $2\lambda_1-2=\lambda_1+\lambda_2-2\Rightarrow\lambda_2=\lambda_1$. This crystal contains only (++$-$) couplings. It coincides with the small crystal.
\end{enumerate}
We also present $5$ additional one-parameter families of infinite crystals which, although arising as specific contractions of \eqref{2inf}, also originate from the two-parameter families of finite crystals. As such, they are important for the complete classification.
\begin{enumerate}
    \item The seed $(0,0,4-\lambda,\lambda)$ generates a crystal with the following fields:
    \begin{equation}
        \pm\{0,2,(2-\lambda)k,(2-\lambda)k\pm 2\},\qquad k\in \mathbb{Z}_{\geq 0}\,.
    \end{equation}
    It comes from \eqref{2pcase1} by imposing $\lambda_1=0$.

    \item The seed $(0,4,-\lambda,\lambda)$ generates a crystal with the following fields:
    \begin{subequations}
    \begin{align}\label{HSSDGRseed}
        &\pm\{0,2,k\}, &&k\in \mathbb{Z}_{\geq 0}\,,\qquad \lambda\;\;\text{odd}\,,\\
        & \pm\{0,2,2k\},&&k\in \mathbb{Z}_{\geq 0}\,,\qquad \lambda\;\;\text{even}\,.
    \end{align}
    \end{subequations}
    It comes from \eqref{2pcase1} by imposing $4-\lambda_1-\lambda_2=-\lambda_1\Rightarrow \lambda_2=4$. This seed leads to HS-SDGR. In particular, starting from an even $\lambda$ yields a truncation of HS-SDGR to even helicities, while an odd $\lambda$ gives the full HS-SDGR spectrum.
    \item The seeds $(2,-2,4-\lambda,\lambda)$, $(2 \lambda-6,\lambda,4-\lambda,6-2 \lambda)$ and $(\lambda,\lambda,-\lambda,4-\lambda)$ generate a crystal with the following fields:
    \begin{equation}
        \pm\{2,(2-\lambda)k\pm 2\}\,,\qquad k\in \mathbb{Z}_{\geq 0}\,.
    \end{equation}
    It comes from \eqref{2pcase2} by imposing $\lambda_2=-2$ and from \eqref{2pcase3} by imposing $2-2\lambda_1=\lambda_2-\lambda_1-2\Rightarrow\lambda_1=4-\lambda_2$ and from \eqref{2pcase5} by imposing $\lambda_2=-\lambda_1$, respectively.
    \item The seed $(2+2 \lambda,\lambda,-\lambda,2-2 \lambda)$ generates a crystal with the following fields:
    \begin{equation}
        \pm\{2,2k\lambda\pm 2,(2k+1)\lambda\}\,,\qquad k\in \mathbb{Z}_{\geq 0}\,.
    \end{equation}
    It comes from \eqref{2pcase3} by imposing $\lambda_2=-\lambda_1$.
    \item The seed $(\lambda,2+\lambda,-\lambda,2-\lambda)$ generates a crystal with the following fields:
    \begin{equation}
        \pm\{0,2,k\lambda\pm 2,k\lambda\}\,,\qquad k\in \mathbb{Z}_{\geq 0}\,.
    \end{equation}
    It comes from \eqref{2pcase4} by imposing $\lambda_2=-\lambda_1$.
\end{enumerate}

\paragraph{Further observations.} 
From the classification above, we find that the only seeds leading to the full HS-SDGR theory are: 
\begin{equation*}
\begin{aligned}
    &(\lambda,-\lambda,1,3)
    &&C^{\lambda,-\lambda,2}C^{-2,1,3},&&C^{\lambda,1,1-\lambda}C^{\lambda-1,-\lambda,3},&& C^{\lambda,3,-1-\lambda}C^{\lambda+1,-\lambda,1}\,,\\
    &(\lambda,-\lambda,0,4)&&
    C^{\lambda,-\lambda,2}C^{-2,0,4},&& C^{\lambda,0,2-\lambda}C^{\lambda-2,-\lambda,4},&&C^{\lambda,4,-2-\lambda}C^{\lambda+2,-\lambda,0}\,,
    &&\lambda\;\text{odd}\,,\\
    &(\lambda,-\lambda,-1,5)&&
    C^{ \lambda,- \lambda,2}C^{-2,-1,5},&& C^{ \lambda,-1,3- \lambda}C^{-3+ \lambda,- \lambda,5},&&C^{ \lambda,5,-3- \lambda} C^{3+ \lambda,- \lambda,-1}\,, &&\lambda\equiv 0\;(\text{mod}\;3)\,,\\
    &(-2,6,\lambda,-\lambda)&&
   C^{-2,6,-2}C^{2,\lambda,-\lambda},&&C^{-2,\lambda,4-\lambda}C^{-4+\lambda,6,-\lambda},&& C^{-2,-\lambda,4+\lambda}C^{-4-\lambda,6,\lambda}\,, &&\lambda\;\text{odd}\,.
\end{aligned}
\end{equation*}
Therefore, HS-SDGR is the unique solution iff at least one of these $C^2C^2$ coupling pairs is turned on.

Another noteworthy observation is that the only finite crystals containing the SDGR coupling $C^{-2,2,2}$ --- and which can thus be interpreted as chiral higher-spin gravity theories --- are those generated by \eqref{2pcase2} and its contractions. In contrast, all infinite crystals include the SDGR coupling.

As we will see later, when solving the system of equations for the couplings associated with a crystal, some solutions may allow certain couplings to vanish. This gives rise to a sub-crystal, which corresponds to a subalgebra in the terminology of \cite{Ponomarev:2017nrr}. As shown in \cite{Ponomarev:2017nrr}, solutions to the light-cone holomorphic constraint correspond, with a few exceptions, to specific Lie algebras.

%%%%%%%%%%%%%%%%%%%%%%%%%%%%%%%%%%%%%%%%%%%%%%%%%%%%%%%%%%%%%
\subsection{One-derivative chiral higher-spin theories}
%%%%%%%%%%%%%%%%%%%%%%%%%%%%%%%%%%%%%%%%%%%%%%%%%%%%%%%%%%%%%
Here, we classify all inequivalent solutions of chiral higher-spin theories involving one-derivative interactions in the presence of a $U(N)$ gauge group. For $SO(N)$ and $USp(N)$, the classification of the crystals remains the same. The only difference is in the system for the couplings, as discussed in Appendix \ref{AppendixD}.
Satisfying the holomorphic constraint \eqref{Unconstraint} requires two basic rules:
\begin{itemize}
    \item Starting from a single pair of couplings $C^1C^1$, four additional ones are required. In particular, the following set of couplings must be present together:
    \begin{equation}\label{small_crystal_1d}
        C^{\lambda_1,\lambda_2,1-\lambda_{12}}C^{\lambda_{12}-1,\lambda_3,\lambda_4}
        ,\;\; C^{\lambda_1,\lambda_3,1-\lambda_{13}}C^{\lambda_{13}-1,\lambda_2,\lambda_4}
        ,\;\; C^{\lambda_1,\lambda_4,1-\lambda_{14}}C^{\lambda_{14}-1,\lambda_2,\lambda_3}\,.
    \end{equation}
    We also impose $\sum_{i=1}^4\lambda_i=2$ to ensure that only one-derivative interactions are present. We refer to the collection of the $6$ couplings above as a small crystal. An exception is the pair $C^{\lambda_1,\lambda_1,0}C^{0,\lambda_2,\lambda_2}$, which is excluded from the small crystal, as it does not participate in the constraint \eqref{LCholocolour}.      
    \item These couplings have to satisfy the following system of equations:
    \begin{align}\label{coupling1d}
    \begin{split}
     &C^{\lambda_1,\lambda_2,1-\lambda_{12}}C^{\lambda_{12}-1,\lambda_3,\lambda_4}=C^{\lambda_4,\lambda_1,1-\lambda_{14}}C^{\lambda_{14}-1,\lambda_2,\lambda_3}\,,\\
    &C^{\lambda_1,\lambda_1,0}C^{0,\lambda_2,\lambda_2}=\;\text{generic}\,.
    \end{split}
    \end{align}
    The same equations also appear for the other colour-ordered constraints \eqref{colour-ordered_terms}.
    While this system does not affect the classification below, it plays a role in determining relations among the couplings in any given theory. Notably, a solution is always guaranteed to exist --- namely the one corresponding to HS-SDYM --- in which all one-derivative couplings in the theory take the same value. However, in general, for smaller theories than HS-SDYM, some couplings remain unconstrained and correspond to free parameters. In what follows, we assume that all such free parameters are kept different from zero.
\end{itemize}

To justify the need for the small crystal in \eqref{small_crystal_1d}, we recall that the constraint \eqref{LCholocolour} applies only to a single colour-ordered case, specifically $[1234]$, and requires just two pairs of couplings. We make a simplifying assumption that $C^{\lambda_1,\lambda_2,\omega}\neq0$ implies $C^{\lambda_2,\lambda_1,\omega}\neq0$. As a result, we also need to satisfy the colour-ordered constraint $[1342]$, which in turn implies that all permutations must be included and all colour-ordered constraints must be fulfilled.

Apart from the change in the number of derivatives of the vertices, the classification strategy remains unchanged. Then we follow the same procedure as in the two-derivative case.
Let us give an example of a crystal generated by the seed $(1,1,-\lambda,\lambda)$, which yields
\begin{equation}\label{example_U(N)}
\begin{aligned}
&C^{1,1,-1}C^{1,-\lambda,\lambda},&&\;\;&& C^{1,-\lambda,\lambda}C^{-\lambda,1,\lambda},&&\;\;&& C^{1,\lambda,-\lambda}C^{\lambda,1,-\lambda}\\
&C^{1,1,-1}C^{1,-1,1},&&\;\;&& C^{1,-1,1}C^{-1,1,1},&&\;\;&&C^{1,1,-1}C^{1,1,-1}\,.
\end{aligned}
\end{equation}
The above defines a specific chiral higher-spin theory with one-derivative interactions. It contains only a finite number of couplings, a finite spectrum, and the SDYM interaction $C^{-1,1,1}$.\\
The notion of equivalence between crystals is the same as in the two-derivative case.

\paragraph{Observations and simple solutions.}

First, HS-SDYM is a well-defined theory. It contains all higher-spin fields and all possible chiral one-derivative couplings, and is thus defined by the largest possible crystal. Moreover, starting from a seed containing only odd helicities, we will never generate even ones. This tells us that HS-SDYM admits a consistent truncation to the odd-helicity sector.

As in the two-derivative case, solutions that include only (++$-$) cubic couplings are allowed. The argument is the same as before, starting from the small crystal \eqref{small_crystal_1d} and by requiring
\begin{equation}
    \lambda_1,\lambda_2,\lambda_3\geq n\geq 1\,,\qquad n\in\mathbb{Z}\,.
\end{equation}
As a consequence
\begin{align}
\begin{split}
    \lambda_4=2-\lambda_{123}\leq -n\leq -1&\,,\\
    1-\lambda_{ij}\leq -n\leq -1&\,,\qquad i,j=\{1,2,3\}\,,\\
    \lambda_{1j}-1\geq n\geq 1&\,,\qquad i,j=\{1,2,3\}\,. 
\end{split}
\end{align}
These define consistent truncations of HS-SDYM, and we can have arbitrary truncations, containing fields of helicities $|\lambda|\geq n$.
As before, there exist solutions that incorporate all possible cubic couplings constructed from the truncated spectrum.

Let us try to include a scalar field. A one-derivative cubic coupling involving a scalar field takes the form $(\lambda,1-\lambda,0)$ and generates the small crystal
\begin{equation}
    C^{\lambda_1,1-\lambda_1,0}C^{0,\lambda_2,1-\lambda_2},\;\;
    C^{\lambda_1,\lambda_2,1-\lambda_{12}}C^{\lambda_{12}-1,1-\lambda_1,1-\lambda_2},\;\;
    C^{\lambda_1,1-\lambda_2,\lambda_2-\lambda_1}C^{\lambda_1-\lambda_2,1-\lambda_1,\lambda_2}\,.
\end{equation}
This crystal contains both (++$-$) and ($--$+) couplings if at least one of the helicities satisfies $\lambda\geq 2$ or $\lambda\leq -1$. Therefore, adding a scalar necessarily introduces ($--$+) couplings as well.\\
The only exceptions occur for $\lambda=0,1$, which gives a crystal with two couplings $C^{0,0,1},C^{-1,1,1}$.

We can now ask which is the simplest theory that includes the SDYM coupling $C^{-1,1,1}$ and a generic coupling containing at least one scalar field $C^{\lambda,1-\lambda,0}$. The smaller crystal containing them is generated by the seed $(0,1,1-\lambda,\lambda)$, which gives the spectrum
\begin{equation}\label{crystal_sdym_scalar}
    \pm\{0,1,\lambda,\lambda-1,1-2\lambda\}\,.
\end{equation}
The cubic couplings are then all the possible ones constructed from the spectrum above:
\begin{equation}
\{C^{0, 0, 1},C^{0,1-\lambda,\lambda},C^{1,-\lambda,\lambda}C^{1,1-\lambda,-1+\lambda},C^{1-\lambda,1-\lambda-1+2 \lambda}, C^{1-2 \lambda, \lambda, \lambda}, C^{-1,1, 1}, C^{1,1 - 2 \lambda, -1 + 2 \lambda}\}\,.
\end{equation}

One final observation is that we can easily recover the minimal coupling to the spin-$1$ gauge field for higher-spin fields, thanks to the small crystal
\begin{align}\label{sdym_universality}
    &C^{1,1,-1}C^{1,\lambda,-\lambda},&
    &C^{1,\lambda,-\lambda}C^{\lambda,1,-\lambda},&
    &C^{1,-\lambda,\lambda}C^{-\lambda,1,\lambda}\,.
\end{align}
This configuration yields the following solution to the system of equations: 
\begin{align}
    &C^{1,1,-1}C^{2,\lambda,-\lambda}=C^{1,\lambda,-\lambda}C^{\lambda,1,-\lambda}&
    &\Rightarrow&
    C^{1,1,-1}=C^{1,\lambda,-\lambda}\,.
\end{align}
This result demonstrates the universality of the coupling to spin-$1$ for higher-spin fields.\\
Both the crystal \eqref{crystal_sdym_scalar} and \eqref{sdym_universality} will appear again in the classification we present below.

\paragraph{All one-derivative solutions.}
We follow the same procedure as in the two-derivative case. Recall that, in this case as well, the spectrum completely determines the cubic couplings.

The first solution is a three-parameter family with the following seed $(\lambda_1,\lambda_2,\lambda_3,2-\lambda_{123})$ and the crystal contains $10$ fields:
\begin{equation}\label{1dgeneral}
    \pm\{1-\lambda_{12},1-\lambda_{13},1-\lambda_{23}\}\cup\{\lambda_1,\lambda_2,\lambda_3,2-\lambda_{123}\}\,.
\end{equation}
There are $4$ two-parameter families of finite crystals:
\begin{enumerate}
    \item The seed $(0, \lambda_1, 2 - \lambda_{12}, \lambda_2)$ generates a crystal with $22$ fields:
    \begin{align}\label{1d2pcase1}
    \begin{split}
\pm\{&0,1-2 \lambda_1,1-\lambda_1,1-2 \lambda_2,3-2 \lambda_{12},2-\lambda_1-2 \lambda_2,1-\lambda_2,\\
&2-2 \lambda_1-\lambda_2,1-\lambda_{12},\lambda_1-\lambda_2\}\cup\{\lambda_1,2-\lambda_{12},\lambda_2\}\,.
    \end{split}
    \end{align}
    It is obtained by imposing $\lambda_3=-\lambda_3\Rightarrow \lambda_3=0$.
    \item The seed $(1, \lambda_1, 1 - \lambda_{12}, \lambda_2 )$ generates a crystal with $8$ fields:
    \begin{equation}\label{1d2pcase2}
        \pm\{1,\lambda_1,1-\lambda_{12},\lambda_2\}\,.
    \end{equation}
    It is obtained by imposing $\lambda_{13}-1=\lambda_1\Rightarrow\lambda_3=1$.
    \item The seed $(1-2 \lambda_1,\lambda_1,1+\lambda_1-\lambda_2,\lambda_2)$ generates a crystal with $13$ fields:
    \begin{equation}\label{1d2pcase3}
        \pm\{\lambda_1,1-2 \lambda_2,1+2 \lambda_1-2 \lambda_2,1-\lambda_{12},2 \lambda_1-\lambda_2\}\cup\{1-2 \lambda_1,1+\lambda_1-\lambda_2,\lambda_2\}\,.
    \end{equation}
    It is obtained by imposing $1-\lambda_{13}=\lambda_1\Rightarrow \lambda_3=1-2\lambda_1$.
    \item The seed $(1-\lambda_1,\lambda_1,1-\lambda_2,\lambda_2)$ generates a crystal with $13$ fields:
    \begin{equation}\label{1d2pcase4}
        \pm\{0,1-2 \lambda_1,1-2 \lambda_2,1-\lambda_{12},\lambda_1-\lambda_2\}\cup\{1-\lambda_1,\lambda_1,1-\lambda_2,\lambda_2\}\,.
    \end{equation}
    It is obtained by imposing $1-\lambda_{13}=\lambda_{13}-1 \Rightarrow \lambda_3=1-\lambda_1$.
\end{enumerate}
There are $4$ one-parameter families of solutions:
\begin{enumerate}
     \item The seed $(0,3-4 \lambda,\lambda,3 \lambda-1)$ generates a crystal with $24$ fields:
    \begin{align}
        \begin{split}
        \pm&\{0,7-12 \lambda,5-9 \lambda,5-8 \lambda,4-7 \lambda,3-6 \lambda,3-5 \lambda,\\
        &2-4 \lambda,2-3 \lambda,1-2 \lambda,1-\lambda\} \cup\{3-4 \lambda,\lambda,-1+3 \lambda\}\,.
        \end{split}
    \end{align}
    It comes from \eqref{1d2pcase1} by imposing $\lambda_2-\lambda_1=2\lambda_1-1\Rightarrow \lambda_2=3\lambda_1-1$.
    \item The seed $(0,2-3 \lambda,\lambda,2 \lambda)$ generates a crystal with $31$ fields:
    \begin{align}
        \begin{split}
        \pm\{&0,5-12 \lambda,4-10 \lambda,4-9 \lambda,3-8 \lambda,3-7 \lambda,2-6 \lambda,3-6 \lambda,2-5 \lambda,\\
        &1-4 \lambda,2-4 \lambda,1-3 \lambda,1-2 \lambda,1-\lambda,-\lambda\}\cup\{2-3 \lambda,2 \lambda\}\,.
        \end{split}
    \end{align}
    It comes from \eqref{1d2pcase1} by imposing $\lambda_2-\lambda_1=\lambda_1\Rightarrow \lambda_2=2\lambda_1$.
        \item The seed  $(2-3 \lambda,1-2 \lambda,\lambda,4 \lambda-1)$ generates a crystal with $16$ fields:
    \begin{equation}
        \pm\{5-12 \lambda,4-9 \lambda,3-8 \lambda,3-6 \lambda,2-5 \lambda,1-2 \lambda,\lambda\}\cup\{4 \lambda-1,2-3 \lambda\}\,.
    \end{equation}
    It comes from \eqref{1d2pcase3} by imposing $\lambda_1=1-2\lambda_2$.
    \item The seed $(6 \lambda-3,\lambda,2-3 \lambda,3-4 \lambda)$ generates a crystal with $13$ fields:
    \begin{equation}
        \pm\{7-12 \lambda,5-8 \lambda,4-7 \lambda,2-3 \lambda,1-2 \lambda\}\cup\{3-4 \lambda,\lambda,6 \lambda-3\}\,.
    \end{equation}
    It comes from \eqref{1d2pcase3} by imposing $\lambda_1+\lambda_2-1=1-2\lambda_2\Rightarrow \lambda_1=2-3\lambda_2$.
\end{enumerate}
There are $2$ particular finite crystals:
\begin{enumerate}
    \item The seed $(0,0,1,1)$ generates a crystal with $3$ fields:
    \begin{equation}
        \pm\{1,0\}\,.
    \end{equation}
    This corresponds to SDYM coupled to a scalar field.
    \item The seed $(-1,1,1,1)$ generates a crystal with $2$ fields:
    \begin{equation}
        \pm\{1\}\,.
    \end{equation}
    This corresponds to SDYM.
\end{enumerate}
There is $1$ two-parameter family of infinite crystals:
\begin{enumerate}
    \item The seed $(-\lambda_1,\lambda_1,2-\lambda_2,\lambda_2)$ generates a crystal with the following fields:
    \begin{equation}\label{1inf}
        \pm\{1,(1-\lambda_2)k\pm 1,(1-\lambda_2)k\pm\lambda_1\}\,,\qquad k\in \mathbb{Z}_{\geq 0}\,.
    \end{equation}
    It is obtained by imposing $\lambda_3=-\lambda_1$.
\end{enumerate}
There is $1$ one-parameter family of infinite crystals:
\begin{enumerate}
     \item The seed $(0,3-2 \lambda,\lambda-1,\lambda)$ generates a crystal with the following fields:
    \begin{equation}
        \pm\{0,1,(1-\lambda)k,(1-\lambda)k\pm 1\}\,,\qquad k\in \mathbb{Z}_{\geq 0}\,.
    \end{equation}
    It comes from \eqref{1d2pcase1} by imposing $\lambda_1=\lambda_2-1$.
\end{enumerate}
There are some particular infinite solutions:
\begin{itemize}
    \item The following seeds generate all integer helicities:
    \begin{align}
        \begin{split}
        \{&(-1,0,1,2), (0,0,0,2), (-2,0,2,2), (-1,0,0,3), (-3,0,2,3),\\
        &(-5,0,2,5), (-4,0,2,4), (-2,-1,0,5), (-2,-1,2,3)\}\,.
         \end{split}
    \end{align}
    \item The following seeds generate all odd helicities:
    \begin{equation}
        \{(-1,-1,1,3), (-5,-1,3,5), (-3,-1,3,3)\}\,.
    \end{equation}
\end{itemize}
In the following, we present additional crystals obtained through specific contractions of the spectra of the previously discussed crystals.\\
There is $1$ two-parameter family of finite crystals:
\begin{enumerate}
    \item The seed $(\lambda_1,\lambda_1,2-2 \lambda_1-\lambda_2,\lambda_2)$ generates a crystal with $7$ fields:
    \begin{equation}\label{1d2pcase5}
        \pm\{ 1 - 2\lambda_1, 1 - \lambda_{12} \}\cup
\{ \lambda_1, 2 - 2\lambda_1 - \lambda_2, \lambda_2\}\,. 
    \end{equation}
    It is a contraction of \eqref{1dgeneral} by imposing $1-\lambda_{12}=1-\lambda_{23}\Rightarrow \lambda_3=\lambda_1$. It coincides with the small crystal.
\end{enumerate}
There are $11$ one-parameter families of finite crystals:
\begin{enumerate}
    \item The seed $(0,1,1-\lambda,\lambda)$ generates a crystal with $9$ fields:
    \begin{equation}
        \pm\{ 0,1,\lambda, 2\lambda-1, \lambda-1\}\,.
    \end{equation}
    It is a contraction of \eqref{1d2pcase1} by imposing $2\lambda_1-1=\lambda_1\Rightarrow \lambda_1=1$.
    \item The seeds $(0,\lambda,\lambda,2-2 \lambda)$ and $(2 \lambda-1,\lambda,1-\lambda,2-2 \lambda)$ generate a crystal with $11$ fields:
    \begin{equation}
        \pm\{0,3-4 \lambda,2-3 \lambda,1-2 \lambda,1-\lambda\}\cup\{2-2 \lambda,\lambda\}\,.
    \end{equation}
    It is a contraction of \eqref{1d2pcase1} by imposing $\lambda_1-\lambda_2=\lambda_2-\lambda_1\Rightarrow \lambda_2=\lambda_1$ and of \eqref{1d2pcase3} by imposing $1-2\lambda_1=2\lambda_2-1\Rightarrow \lambda_1=1-\lambda_2$, respectively.
    \item The seed $(0,3-3 \lambda,\lambda,2 \lambda-1)$ generates a crystal with $17$ fields:
    \begin{equation}
       \pm\{0, 5 - 6 \lambda, 4 - 5 \lambda, 3 - 4 \lambda, 
  2 - 3 \lambda, 1 - 2 \lambda, 2 - 2 \lambda, 
  1 - \lambda\}\cup\{3 - 3 \lambda, \lambda\}\,.
    \end{equation}
    It is a contraction of \eqref{1d2pcase1} by imposing $\lambda_2=2\lambda_1-1$.
    \item The seed $(1,1,-\lambda,\lambda)$ generates a crystal with $4$ fields:
    \begin{equation}
        \pm\{1,\lambda\}\,.
    \end{equation}
    It is a contraction of \eqref{1d2pcase2} by imposing $\lambda_2=1$. This crystal contains only (++$-$) couplings. It coincides with the small crystal.
    \item The seed $(1,1-2 \lambda,\lambda,\lambda)$ generates a crystal with $6$ fields:
    \begin{equation}
        \pm\{1,1-2 \lambda,\lambda\}\,.
    \end{equation}
    It is a contraction of \eqref{1d2pcase2} by imposing $\lambda_2=\lambda_1$. This crystal contains only (++$-$) couplings.
    \item The seed $(3-4 \lambda,\lambda,\lambda,2 \lambda-1)$ generates a crystal with $6$ fields:
    \begin{equation}
        \pm\{2-3 \lambda,1-2 \lambda\}\cup\{3-4 \lambda,\lambda\}\,.
    \end{equation}
    It is a contraction of \eqref{1d2pcase3} by imposing $\lambda_1=2\lambda_2-1$. This crystal contains only (++$-$) couplings. It coincides with the small crystal.
    \item The seed $(1-2 \lambda,1-2 \lambda,\lambda,3 \lambda)$ generates a crystal with $8$ fields:
    \begin{equation}
        \pm\{1-6 \lambda,1-4 \lambda,\lambda\}\cup\{3 \lambda,1-2 \lambda\}\,.
    \end{equation}
    It is a contraction of \eqref{1d2pcase3} by imposing $\lambda_2=1-2\lambda_1$.
    \item The seed $(5-6 \lambda,\lambda,2 \lambda-1,3 \lambda-2)$ generates a crystal with $10$ fields:
    \begin{equation}
        \pm\{4-5 \lambda,3-4 \lambda,2-3 \lambda,1-2 \lambda\}\cup\{5-6 \lambda,\lambda\}\,.
    \end{equation}
    It is a contraction of \eqref{1d2pcase3} by imposing $1+\lambda_1-\lambda_2=2\lambda_2-1\Rightarrow \lambda_1=3\lambda_2-2$. This crystal contains only (++$-$) couplings.
    \item The seed $(\lambda,\lambda,1-\lambda,1-\lambda)$ generates a crystal with $5$ fields:
    \begin{equation}
        \pm\{0,1-2 \lambda\}\cup\{1-\lambda,\lambda\}\,.
    \end{equation}
    It is a contraction of \eqref{1d2pcase5} by imposing $\lambda_1+\lambda_2-1=1-\lambda_1-\lambda_2\Rightarrow \lambda_2=1-\lambda_1$. It coincides with the small crystal.
    \item The seed $(3 \lambda-1,\lambda,1-\lambda,2-3 \lambda\}$ generates a crystal with $11$ fields:
    \begin{equation}
    \pm\{0,3-6 \lambda,2-4 \lambda,1-2 \lambda\}\cup\{2-3 \lambda,1-\lambda,\lambda,-1+3 \lambda\}\,.
    \end{equation}
    It is a contraction of \eqref{1d2pcase4} by imposing $\lambda_2-\lambda_1=2\lambda_1-1\Rightarrow \lambda_2=3\lambda_1-1$.
    \item The seed $(2-3 \lambda,\lambda,\lambda,\lambda)$ generates a crystal with $4$ fields:
    \begin{equation}
        \pm\{1-2 \lambda\}\cup\{2-3 \lambda,\lambda\}\,.
    \end{equation}
    It is a contraction of \eqref{1d2pcase5} by imposing $2\lambda_1-1=\lambda_1+\lambda_2-1\Rightarrow \lambda_2=\lambda_1$. This crystal contains only (++$-$) couplings. It coincides with the small crystal.
    \end{enumerate}
There are $7$ one-parameter families of infinite crystals which, although arising as specific contractions of \eqref{1inf}, also originate from the two-parameter families of finite crystals:
\begin{enumerate}
    \item The seed $(0,0,2-\lambda,\lambda)$ generates a crystal with the following fields: 
    \begin{equation}
        \pm\{0,1,(1-\lambda)k,(1-\lambda)k\pm 1\}\,,\qquad k\in \mathbb{Z}_{\geq 0}\,.
    \end{equation}
    It comes respectively from \eqref{1d2pcase1} by imposing $\lambda_1=0$.
    \item The seed $(0,2,-\lambda,\lambda)$ generates a crystal with the following fields: 
    \begin{equation}
        \pm\{0,1,k\}\,,\qquad k\in \mathbb{Z}_{\geq 0}\,.
    \end{equation}
    It comes from \eqref{1d2pcase1} by imposing $2-\lambda_1-\lambda_2=-\lambda_1\Rightarrow \lambda_2=2$. This seed leads to HS-SDYM.
    \item The seeds $(1,-1,2-\lambda,\lambda)$ and $(\lambda,\lambda,-\lambda,2-\lambda)$ generate a crystal with the following fields:
    \begin{equation}
        \pm\{1,(1-\lambda)k\pm 1\}\,,\qquad k\in \mathbb{Z}_{\geq 0}\,.
    \end{equation}
    It comes from \eqref{1d2pcase2} by imposing $\lambda_2=-1$ and from \eqref{1d2pcase5} by imposing $\lambda_2=-\lambda_1$, respectively.
    \item The seed $(1-2 \lambda,2-\lambda,\lambda,2 \lambda-1)$ generates a crystal with the following fields:
    \begin{equation}
        \pm\{1,(1+\lambda)k\pm 1\}\,,\qquad k\in \mathbb{Z}_{\geq 0}\,.
    \end{equation}
    It comes from \eqref{1d2pcase3} by imposing $1+\lambda_1-\lambda_2=2\lambda_1-1\Rightarrow\lambda_2=2-\lambda_1$.
    \item The seed $(1+2 \lambda,\lambda,-\lambda,1-2 \lambda)$ generates a crystal with the following fields:
    \begin{equation}
        \pm\{1,2k\lambda\pm 1,(2k+1)\lambda\}\,,\qquad k\in \mathbb{Z}_{\geq 0}\,.
    \end{equation}
    It comes from \eqref{1d2pcase3} by imposing $\lambda_2=-\lambda_1$.
    \item The seed $(-1,3,-\lambda,\lambda)$ generates a crystal with the following fields: 
    \begin{subequations}
    \begin{align}
        &\pm\{0,1,k\}\,,&&k\in \mathbb{Z}_{\geq 0}\,,\qquad \lambda\;\;\text{even}\,,\\
        & \pm\{1,2k+1\}\,,&&k\in \mathbb{Z}_{\geq 0}\,,\qquad \lambda\;\;\text{odd}\,.
    \end{align}
     \end{subequations}
    It comes from \eqref{1d2pcase3} by imposing $\lambda_2=\lambda_2-\lambda_1-1\Rightarrow\lambda_1=-1$. This leads to HS-SDYM. In particular, starting from an odd $\lambda$ yields a truncation of HS-SDYM to odd helicities, while an even $\lambda$ gives the full HS-SDYM spectrum.
    \item The seed $(\lambda,1+\lambda,-\lambda,1-\lambda)$ generates a crystal with the following fields:
    \begin{equation}
        \pm\{0,1,k\lambda\pm 1,k\lambda\}\,,\qquad k\in \mathbb{Z}_{\geq 0}\,.
    \end{equation}
    It comes from \eqref{1d2pcase4} by imposing $\lambda_2=-\lambda_1$.
\end{enumerate}
\paragraph{Further observations.}
From the classification above, we find that the only seeds leading to HS-SDYM are:
\begin{equation*}
\begin{aligned}
    &(\lambda,-\lambda,2,0)&&
    C^{\lambda,-\lambda,1}C^{-1,2,0},&& C^{\lambda,2,-1-\lambda}C^{1+\lambda,-\lambda,0},&& C^{\lambda,0,1-\lambda}C^{-1+\lambda,-\lambda,2}\,,\\
    &(\lambda,-\lambda,3,-1)&& C^{\lambda,-\lambda,1}C^{-1,3,-1},&& C^{\lambda,3,-2-\lambda}C^{2+\lambda,-\lambda,-1},&&C^{\lambda,-1,2-\lambda}C^{-2+\lambda,-\lambda,3}\,,&& \lambda\;\text{even}\,,\\
    &(\lambda,-\lambda,-2,4)&& C^{\lambda,-\lambda,1}C^{-1,-2,4},&& C^{\lambda,-2,3-\lambda}C^{-3+\lambda,-\lambda,4},&&C^{\lambda,4,-3-\lambda}C^{3+\lambda,-\lambda,-2} \,, &&\lambda\equiv 0\;(\text{mod}\;3)\,,\\
    &(-2,-1,0,5)&& C^{-2,-1,4}C^{-4,0,5},&& C^{-2,0,3}C^{-3,-1,5},&& C^{-2,5,-2}C^{2,-1,0}\,.
\end{aligned}
\end{equation*}
Therefore, HS-SDYM is the unique solution iff at least one of these pairs of $C^1C^1$ couplings is turned on.

Another interesting observation is that the only finite crystals containing the SDYM coupling $C^{-1,1,1}$ are those arising from \eqref{1d2pcase2} and its contractions. In contrast, all infinite crystals do contain the SDYM coupling.
%%%%%%%%%%%%%%%%%%%%%%%%%%%%%%%%%%%%%%%%%%%%%%%%%%%%%%%%%%%%%
\subsection{Higher-derivative}
%%%%%%%%%%%%%%%%%%%%%%%%%%%%%%%%%%%%%%%%%%%%%%%%%%%%%%%%%%%%%
Here, we explore the various features that emerge in the higher-derivative case, which also makes its classification more involved.
We focus on the case without a gauge algebra. When a gauge algebra is present, one simply needs to include odd-derivative couplings as well; the rest of the analysis remains unchanged. Satisfying the holomorphic constraint \eqref{holo2}, requires respecting two basic rules:
\begin{itemize}
    \item Starting from a single pair of couplings $C^nC^m$, with $n+m=k+2$ being the total number of derivatives, the following set of $3k$ couplings must be present together:\footnote{Sometimes, for convenience, we indicate the number of derivatives of a coupling on top of it.}
    \begin{equation*}
    \begin{aligned}\label{symHD}
&\overset{k}{C^{\lambda_1,\lambda_2,k-\lambda_{12}}}\overset{2}{C^{\lambda_{12}-k,\lambda_3,\lambda_4}},&&\;\;&&\overset{k}{C^{\lambda_1,\lambda_3,k-\lambda_{13}}}\overset{2}{C^{\lambda_{13}-k,\lambda_2,\lambda_4}},&&\;\;&& \overset{k}{C^{\lambda_2,\lambda_3,k-\lambda_{23}}}\overset{2}{C^{\lambda_{23}-k,\lambda_1,\lambda_4}}\\
    &\overset{k-2}{C^{\lambda_1,\lambda_2,k-2-\lambda_{12}}}\overset{4}{C^{\lambda_{12}+2-k,\lambda_3,\lambda_4}},&&\;\;&&\overset{k-2}{C^{\lambda_1,\lambda_3,k-2-\lambda_{13}}}\overset{4}{C^{\lambda_{13}+2-k,\lambda_2,\lambda_4}},&&\;\;&& \overset{k-2}{C^{\lambda_2,\lambda_3,k-2-\lambda_{23}}}\overset{4}{C^{\lambda_{23}+2-k,\lambda_1,\lambda_4}}\\
    &\cdots,&&\;\;&&\cdots,&&\;\;&&\cdots\\
    &\overset{2}{C^{\lambda_1,\lambda_2,2-\lambda_{12}}}\overset{k}{C^{\lambda_{12}-2,\lambda_3,\lambda_4}},&&\;\;&& \overset{2}{C^{\lambda_1,\lambda_3,2-\lambda_{13}}}\overset{k}{C^{\lambda_{13}-2,\lambda_2,\lambda_4}},&&\;\;&& \overset{2}{C^{\lambda_2,\lambda_3,2-\lambda_{23}}}\overset{k}{C^{\lambda_{23}-2,\lambda_1,\lambda_4}}\,.
    \end{aligned}
 \end{equation*}
We also impose $\sum_{i=1}^4\lambda_i=k+2$ to ensure that only couplings with up to $k$ derivatives are present. We refer to the collection of the $3k$ couplings above as a small crystal. An exception is the pair $C^{\lambda_1,\lambda_1,0}C^{0,\lambda_2,\lambda_2}$, which is excluded from the small crystal, as it does not participate in the constraint \eqref{LCholo}.
\item These couplings have to satisfy the following system of equations:
    \begin{align}
\begin{split} 
     &C^{\lambda_1,\lambda_2,\omega}C^{-\omega,\lambda_3,\lambda_4}=\frac{k^{1234}_+(\Lambda-2)!}{2^{\Lambda-3}(\lambda_{12}+\omega-1)!(\lambda_{34}-\omega-1)!}\qquad 
    \forall\;\omega\,,\\
    &\text{same for $(1324)$ and $(1423)$}\,,\\
    &k_+^{1234}= k_+^{1324}= k_+^{1423}\,,\qquad C^{\lambda_1,\lambda_1,0}C^{0,\lambda_2,\lambda_2}=\;\text{generic}\,.
\end{split}
    \end{align}
    While this system does not affect the classification, it plays a role in determining relations among the couplings in any given theory. Notably, a solution is always guaranteed to exist --- namely the one corresponding to full chiral higher-spin theory, with the Metsaev solution \eqref{alleven}.
\end{itemize}

Let us examine a specific example to highlight the key differences from the lower-derivative case. The seed $(6,-4,2,2)$, with $\Lambda=6$, generates the higher-derivative crystal
\begin{equation}\label{L=6_crystal}
\begin{aligned}
    &C^{6,-4,[2,0]}C^{[-2,0],2,2},&&\;\;&& C^{6,2,[-4,-6]}C^{[4,6],-4,2},&&\;\;&& C^{6,2,[-4,-6]}C^{[4,6],-4,2}\\
    &C^{2,6,-6}C^{6,-6,2},&&\;\;&& C^{2,-6,6}C^{-6,6,2},&&\;\;&& C^{2,2,-2}C^{2,6,-6}\\
    &C^{2,6,-6}C^{6,-4,0},&&\;\;&& C^{2,-4,4}C^{-4,6,0},&&\;\;&&C^{2,0,0}C^{0,6,-4}\\
    &C^{-4,6,0}C^{0,-4,6},&&\;\;&&C^{-4,-4,10}C^{-10,6,6},&&\;\;&&C^{-4,6,0}C^{0,6,-4}\\
    &C^{-10,6,6}C^{-6,2,6},&&\;\;&&C^{-10,2,10}C^{-10,6,6},&&\;\;&& C^{-10,6,6}C^{-6,6,2}\\
    &C^{2,10,-10}C^{10,-10,2},&&\;\;&& C^{2,-10,10}C^{-10,10,2},&&\;\;&& C^{2,2,-2}C^{2,10,-10}\\
    &C^{2,10,-10}C^{10,-4,-4},&&\;\;&&C^{2,-4,4}C^{-4,10,-4},&&\;\;&& C^{2,-4,4}C^{-4,10,-4}\\
    &C^{2,4,-4}C^{4,-4,2},&&\;\;&& C^{2,-4,4}C^{-4,4,2},&&\;\;&& C^{2,2,-2}C^{2,4,-4}\\
    &C^{2,2,-2}C^{2,-2,2},&&\;\;&& C^{2,-2,2}C^{-2,2,2},&&\;\;&& C^{2,2,-2}C^{2,2,-2}\\
    &C^{2,2,-2}C^{2,0,0},&&\;\;&& C^{2,0,0}C^{0,2,0},&&\;\;&& C^{2,0,0}C^{0,2,0}\\
    &C^{2,2,[0,-2]}C^{[0,2],0,2},&&\;\;&&C^{2,0,[2,0]}C^{[-2,0],2,2},&&\;\;&& C^{2,2,[0,-2]}C^{[0,2],2,0}\,,
\end{aligned}
\end{equation}
where the square bracket $[-,-]$ is a notation to indicate the range of exchanged helicities, incremented by steps of $2$.\footnote{If we want to include both even- and odd-derivative interactions and/or in the presence of a gauge group, the step would be of $1$.}

The above crystal contains the following $10$ couplings:
\begin{equation}
    \{\overset{2}{C^{-2, 2, 2}}, \overset{2}{C^{-4, 2, 4}}, \overset{2}{C^{-4, 0, 6}}, \overset{2}{C^{-6, 2, 6}}, \overset{2}{C^{0, 0, 2}}, \overset{2}{C^{-4, -4, 10}}, \overset{2}{C^{-10, 6, 6}}, \overset{2}{C^{-10, 2, 10}},\overset{4}{C^{0, 2, 2}},\overset{4}{C^{-4, 2, 6}}\}\,.
\end{equation}
It includes SDGR, but does not contain all possible couplings that can be constructed from the spectrum up to $4$-derivatives. In particular, it omits the following $12$ couplings:
\begin{align}
    \begin{split}
    &\{\overset{2}{C^{-6, -2, 10}},\overset{2}{C^{-6, 4,4}}, \overset{2}{C^{-2, -2, 6}}, \overset{2}{C^{-2, 0, 
  4}},\overset{4}{C^{-10, 4, 10}}, \overset{4}{C^{-6, 0, 10}},\\
  &\overset{4}{C^{-6, 4, 6}}, \overset{4}{C^{-4, -2, 10}}, \overset{4}{C^{-4,4, 4}}, \overset{4}{C^{-2, 0, 6}}, \overset{4}{C^{-2, 2, 4}}, \overset{4}{C^{0, 0, 4}}\}\,.
  \end{split}
\end{align}
The solution for the couplings in the crystal \eqref{L=6_crystal} is:
\begin{align}
    \begin{split}
&\{C^{-2, 2, 2}=C^{0, 0, 2}=C^{-4, 2, 4}=C^{-6, 2, 6}=C^{-10, 2, 10}\,,\\
    &C^{-4, -4, 10}=\frac{(C^{-4, 0, 6})^2}{C^{-10, 6, 6}},\qquad 
C^{0, 2, 2}=\frac{C^{-2, 2, 2} C^{-4, 2, 6}}{C^{-4, 0, 6}}\}\,.
    \end{split}
\end{align}
The first line is a manifestation of the universality of gravitational interactions.

It is worth stressing that, without the freedom of having an arbitrary value for the product of couplings $C^{\lambda_1,\lambda_1,0}C^{0,\lambda_2,\lambda_2}$, we would get an additional constraint:
\begin{equation}
    C^{2, 2, 2}=\frac{10}{3} \frac{(C^{0,2,2})^2}{C^{-2,2,2}}\,.
\end{equation}
This is the same constraint encountered in \eqref{wrong_condition} for the higher-derivative theory for lower spins, and also here it would lead to a vanishing amplitude.

The example above illustrates all the new features of the higher-derivative case:
\begin{itemize}
    \item The couplings are no longer fully determined by the spectrum. In the example above, some $2$- and $4$-derivative couplings are missing. This complicates the classification of higher-derivative chiral theories.
    \item The presence or absence of a gauge group becomes more relevant. In its presence, odd-derivative vertices must also be included, leading to potentially very different solutions, unlike the similarity observed in the lower-derivative case.
    \item As already seen in  \eqref{HDcolourtheory} and in the example above, the freedom to choose the product $C^{\lambda_1,\lambda_1,0}C^{0,\lambda_2,\lambda_2}$ becomes relevant. It is also tied to the possibility of obtaining non-vanishing amplitudes.
\end{itemize}

\paragraph{Simple observations.}
Full chiral higher-spin theory is a well-defined theory. It contains all higher-spin fields and all possible chiral even-derivative cubic couplings, and is thus defined by the largest possible crystal. Moreover, starting from a seed containing only even helicities, we will never generate odd ones. This tells us that full chiral higher-spin admits a consistent truncation to the even-helicity sector.

As in the low-derivative cases, we have solutions that include only (++$-$) cubic couplings. Starting from a small crystal, we impose that the newly generated couplings never produce any ($--$+) terms, by requiring
\begin{equation}
    \lambda_1,\lambda_2,\lambda_3\geq n\geq k\,,\qquad n\in\mathbb{Z}\,.
\end{equation}
As a consequence
\begin{align}
    \begin{split}
    \lambda_4=k+2-\lambda_{123}\leq -n\leq -k&\,,\\
    \ell-\lambda_{ij}\leq -n\leq -k&\,,\qquad i,j=\{1,2,3\}\,,\quad \ell=\{2,4,...,k\}\,,\\
    \lambda_{ij}-\ell\geq n\geq k&\,,\qquad i,j=\{1,2,3\}\,,\quad \ell=\{2,4,...,k\}\,.
    \end{split}
\end{align}
These define consistent truncations of the full chiral higher-spin theory and can have arbitrary truncations containing fields of helicities $|\lambda|\geq k$. There also exist solutions that include all possible cubic couplings constructed from the truncated spectrum.

Note that a theory that includes both higher-derivative and the SDGR coupling $C^{-2,2,2}$, or one that includes a scalar field, inevitably leads to the presence of both types of couplings (++$-$) and ($--$+).

\paragraph{Metsaev solution.}
Including the self-interacting higher-derivative coupling $C^{-\lambda,\lambda,\lambda}$, for $|\lambda|>2$, uniquely leads to the Metsaev solution. We now present a proof of this statement.

We begin by using the coupling $C^{-\lambda,\lambda,\lambda}$ to form the small crystal
\begin{equation}
    \overset{2\lambda}{C^{\lambda,-\lambda,[2\lambda-2,2]}C^{[2-2\lambda,-2],\lambda,\lambda}},\;\;\overset{2\lambda}{C^{\lambda,\lambda,[-2,2-2\lambda]}C^{[2,2\lambda-2],-\lambda,\lambda}},\;\;\overset{2\lambda}{C^{\lambda,\lambda,[-2,2-2\lambda]}C^{[2,2\lambda-2],-\lambda,\lambda}}\,.
\end{equation}
From this point onward, we write only the necessary
permutations of the small crystal and, for clarity, indicate the total number of derivatives above each pair.

Starting from the small crystal above, we can generate further ones, forming a chain of small crystals:
\begin{subequations}\label{Metsaev_chain}
\begin{align}
&\overset{2\lambda}{C^{\lambda,-\lambda,[2\lambda-2,2]}C^{[2-2\lambda,-2],\lambda,\lambda}},\;\;\overset{2\lambda}{C^{\lambda,\lambda,[-2,2-2\lambda]}C^{[2,2\lambda-2],-\lambda,\lambda}}\,,\\
&\overset{4\lambda-4}{C^{\lambda,2\lambda-2,[\lambda-4,4-3\lambda]}C^{[4-\lambda,3\lambda-4],-\lambda,2\lambda-2}},\;\;\overset{4\lambda-4}{C^{\lambda,-\lambda,[4\lambda-6,2]}C^{[6-4\lambda,-2],2\lambda-2,2\lambda-2}}\,,\\
&\overset{8\lambda-12}{C^{\lambda,4\lambda-6,[3\lambda-8,8-5\lambda]}C^{[8-3\lambda,5\lambda-8],-\lambda,4\lambda-6}},\;\;\overset{8\lambda-12}{C^{\lambda,-\lambda,[8\lambda-14,2]}C^{[14-8\lambda,-2],4\lambda-6,4\lambda-6}}\,,\\
\nonumber
&\cdots,\;\;\cdots\,.
\end{align}
\end{subequations}
Since $|\lambda|>2$,\footnote{Indeed, for $\lambda=2$, we have the SDGR coupling $C^{-2,2,2}$, which does not lead to the full chiral higher-spin theory.} we can reach a pair of arbitrary numbers of total derivatives, and generate the coupling $C^{\lambda,-\lambda,n}$ for arbitrary $n\geq 2$. Taking two such couplings, we can generate the small crystal
\begin{equation}
    \overset{n_{12}}{C^{\lambda,n_1,[n_2-\lambda-2,2-n_1-\lambda]}C^{[\lambda-n_2+2,\lambda+n_1-2],-\lambda,n_2}},\;\;\overset{n_{12}}{C^{\lambda,-\lambda,[n_{12}-2,2]}C^{[2-n_{12},-2],n_1,n_2}}\,.
\end{equation}
The coupling $C^{[2-n_{12},-2],n_1,n_2}$ has now $n_1,n_2$ as arbitrary helicities. Looking again at \eqref{Metsaev_chain}, we see that couplings of the form $C^{\lambda,n_3,2}$ can always be generated. In particular, they can be generated from $C^{\lambda,-\lambda,n}$ and $C^{[2-n_{23},-2],n_3,n_2}$ in this way:
\begin{equation}
    \overset{n_{23}}{C^{\lambda,-\lambda,[n_{23}-2,2]}C^{[2-n_{23},-2],n_3,n_2}},\;\;\overset{n_{23}}{C^{\lambda,n_3,[n_2-\lambda-2,2-n_3-\lambda]}C^{[2+\lambda-n_2,\lambda+n_3-2],-\lambda,n_2}}\,.
\end{equation}
Then by suitably choosing $n_2$, we can always generate $C^{\lambda,n_3,2}$ with arbitrary $n_3$.\\
Now we can pair $C^{\lambda,n_3,2}$ with $C^{n_1,n_2,[2-n_{12},-2]}$ by choosing $n_3=k+2-\lambda-n_{12}$ and generate
\begin{equation}
    \overset{k+2}{C^{n_1,n_2,[k-n_{12},2-n_{12}]}C^{[n_{12}-k,n_{12}-2],\lambda,n_3}},\;\;\overset{k+2}{C^{n_1,n_3,[k-n_{13},2-n_{13}]}C^{[n_{13}-k,n_{13}-2],n_2,\lambda}}\,.
\end{equation}
Then $C^{n_1,n_2,[k-n_{12},2-n_{12}]}$ is a generic cubic coupling $C^{n_1,n_2,n_3}$ involving up to $k$ derivatives with arbitrary $k$. This proves that all couplings must be present.

Therefore, following Appendix A of \cite{Ponomarev:2016lrm} we conclude that the Metsaev solution is the unique one in the presence of at least one higher-derivative coupling of the form $C^{-\lambda,\lambda,\lambda}$.
While this result has long been known in the higher-spin literature, to the best of our knowledge, it has not been rigorously proven before.

As a consequence of the result above, we can make two additional observations.
Firstly, in the presence of a coupling of the type $C^{\lambda,-\lambda,n}$ with $2n-\lambda\geq 2$, one has to add $C^{\lambda,-\lambda,\lambda}$, thereby recovering the Metsaev solution. Secondly, if we assume the existence of at least one nonabelian higher-derivative coupling, the following reasoning applies. Higher derivatives imply that many values of $\omega$ can appear in the exchange. Therefore, starting from a generic pair (assuming $\lambda_1\geq \lambda_2$), we can always recover the specific type of coupling mentioned above:
\begin{align}
    &C^{\lambda_1,\lambda_2,\omega}C^{-\omega,\lambda_3,\lambda_4}&
    &\Longrightarrow &
    &C^{\lambda_1,\lambda_2,-\lambda_2}C^{-\lambda_2,\lambda_3,\lambda_4}\,.
\end{align}
This again leads to the Metsaev solution. In this sense, one can view the emergence of the Metsaev solution as more tied to the presence of high-derivative vertices rather than to the presence of higher-spin fields themselves.

Note that in the presence of a gauge group, we can follow the same logic as above by replacing the condition on $C^{-\lambda,\lambda,\lambda}$ with the requirement $|\lambda|>1$. Consequently, a colour-graviton can self-interact through the coupling $C^{-2,2,2}$ only if all cubic couplings are present.

%%%%%%%%%%%%%%%%%%%%%%%%%%%%%%%%%%%%%%%%%%%%%%%%%%%%%%%%%%%%%
\subsection{Solutions for the couplings}
%%%%%%%%%%%%%%%%%%%%%%%%%%%%%%%%%%%%%%%%%%%%%%%%%%%%%%%%%%%%%
Now that we have the classification, we can solve the system for the couplings for each case. This also provides information about the sub-crystals contained within each inequivalent crystal, as well as explicit solutions for the couplings. The couplings generally have more freedom than in HS-SDGR, HS-SDYM, and the full chiral higher-spin theory.
We do not present all cases here --- only some examples --- to clarify the logic and display explicit solutions.

We begin by noting that the crystal in \eqref{2d_lower_spin} correctly reproduces the result of \eqref{lower_der_lower_spin}. The simplest example involving higher-spin fields is given by \eqref{simplestcase}, which admits the following solutions for the couplings:
\begin{subequations}
\begin{align}
    &\{C^{2,\lambda,-\lambda}=C^{2,2,-2}\}\,,\\
    & \{C^{-2,2,2}\}\,,
\end{align}
\end{subequations}
where we list only the nonzero couplings. The first line corresponds to the full crystal, and the second to its sub-crystal, which coincides with SDGR.

The crystal \eqref{2pcase2} admits the solutions:
\begin{subequations}
\begin{align}
    &\{C^{2,-\lambda_2,\lambda_2}=C^{2,2-\lambda_1-\lambda_2,-2+\lambda_1+\lambda_2}=C^{2,-\lambda_1,\lambda_1}=C^{-2,2,2},\quad C^{\lambda_1,\lambda_2, 2 - \lambda_1 - \lambda_2}\}\,,\\
    &\{C^{2,2,-2}=C^{2,-\lambda_2,\lambda_2}=C^{2,-\lambda_1,\lambda_1}\}\,,\\
    &\{C^{2,2,-2}=C^{2,-\lambda_1,\lambda_1}=C^{2,2-\lambda_1-\lambda_2,-2+\lambda_1+\lambda_2}\}\,,\\
    &\{C^{2,2,-2}=C^{2,-\lambda_2,\lambda_2}=C^{2,2-\lambda_1-\lambda_2,-2+\lambda_1+\lambda_2}\}\,,\\
    &\{C^{2,2,-2}=C^{2,-\lambda_1,\lambda_1}\}\,,\\
    &\{C^{2,2,-2}=C^{2,-\lambda_2,\lambda_2}\}\,,\\
    &\{C^{2,2,-2}=C^{2,2-\lambda_1-\lambda_2,-2+\lambda_1+\lambda_2}\}\,,\\
    &\{C^{2,2,-2}, C^{\lambda_1, \lambda_2, 2 - \lambda_1 - \lambda_2}\}\,.
\end{align}
\end{subequations}
Thus, both \eqref{simplestcase} and SDGR are contained as sub-crystals. Note that the coupling $C^{\lambda_1, 2 - \lambda_1 - \lambda_2, \lambda_2}$ is free and is what makes this crystal more than just a sum of \eqref{simplestcase}.

We give two other examples, where the couplings differ from the standard Metsaev-type solution. The crystal \eqref{example3} has the solution:
\begin{align}
\begin{split}
\{&C^{-2, 2, 2}=C^{0, 0, 2}=C^{2, 2 - 2 \lambda, -2 + 2 \lambda}=C^{2, 2 - \lambda, -2 + \lambda}=C^{2, -\lambda, \lambda},\\
&C^{2 - \lambda, 2 - \lambda, -2 + 2 \lambda}=\frac{(C^{0, 2 - \lambda, \lambda})^2}{C^{2 - 2 \lambda, \lambda, \lambda}},C^{0, 2 - \lambda, \lambda},C^{2 - 2  \lambda, \lambda, \lambda}\}\,.
\end{split}
\end{align}
Here, we only illustrate the solution involving all the couplings, not all the possible truncations. As we can see, there is more freedom compared to HS-SDGR, where all couplings must be equal.

The crystal \eqref{example4} has the solution:
\begin{align}
\begin{split}
    &\{C^{4 - 3 \lambda, 4 - 3 \lambda, -6 + 6 \lambda}=\frac{(C^{0, 4 - 3 \lambda, -2 + 3 \lambda})^2}{C^{6 - 6 \lambda, -2 + 3 \lambda, -2 + 3 \lambda}},
C^{4 - 3 \lambda, 2 - \lambda, -4 + 4 \lambda}=\frac{C^{0, 4 - 3 \lambda, -2 + 3 \lambda} C^{0, 
   2 - \lambda, \lambda}}{C^{4 - 4 \lambda, \lambda, -2 + 3 \lambda}},\\
&C^{2 - 2 \lambda, 2 - \lambda, -2 + 3 \lambda}=\frac{C^{0, 4 - 3 \lambda, -2 + 3 \lambda}, C^{0, 
   2 - \lambda, \lambda}}{C^{4 - 3 \lambda, \lambda, -2 + 2 \lambda}}, 
C^{2 - \lambda, 2 - \lambda, -2 + 2 \lambda}=\frac{(C^{0, 2 - \lambda, \lambda})^2}{ C^{2 - 2 \lambda, \lambda, \lambda}},\\
&C^{4 - 4 \lambda, \lambda, -2 + 3 \lambda}, C^{
 4 - 3 \lambda, \lambda, -2 + 2  \lambda},C^{0, 4 - 3 \lambda, -2 + 3 \lambda}, C^{0, 
 2 - \lambda, \lambda},C^{6 - 6 \lambda, -2 + 3 \lambda, -2 + 3  \lambda},C^{2 - 2 \lambda, \lambda, \lambda}\}\,.
 \end{split}
\end{align}

%%%%%%%%%%%%%%%%%%%%%%%%%%%%%%%%%%%%%%%%%%%%%%%%%%%%%%%%%%%%%
\section{Amplitudes}\label{section5}
%%%%%%%%%%%%%%%%%%%%%%%%%%%%%%%%%%%%%%%%%%%%%%%%%%%%%%%%%%%%%
In \cite{Skvortsov:2020wtf, Skvortsov:2020gpn}, it was shown that the tree-level and one-loop amplitudes of the full chiral higher-spin theory constructed from the Metsaev couplings vanish. A more general statement about solutions of the holomorphic constraints was made in \cite{Ponomarev:2017nrr}, see also \cite{Monteiro:2022lwm,Monteiro:2022xwq}. Here, we show that this result can be extended to all chiral higher-spin theories, with a caveat.

Let us compute the $4$-pt amplitude for a generic chiral higher-spin theory. We use identities from Appendix \ref{AppendixB}, which hold on-shell $(H_2=0)$. The amplitude can be written as
\begin{align}\label{OPE_Associativity}
\begin{split}
\mathcal{A}&=\mathcal{A}_s+\mathcal{A}_t+\mathcal{A}_u=
    \sum_{\omega}\mathcal{C}^{1234\omega}\frac{\PPb_{12}^{\lambda_{12}+\omega}}{\beta_1^{\lambda_1}\beta_2^{\lambda_2}}\frac{1}{(q_1+q_2)^2}\frac{\PPb_{34}^{\lambda_{34}-\omega}}{\beta_3^{\lambda_3}\beta_4^{\lambda_4}}+2\leftrightarrow 4+2\leftrightarrow 3\\
    &=\frac{\PPb_{12}\PPb_{34}}{(q_1+q_2)^2\prod_{i=1}^4\beta_i^{\lambda_i}}\sum_{\omega}\Big(\mathcal{C}^{1234\omega}\PPb_{12}^{\lambda_{12}+\omega-1}\PPb_{34}^{\lambda_{34}-\omega-1}-\mathcal{C}^{1423\omega}\PPb_{14}^{\lambda_{14}+\omega-1}\PPb_{32}^{\lambda_{23}-\omega-1}\\
    &\;\;\;\;-\mathcal{C}^{1324\omega}\PPb_{13}^{\lambda_{13}+\omega-1}\PPb_{24}^{\lambda_{24}-\omega-1}\Big)=\frac{\PPb_{12}\PPb_{34}}{(q_1+q_2)^2\prod_{i=1}^4\beta_i^{\lambda_i}}\sum_{\omega}\Big(\mathcal{C}^{1234\omega}\PPb_{12}^{\lambda_{12}+\omega-1}\PPb_{34}^{\lambda_{34}-\omega-1}\\
&\;\;\;\;+\mathcal{C}^{2314\omega}\PPb_{23}^{\lambda_{23}+\omega-1}\PPb_{14}^{\lambda_{14}-\omega-1}
+\mathcal{C}^{1324\omega}\PPb_{31}^{\lambda_{13}+\omega-1}\PPb_{24}^{\lambda_{24}-\omega-1}\Big)=0\,,
\end{split}
\end{align}
where in the last equality we assume only even couplings, so the sum is over even or odd values of $\omega$ only.
Plugging in the general solution \eqref{symfinalsystem}, we obtain
\begin{align}
\begin{split}
\mathcal{A}=&\frac{k^{1234}_+\PPb_{12}\PPb_{34}}{2^{\Lambda-2}\prod_{i=1}^4\beta_i^{\lambda_i}(q_1+q_2)^2}\Big((\PPb_{12}+\PPb_{34})^{\Lambda-2}-(\PPb_{34}-\PPb_{12})^{\Lambda-2}\\
&+(\PPb_{23}+\PPb_{14})^{\Lambda-2}-(\PPb_{14}-\PPb_{23})^{\Lambda-2}+(\PPb_{31}+\PPb_{24})^{\Lambda-2}-(\PPb_{24}-\PPb_{31})^{\Lambda-2}\Big)\,,
\end{split}
\end{align}
which, using the variables $A,B,C$, becomes
\begin{equation}
    \mathcal{A}=\frac{k^{1234}_+\PPb_{12}\PPb_{34}}{\prod_{i=1}^4\beta_i^{\lambda_i}(q_1+q_2)^2}\Big((A)^{\Lambda-2}-(B)^{\Lambda-2}+(C)^{\Lambda-2}-(-A)^{\Lambda-2}+(-B)^{\Lambda-2}-(-C)^{\Lambda-2}\Big)=0\,.
\end{equation}
The same conclusion holds if we include odd-derivative couplings, as long as the conditions \eqref{case4system} are satisfied.

There is a caveat: in the case discussed above, we did not account for the fact that the product $C^{\lambda_1,\lambda_1,0}C^{0,\lambda_2,\lambda_2}$ can be arbitrary. When this is properly taken into account, certain exceptions to the vanishing of the amplitudes may arise.
For example, we can examine the amplitude in the simplest higher-derivative theory that includes the abelian $R^2\phi$ and/or $R^3$ terms \eqref{HDlowerspin}, and compute the $(2,2,2,2)$ scattering amplitude
\begin{align}
\begin{split}
    \mathcal{A}&(2,2,2,2)=\frac{\PPb_{12}\PPb_{34}}{(q_1+q_2)^2\prod_{i=1}^4\beta_i^2}\sum_{\omega}C^{2,2,\omega}C^{-\omega,2,2}\Big(\PPb_{12}^{3+\omega}\PPb_{34}^{3-\omega}+\PPb_{23}^{3+\omega}\PPb_{14}^{3-\omega}+\PPb_{31}^{3+\omega}\PPb_{24}^{3-\omega}\Big)\\
    &=\frac{\PPb_{12}\PPb_{34}}{(q_1+q_2)^2\prod_{i=1}^4\beta_i^2}(C^{2,2,0}C^{0,2,2}-\frac{10}{3}C^{2,2,-2}C^{2,2,2})\Big(\PPb_{12}^3\PPb_{34}^3+\PPb_{23}^3\PPb_{14}^3+\PPb_{31}^3\PPb_{24}^3\Big)\,.
    \end{split}
\end{align}
Here the freedom in choosing $C^{2,2,0}C^{0,2,2}$ allows for a non-vanishing amplitude. Notice that this does not contradict the Weinberg no-go theorem \cite{Weinberg:1964ev}, since these are abelian (Born-Infeld type) vertices, not constrained by the Weinberg low-energy theorem.\\
Note that in a larger theory (e.g. one including all higher-spin fields), the products $C^{\lambda_1,\lambda_1,0}C^{0,\lambda_2,\lambda_2}$ may be subject to constraints.

We can also demonstrate the vanishing of amplitudes in the presence of a gauge group. Remember that in this case we can separate the colour factor and rewrite it in the form
\begin{equation}
    \mathcal{A}(12\cdots n)=\sum_{\sigma\in S_n/\mathbb{Z}_n}\mathrm{Tr}(T^{a_{\sigma_1}}T^{a_{\sigma_2}}\cdots T^{a_{\sigma_n}})\tilde{\mathcal{A}}(\sigma_1\sigma_2\cdots\sigma_n)\,.
\end{equation}
By considering a single colour-ordered amplitude, namely $[1234]$, since the same is valid for the others, we obtain
\begin{align}
\begin{split}
    \tilde{\mathcal{A}}&(1234)=\frac{\PPb_{12}\PPb_{34}}{(q_1+q_2)^2\prod_{i=1}^4\beta_i^{\lambda_i}}\sum_{\omega}\Big(\mathcal{C}^{1234\omega}\PPb_{12}^{\lambda_{12}+\omega-1}\PPb_{34}^{\lambda_{34}-\omega-1}-\mathcal{C}^{2341\omega}\PPb_{23}^{\lambda_{23}+\omega-1}\PPb_{41}^{\lambda_{41}-\omega-1}\Big)\\
    &=\frac{\PPb_{12}\PPb_{34}}{(q_1+q_2)^2\prod_{i=1}^4\beta_i^{\lambda_i}}\sum_{\omega}\Big(\mathcal{C}^{1234\omega}\PPb_{12}^{\lambda_{12}+\omega-1}\PPb_{34}^{\lambda_{34}-\omega-1}+(-)^{\omega}\mathcal{C}^{1423\omega}\PPb_{14}^{\lambda_{14}+\omega-1}\PPb_{23}^{\lambda_{23}-\omega-1}\Big)\\
    &=\frac{\PPb_{12}\PPb_{34}}{2^{\Lambda-2}(q_1+q_2)^2\prod_{i=1}^4\beta_i^{\lambda_i}}\Big((\PPb_{12}+\PPb_{34})^{\Lambda-2}+(\PPb_{14}-\PPb_{23})^{\Lambda-2}\Big)\sim (A^{\Lambda-2}-A^{\Lambda-2})=0\,.
\end{split}
\end{align}
Here, we include both even- and odd-derivative vertices in the sum.

Again, examples exist where the amplitude does not vanish. For instance, in the lower-spin theory with an $F^2\phi$ and/or $F^3$ interactions, the amplitude $A(1111)$ can be nonzero, thanks to the freedom of the product $C^{1,1,0}C^{0,1,1}$.

In principle, there seems to be no obstruction to extending the analysis of \cite{Skvortsov:2018jea,Skvortsov:2020wtf,Skvortsov:2020gpn} to show that all tree-level amplitudes vanish, up to the small caveat we discussed, for any generic $n$-pt scattering in all chiral higher-spin theories.

%%%%%%%%%%%%%%%%%%%%%%%%%%%%%%%%%%%%%%%%%%%%%%%%%%%%%%%%%%%%%
\section{Conclusions and discussion}\label{section6}
%%%%%%%%%%%%%%%%%%%%%%%%%%%%%%%%%%%%%%%%%%%%%%%%%%%%%%%%%%%%%
One of the main results of the paper is that we have shown that, contrary to common belief, there are nontrivial higher-spin theories with finite spectra of fields. We have found a great number of nontrivial solutions to the quartic consistency condition in the light-cone gauge, which enriches the space of higher-spin theories in $4d$. However, we have not found all of them since we omitted the higher-derivative interactions, and this is the first obvious direction for future work. In addition, the choice of representations for the colour cases has been very restrictive and we have not studied theories that have both coloured and singlet fields under some gauge algebra as well. Lastly, we have completely omitted fermions and, as a result, supersymmetric theories.

So far, we have directly analysed the quartic consistency condition in the light-cone gauge. It was shown in \cite{Ponomarev:2017nrr} that it entails the Jacobi identity for a Lie algebra, save for some degenerate cases. Therefore, it seems plausible that the problem can be solved by the algebraic tools and one can find a class of finite- and infinite-dimensional Lie algebras that lead to consistent higher-spin theories.\footnote{Not all of the new theories feature a spin-two field, i.e. not all of them are ``gravities''.} The latter would make the parallel to the case $3d$ higher-spin gravities even more pronounced.

It is worth discussing to what extent the new theories are ``nontrivial''. Obviously, one can take any number of abelian couplings and add them up, which imposes no constraints in the light-cone gauge. One can also activate some $C^{\lambda_1,\lambda_2,\lambda_3}$ that do not ``talk'' to each other, i.e. no exchange diagram can be formed. These cases are trivial. Whenever an exchange diagram exists, there is a nontrivial equation for the couplings, which quite often forces us to introduce further couplings that contribute to the same exchange diagram to make it Poincaré invariant. From what we can see, there does not seem to be any way to discriminate such cases further, i.e. to divide couplings into ``more nonabelian'' and ``less nonabelian''. Nevertheless, the final results, i.e. the set of couplings that need to be turned on, can be rather different. For example, $C^{+s,+s,-s}$, $s>2$ leads to the full chiral higher-spin gravity and activates all couplings save for $C^{0,0,0}$. Also, the presence of scalar couplings $C^{\lambda_1,\lambda_2,0}$ is very important.

It is also important to note that in the classification we presented we assume that all coupling constants, if more than just one, are nonvanishing. First of all, this gives a simple way to list all cubic vertices that belong to a given theory, which is what we call ``spectrum determines couplings/cubic vertices''. Secondly, the equations for the couplings $C_i$ have the schematic form $C_1C_2+C_3C_4+...=0$ with many degenerate points. Nevertheless, if some of the coupling constants are set to zero, we will have to end up with another theory that belongs to the classification. It would be interesting to study the ``degeneration tree'' that determines the relation between various theories in the classification when some of the coupling constants are set to zero.

There are a number of natural future directions triggered by the results obtained in the paper. 
\begin{description}
    \item[Covariantization.] So far, it has always been possible to covariantize the results obtained in the light-cone gauge, e.g. \cite{Krasnov:2021nsq}, and we expect this to be true for the new theories. Any cubic interaction can easily be written in a covariant way, e.g. one can extract such couplings from the equations of motion of chiral theory in flat space \cite{Sharapov:2022faa,Sharapov:2022wpz,Sharapov:2022awp,Sharapov:2022nps,Sharapov:2023erv}. However, the covariantization of chiral theory forces one to introduce higher-order interactions. Therefore, the covariantization of the new theories may not be straightforward.

    We also note that the covariant form of HS-SDYM and HS-SDGR obtained in \cite{Krasnov:2021nsq} does not contain vertices of type ($--$+) and the scalar field as in \cite{Ponomarev:2017nrr}, see also \cite{Monteiro:2022xwq}.

    \item[Twistor space.] Self-dual theories admit a simple description on twistor space. This applies to HS-SDYM and HS-SDGR as well, \cite{Tran:2021ukl,Herfray:2022prf,Tran:2022tft,Adamo:2022lah}, see also \cite{Adamo:2016ple} for the self-dual conformal higher-spin gravity. However, the twistor formulation of chiral higher-spin gravity was lacking, see \cite{Tran:2022tft} for the first steps and \cite{Mason:2025pbz} for the very recent update. 

    \item[(Anti)-de Sitter space.] All known higher-spin gravities are smooth in the cosmological constant, and one could expect this to be true for the new solutions. However, the light-cone analysis becomes more complicated, see \cite{Metsaev:2018xip}. In particular, there seem to be more constraints, because $AdS_4$ cubic vertices have a more complicated structure compared to their flat space limits, but some interesting results can still be obtained \cite{Skvortsov:2018uru, Neiman:2023bkq, Neiman:2024vit, Lipstein:2023pih, Chowdhury:2024dcy}. Therefore, covariantization of the new theories seems a necessary step to proceed. 

    \item[Quantum corrections.] One can quite generally show that most holomorphic solutions of the quartic light-cone consistency condition have vanishing tree level amplitudes, \cite{Ponomarev:2017nrr} and also the present paper. This implies that there are no UV-divergences in these theories at least at one-loop. However, for the theories that have a finite spectrum, one cannot expect the additional mechanism (regularized sum over the spectrum) that would make them vanish \cite{Skvortsov:2018jea,Skvortsov:2020wtf,Skvortsov:2020gpn}. After all, the scattering of higher-spin fields might have nontrivial S-matrix \cite{Tran:2022amg}. For a more recent analysis of anomalies for higher-spin theories in twistor space, see \cite{Tran:2025uad}.

    \item[Celestial holography.] It was shown in \cite{Ren:2022sws} that chiral higher-spin gravity solves the celestial OPE associativity condition. We expect that all of the theories constructed here (and other, higher derivative, solutions that are yet to be found) do solve the same constraints. On the other hand, the study of flat space holography for chiral higher-spin gravity has just begun in  \cite{Ponomarev:2022atv,Ponomarev:2022ryp,Ponomarev:2022qkx,Monteiro:2022xwq}. It would be interesting to see if similar statements can be made about the new theories.

    \item[Unitary higher-spin theory.] It would be interesting to investigate whether some of these chiral higher-spin theories --- especially the higher-derivative ones --- admit a local unitary completion, given that they display more freedom in their couplings. In any case, studying their unitary completions could provide valuable insights and better control over the non-localities. We have also found that there exist some exotic low-spin couplings, e.g.\footnote{If fermions are included, there is also $C^{0,-1/2,3/2}$. } $C^{-2,1,2}$ that pass certain consistency checks and, in general, more freedom (holomorphic and anti-holomorphic vertices can be considered independent) is available in the choice of vertices as compared to covariant approaches \cite{Bengtsson:1986kh,Metsaev:1991nb,Metsaev:1991mt,Metsaev:1993ap,Bengtsson:2014qza}. Therefore, it is worth investigating if more consistent theories can be constructed with the usual, low-spin fields.

    \item[Higher dimensions.] Recently, some of the chiral higher-spin ``magic'' has been extended to $6d$ and, more generally, to all even dimensions \cite{Basile:2024raj}. One can hope that some of the results of the present paper can be extended to even dimensions.
\end{description}

%%%%%%%%%%%%%%%%%%%%%%%%%%%%%%%%%%%%%%%%%%%%%%%%%%%%%%%%%%
\section*{Acknowledgments}
\label{sec:Aknowledgements}
%%%%%%%%%%%%%%%%%%%%%%%%%%%%%%%%%%%%%%%%%%%%%%%%%%%%%%%%%%
I would like to thank Zhenya Skvortsov for the guidance and numerous discussions during this project.  I am grateful to Dmitry Ponomarev for the many useful comments on the preliminary version of the paper. This project has received funding from the European Research Council (ERC) under the European Union’s Horizon 2020 research and innovation programme (grant agreement No 101002551).

\appendix
%%%%%%%%%%%%%%%%%%%%%%%%%%%%%%%%%%%%%%%%%%%%%%%%%%%%%%%%%%%%%
\section{Standard results and notations}\label{AppendixA}
%%%%%%%%%%%%%%%%%%%%%%%%%%%%%%%%%%%%%%%%%%%%%%%%%%%%%%%%%%%%%
The $4d$ Poincaré algebra $iso(3,1)$ is defined as
\begin{subequations}
\begin{align}
    [P^A,P^B]=&\,0\,,\\
    [J^{AB},P^C]=&\,P^A\eta^{BC}-P^B\eta^{AC}\,,\\
    [J^{AB},J^{CD}]=&\,J^{AD}\eta^{BC}-J^{BD}\eta^{AC}-J^{AC}\eta^{BD}+J^{BC}\eta^{AD}\,,
\end{align}
\end{subequations}
where $P^A$ are the generators of translations and $J^{AB}$ of Lorentz transformations.

To perform canonical quantisation in quantum field theory, one has to choose a hypersurface. A standard approach is to select equal time slices, $t=t_0$, and define the Hamiltonian $H=P^0$ as the operator responsible for evolving the system in the time direction.

In this work, however, we adopt light-front quantisation, where we fix a light-like hypersurface and treat $x^+$ as our time direction. The corresponding Hamiltonian is then given by $H\equiv P^-$.\footnote{Unlike standard equal time slices of the spacetime $\mathcal{M}$ at fixed $x^0$, we consider surfaces of constant $x^+$, and for simplicity we can fix $x^+=0$. Choosing initial data on a light-like surface changes the standard Cauchy problem, requiring careful treatment of boundary conditions. A detailed discussion can be found in \cite{Neville:1971zk,Heinzl:2000ht}.}

We consider general massless higher-spin fields, which in $4d$ possess two degrees of freedom. These are denoted $\phi^{\pm s}$, representing helicities $\pm s$, and are complex conjugates of each other.

We use both light-cone coordinates and the light-cone gauge.
In flat spacetime, we adopt the $4d$ Minkowski metric with mostly plus signature as
\begin{equation}
    ds^2=-(dx^0)^2+\sum^{3}_{i=1}\,(dx^i)^2\,.
\end{equation}
We define light-cone coordinates as
\begin{align}
    &x^+=\frac{x^3+x^0}{\sqrt{2}}\,,&
    &x^-=\frac{x^3-x^0}{\sqrt{2}}\,,&
    &z=\frac{x^1-ix^2}{\sqrt{2}}\,,&
    &\bar{z}=\frac{x^1+ix^2}{\sqrt{2}}\,,
\end{align}
and the metric becomes
\begin{equation}
    ds^2=2\,dx^+ dx^- + 2\,dz d\bar{z}\,.
\end{equation}

We work with Fourier transformed fields with respect to $x^-$ and the transverse directions. The Dirac bracket is given by
\begin{equation}
    [\phi_p^{\mu}(x^+),\phi_q^{\nu}(x^+)]=\delta^{\mu,-\nu}\frac{\delta^3(p+q)}{2p^+}\,.
\end{equation}
The free field realisation of the kinematical Poincaré generators is\footnote{Following standard notations in the light-cone, we rename $\beta=p^+$.}
\begin{subequations}
\begin{align}
    &P^+=\beta\,,&
    &P=q\,,&
    &\bar{P}=\bar{q}\,,\\
    &J^{z+}=-\beta\frac{\partial}{\partial \bar{q}}\,,&
    &J^{\bar{z}+}=-\beta\frac{\partial}{\partial q}\,,&
    &J^{-+}=-N_{\beta}-1\,,\\
    &J^{z\bar{z}}=N_q-N_{\bar{q}}-\lambda\,,
\end{align}
\end{subequations}
where $N_q=q\partial_q$ is the Euler operator.
The dynamical generators are
\begin{align}
    &H_2=-\frac{q\bar{q}}{\beta}\,,&
    &J_2^{z-}=\frac{\partial}{\partial\bar{q}}\frac{q\bar{q}}{\beta}+q\frac{\partial}{\partial\beta}+\lambda\frac{q}{\beta}\,,&
    &J_2^{\bar{z}-}=\frac{\partial}{\partial q}\frac{q\bar{q}}{\beta}+\bar{q}\frac{\partial}{\partial\beta}-\lambda\frac{\bar{q}}{\beta}\,.
\end{align}
We now deform the dynamical generators using a local ansatz:\footnote{Only the dynamical generators are deformed. One advantage of working in light-front quantisation is the reduced number of dynamical generators. Out of $10$ Poincaré generators only $3$ are dynamical ($H,J^{z-},J^{\bar{z}-}$), compared to $4$ in the usual equal-time quatisation ($H^0,J^{0a}$).}
\begin{align}\label{hamiltonian}
    H=&\,H_2+\sum_n\int d^{3n}q\;\delta\Big(\sum_i q_i\Big)h^{q_1,...,q_n}_{\lambda_1,...,\lambda_n}\phi^{\lambda_1}_{q_1}\cdots\phi^{\lambda_n}_{q_n}\,,\\\label{boostz}
    J^{z-}=&\,J_2^{z-}+\sum_n\int d^{3n}q\;\delta\Big(\sum_i q_i\Big)\Big[j^{q_1,...,q_n}_{\lambda_1,...,\lambda_n}-\frac{1}{n}\,h^{q_1,...,q_n}_{\lambda_1,...,\lambda_n}\Big(\sum_j\frac{\partial}{\partial \bar{q}_j}\Big)\Big]\phi^{\lambda_1}_{q_1}\cdots\phi^{\lambda_n}_{q_n}\,,\\ \label{boostzbar}
    J^{\bar{z}-}=&\,J_2^{\bar{z}-}+\sum_n\int d^{3n}q\;\delta\Big(\sum_i q_i\Big)\Big[\bar{j}^{q_1,...,q_n}_{\lambda_1,...,\lambda_n}-\frac{1}{n}\,h^{q_1,...,q_n}_{\lambda_1,...,\lambda_n}\Big(\sum_j\frac{\partial}{\partial q_j}\Big)\Big]\phi^{\lambda_1}_{q_1}\cdots\phi^{\lambda_n}_{q_n}\,.
\end{align}
Let us define the momentum combinations
\begin{align}
    &\PP_{ij}=q_i\beta_j-q_j\beta_i\,,&
    &\PPb_{ij}=\bar{q}_i\beta_j-\bar{q}_j\beta_i\,,
\end{align}
where $\PPb_{ij}=-\PPb_{ji}$ and $\PP_{ij}=-\PP_{ji}$. Solving order by order all constraints required by the closure of the Poincaré algebra \cite{Ponomarev:2016lrm}, leads to the classification of cubic vertices:
\begin{align}
h_{\lambda_1,\lambda_2,\lambda_3}=&\,C^{\lambda_1,\lambda_2,\lambda_3}\frac{\PPb^{\lambda_{123}}}{\beta_1^{\lambda_1}\beta_2^{\lambda_2}\beta_3^{\lambda_3}}+\bar{C}^{-\lambda_1,-\lambda_2,-\lambda_3}\frac{\PP^{-\lambda_{123}}}{\beta_1^{-\lambda_1}\beta_2^{-\lambda_2}\beta_3^{-\lambda_3}}\,,\\
    j_{\lambda_1,\lambda_2,\lambda_3}=&\,\frac{2}{3}\,C^{\lambda_1,\lambda_2,\lambda_3}\frac{\PPb^{\lambda_{123}-1}}{\beta_1^{\lambda_1}\beta_2^{\lambda_2}\beta_3^{\lambda_3}}\Lambda^{\lambda_1,\lambda_2,\lambda_3}\,,\\
    \bar{j}_{\lambda_1,\lambda_2,\lambda_3}=&\,-\frac{2}{3}\,\bar{C}^{-\lambda_1,-\lambda_2,-\lambda_3}\frac{\PP^{-\lambda_{123}-1}}{\beta_1^{-\lambda_1}\beta_2^{-\lambda_2}\beta_3^{-\lambda_3}}\Lambda^{\lambda_1,\lambda_2,\lambda_3}\,,
\end{align}
with $\lambda_{123}=\lambda_1+\lambda_2+\lambda_3$ and where we define
\begin{align}
    &\PP^a_{12}=\PP^a_{23}=\PP^a_{31}=\PP^a=\frac{1}{3}\,\Big[(\beta_1-\beta_2)q_3^a+(\beta_2-\beta_3)q_1^a+(\beta_3-\beta_1)q_2^a\Big]\,,\\
    &\Lambda^{\lambda_1,\lambda_2,\lambda_3}=\,\beta_1(\lambda_2-\lambda_3)+\beta_2(\lambda_3-\lambda_1)+\beta_3(\lambda_1-\lambda_2)\,.
\end{align}
Note that due to momentum conservation $\PP$ is cyclic invariant, then $\sigma_{123}\PP=\PP\,,$ same for $\PPb$.

Here $C^{\lambda_1,\lambda_2,\lambda_3}$ and $\bar{C}^{\lambda_1,\lambda_2,\lambda_3}$ are a priori independent coupling constants.
For dimensional reasons, we can introduce a length parameter to compensate for the powers of momenta. We denote it by $\ell_P$, which can naturally be associated with the Planck length. The couplings take the form
\begin{align}
    &C^{\lambda_1,\lambda_2,\lambda_3}=(\ell_P)^{\lambda_{123}-1}c^{\lambda_1,\lambda_2,\lambda_3}\,,&
    &\bar{C}^{\lambda_1,\lambda_2,\lambda_3}=(\ell_P)^{\lambda_{123}-1}\bar{c}^{\lambda_1,\lambda_2,\lambda_3}\,.
\end{align}
If we demand unitarity, in the case with no internal symmetry, we have to impose
\begin{equation}
    \bar{C}^{\lambda_1,\lambda_2,\lambda_3}=(C^{\lambda_1,\lambda_2,\lambda_3})^*\,.
\end{equation}
If we include a gauge group, unitarity imposes
\begin{align}
     \bar{C}^{\lambda_1,\lambda_2,\lambda_3}&= (-)^{\lambda_{123}}(C^{\lambda_1,\lambda_2,\lambda_3})^*&&\text{for:}\;SO(N),USp(N)\,.
\end{align}
%%%%%%%%%%%%%%%%%%%%%%%%%%%%%%%%%%%%%%%%%%%%%%%%%%%%%%%%%%%%%
\section{Useful relations}\label{AppendixB}
%%%%%%%%%%%%%%%%%%%%%%%%%%%%%%%%%%%%%%%%%%%%%%%%%%%%%%%%%%%%%
In $4d$ and light-cone gauge, for a given $N$-pt function, there are $N-2$ independent $\PPb_{ij}$ variables.
In particular, for $4$-pt scattering, we have $2$ independent $\PPb$ variables, which, for example, can be chosen to be $\PPb_{12}$, $\PPb_{34}$. Additionally, there are three independent $\beta$'s. All other $\PPb_{ij}$ can be expressed as
\begin{subequations}
\begin{align}
    \PPb_{13}&=\frac{\beta_3 \PPb_{12}+\beta_1 \PPb_{34}}{\beta_1+\beta_2}\,,
    &&\PPb_{14}=-\frac{\PPb_{12} (\beta_1+\beta_2+\beta_3)+\beta_1 \PPb_{34}}{\beta_1+\beta_2}\,,\\
    \PPb_{23}&=\frac{\beta_2 \PPb_{34}-\beta_3 \PPb_{12}}{\beta_1+\beta_2}\,,
    &&\PPb_{24}=\frac{\PPb_{12} (\beta_1+\beta_2+\beta_3)-\beta_2 \PPb_{34}}{\beta_1+\beta_2}\,.
\end{align}
\end{subequations}
Below, we collect some useful relations for $n > 0$ that are employed in the main text:
\begin{subequations}
\begin{align}
    \sum_{k=0}^{n}\begin{pmatrix}
        n\\
        k
\end{pmatrix}k\,(k-1)\cdots (k-\ell+1)=&\,n\,(n-1)\cdots (n-\ell+1)\,2^{n-\ell}\,,\\
\sum_{k\in\text{even/odd}}^{n}\begin{pmatrix}
        n\\
        k
\end{pmatrix}k\,(k-1)\cdots (k-\ell+1)=&\,
n\,(n-1)\cdots (n-\ell+1)\,2^{n-\ell-1}\,.
\end{align}
\end{subequations}
These identities can be derived by taking $\ell$ derivatives of the binomial theorem. In particular, the following relations will also prove useful:
\begin{align}
    \sum_{k=0}^n
    &\begin{pmatrix}
        n\\
        k
    \end{pmatrix}
    (-)^k(A-B)^k(A+B)^{n-k}=2^nB^n\,,\\
 \sum_{k=0}^n
    &\begin{pmatrix}
        n\\
        k
    \end{pmatrix}
    (A-B)^k(A+B)^{n-k}=2^nA^n\,,\\
    \sum_{k\in\text{even}}^n
    &\begin{pmatrix}
        n\\
        k
    \end{pmatrix}
    (A-B)^k(A+B)^{n-k}=2^{n-1}(A^n+B^n),\\
    \sum_{k\in\text{odd}}^n
    &\begin{pmatrix}
        n\\
        k
    \end{pmatrix}
    (A-B)^k(A+B)^{n-k}=2^{n-1}(A^n-B^n)\,,
\end{align}
\vspace{-2em}  % Adjust this value to reduce vertical spacing
\begin{align}
    \begin{split}
    \sum_{k=0}^n
    \begin{pmatrix}
        n\\
        k
    \end{pmatrix}
    (-)^k k\,(k-&1)\cdots (k-\ell+1)(A-B)^k(A+B)^{n-k}\\
    &=n\,(n-1)\cdots (n-\ell+1)(B-A)^{\ell}(2B)^{n-\ell}\,,
    \end{split}
\end{align}
\vspace{-2em}  % Adjust this value to reduce vertical spacing
\begin{align}
\begin{split}
    \sum_{k=0}^n
    \begin{pmatrix}
        n\\
        k
    \end{pmatrix}
    k\,(k-1)\cdots& (k-\ell+1)(A-B)^k(A+B)^{n-k}\\
    &=n\,(n-1)\cdots (n-\ell+1)(A-B)^{\ell}(2A)^{n-\ell}\,.
     \end{split}
\end{align}
Finally, for $4$-pt scattering, the following relations hold on-shell:
\begin{align}
    &(q_i+q_j)^2=-\frac{2}{\beta_i\beta_j}\PP_{ij}\PPb_{ij}\,,& 
    &\sum_{j=1}^4\frac{\PP_{ij}\PPb_{jk}}{\beta_j}=0\,.
\end{align}
From these, we deduce
\begin{equation}
    \frac{\PPb_{12}\PPb_{34}}{(q_1+q_2)^2}=\frac{\PPb_{31}\PPb_{24}}{(q_1+q_3)^2}=\frac{\PPb_{14}\PPb_{23}}{(q_1+q_4)^2}\,.
\end{equation}
%%%%%%%%%%%%%%%%%%%%%%%%%%%%%%%%%%%%%%%%%%%%%%%%%%%%%%%%%%%%%
\section{Low-derivative constraints}\label{AppendixC}
%%%%%%%%%%%%%%%%%%%%%%%%%%%%%%%%%%%%%%%%%%%%%%%%%%%%%%%%%%%%%
We begin by analysing the cases $\Lambda = 2$ and $\Lambda = 3$ for the constraint \eqref{LCholo}. For simplicity, we adopt the following notation in both cases:
\begin{equation}\label{simple_notation}
    C^{1234}_i\equiv C^{\lambda_1,\lambda_2,i-\lambda_{12}}C^{\lambda_{12}-i,\lambda_3,\Lambda-\lambda_{123}}\,,
\end{equation}
with analogous definitions for the orderings $(1324)$ and $(1423)$.

\paragraph{Case $\Lambda=2$.} In this case, only one-derivative interactions are present, and the constraint is
\begin{subequations}
\begin{align}
    &(\lambda_{13}-1)(C^{1234}_1-C^{1324}_1+C^{1423}_1)=0\,,\\
    &(1-\lambda_{23})(C^{1234}_1-C^{1324}_1+C^{1423}_1)=0\,,\\
    &(\lambda_{12}-1)(C^{1234}_1-C^{1324}_1+C^{1423}_1)=0\,.
\end{align}
\end{subequations}
These equations are consistent with the antisymmetry of the couplings, and they admit the solution
\begin{equation}\label{L=2_constraint}
\boxed{
    C^{1234}_1=C^{1324}_1-C^{1423}_1\,.
    }
\end{equation}
Therefore, in one-derivative theories, it is possible to find solutions even when only odd-derivative vertices are present. This stands in contrast to the general case, where odd-derivative vertices must be accompanied by even-derivative ones for consistent solutions to exist.

As expected, this solution does not lead to a self-interacting photon. For example, assuming a theory with the coupling $C^{-1,1,1}$, substituting it into \eqref{L=2_constraint} gives $C^{-1,1,1}=0$. This result also follows directly from symmetry arguments --- in particular, $C^{\lambda,\lambda,\lambda'} \equiv 0$ for odd-derivative vertices, with the sole exception of $C^{0,0,1}$, which gives rise to the self-dual sector of scalar QED. In this case, the symmetry argument no longer applies because the two scalar fields need not be identical; rather, they can be complex conjugates of one another. Accordingly, it is more appropriate to write the coupling as $C^{0,\bar{0},1}$ corresponding to the cubic interaction $\phi\bar{\phi}A^+$. Furthermore, the constraint \eqref{L=2_constraint} leaves $C^{0,\bar{0},1}$ unconstrained.

\paragraph{Case $\Lambda=3$.} In this case, the sum runs over two values of $\omega$ and includes two-derivative interactions. The constraint becomes
\begin{subequations}
\begin{align}
    (2\lambda_{13}-3)((-)^{\lambda_{12}}C_1^{1234}+(-)^{\lambda_{12}+1}C_2^{1234}+(-)^{\lambda_{14}}C_1^{1423}+(-)^{\lambda_{14}}C_2^{1423})=&\,0\,,\\
    (3-2\lambda_{23})((-)^{\lambda_{12}}C_1^{1234}+(-)^{\lambda_{12}}C_2^{1234}+(-)^{\lambda_{13}}C_1^{1324}+(-)^{\lambda_{13}}C_2^{1324})=&\,0\,,\\
    (2\lambda_{12}-3)((-)^{\lambda_{13}}C_1^{1324}+(-)^{\lambda_{13}+1}C_2^{1324}+(-)^{\lambda_{14}}C_1^{1423}+(-)^{\lambda_{14}+1}C_2^{1423})=&\,0\,.
\end{align}
\end{subequations}
Using the definitions in \eqref{definitions1}, this can be rewritten as
\begin{subequations}
\begin{align}
    (2\lambda_{13}-3)(k^{1234}_--k^{1423}_+)=&\,0\,,\qquad
    (3-2\lambda_{23})(k^{1234}_++k^{1324}_+)=\,0\,,\\
    (2\lambda_{12}-3)(k^{1324}_-+k^{1423}_-)=&\,0\,.
\end{align}
\end{subequations}
These are the same conditions found for the general case \eqref{case4system}.
\paragraph{Case $\Lambda=2$ with a gauge group.} We now consider the constraint \eqref{LCholocolour} at $\Lambda=2$ with a $U(N)$ gauge group. The constraint for the $[1234]$ colour-ordering takes the form
\begin{subequations}
\begin{align}
    (\lambda_{13}-1)((-)^{1-\lambda_{12}}\theta_{1-\lambda_{12}}C^{1234}_1+(-)^{\lambda_{14}}\theta_{1-\lambda_{14}}C^{4123}_1)=&\,0\,,\\
    (1-\lambda_{23})((-)^{1-\lambda_{12}}\theta_{1-\lambda_{12}}C^{1234}_1+(-)^{1-\lambda_{14}}\theta_{\lambda_{14}}C^{4123}_1)=&\,0\,,\\
    (\lambda_{12}-1)((-)^{1-\lambda_{12}}\theta_{1-\lambda_{12}}C^{1234}_1+(-)^{\lambda_{14}}\theta_{1-\lambda_{14}}C^{4123}_1)=&\,0\,,
\end{align}
\end{subequations}
and gives the solution
\begin{align}\label{U(N)_1d_constraint}
    &(-)^{1-\lambda_{12}}\theta_{1-\lambda_{12}}C^{1234}_1=(-)^{1-\lambda_{14}}\theta_{1-\lambda_{14}}C^{4123}_1&
    &\Rightarrow&
    &C^{1234}_1=(-)^{\lambda_{24}}\theta_{1-\lambda_{12}}\theta_{1-\lambda_{14}}C^{4123}_1\,.
\end{align}
Meanwhile, the constraint corresponding to the colour ordering $[2341]$ leads to
\begin{equation}
    C^{2341}_1=(-)^{\lambda_{13}}\theta_{1-\lambda_{23}}\theta_{1-\lambda_{14}}C^{1234}_1\,.
\end{equation}
Using the identity $C^{4123}_1=C^{2341}_1$, we find $\theta_{1-\lambda_{23}}\theta_{1-\lambda_{14}}=1$, which implies four possible choices
\begin{align}
    \theta_{\lambda}=+1,-1,(-)^{\lambda},(-)^{\lambda+1}\,.
\end{align}
The cases $\theta_{\lambda}=+1,-1$ and $\theta_{\lambda}=(-)^{\lambda},(-)^{\lambda+1}$ yield, respectively
\begin{equation}
\boxed{
    C_1^{1234} = (-)^{\lambda_{24}} C_1^{4123}\,,
}
\qquad
\boxed{
    C_1^{1234} = C_1^{4123}\,.
}
\end{equation}
In the main text, we consider $C_1^{1234}=C^{4123}_1$, in analogy with the higher-derivative case. Moreover, for the classification of crystals, the specific values of the couplings are not important.

As we will see in Appendix \ref{AppendixE}, the special solutions with $\theta_{\omega}=\pm 1$ are closely tied to the existence of solutions in which all fields transform in the adjoint representation. 

For the gauge groups $SO(N)$ and $USp(N)$, by using the symmetry properties given in \eqref{SO(n)_sym} and \eqref{USp(n)_sym} of Appendix \ref{AppendixD}, we arrive at the same constraint \eqref{U(N)_1d_constraint} as in the $U(N)$ case. However, here the solutions to the various colour-ordered constraints mix due to these symmetry properties, which also determine the symmetry of the couplings. In particular, we find the following solutions.

For $\theta_{\lambda}=(-)^{\lambda},(-)^{\lambda+1}$ the solution is given by
\begin{equation}
\boxed{
    C_1^{1234}=C^{4123}_1=C^{1324}_1\,,
    }
\end{equation}
and for $SO(N)$ the symmetry of the couplings is
\begin{align}
&C^{\lambda_1,\lambda_2,\lambda_3}=C^{\lambda_{\sigma_1},\lambda_{\sigma_2},\lambda_{\sigma_3}}\quad(\theta_{\omega}=(-)^{\lambda})\,,&
    &C^{\lambda_1,\lambda_2,\lambda_3}=-C^{\lambda_{\sigma_1},\lambda_{\sigma_2},\lambda_{\sigma_3}}\quad(\theta_{\omega}=(-)^{\lambda+1})\,,
\end{align}
while for $USp(N)$ the symmetry of the couplings is
\begin{align}
&C^{\lambda_1,\lambda_2,\lambda_3}=C^{\lambda_{\sigma_1},\lambda_{\sigma_2},\lambda_{\sigma_3}}\quad(\theta_{\omega}=(-)^{\lambda+1})\,,&
    &C^{\lambda_1,\lambda_2,\lambda_3}=-C^{\lambda_{\sigma_1},\lambda_{\sigma_2},\lambda_{\sigma_3}}\quad(\theta_{\omega}=(-)^{\lambda})\,.
\end{align}
For $\theta_{\omega}=+1,-1$ the solution is given by
\begin{equation}
\boxed{
    C_1^{1234}=(-)^{\lambda_{24}} C^{4123}_1=(-)^{\lambda_{14}} C^{1324}_1\,,
    }
\end{equation}
and for $SO(N)$ the symmetry of the couplings is
\begin{align}
&C^{\lambda_1,\lambda_2,\lambda_3}=C^{\lambda_{\sigma_1},\lambda_{\sigma_2},\lambda_{\sigma_3}}\quad(\theta_{\omega}=-1)\,,&
    &C^{\lambda_1,\lambda_2,\lambda_3}=-C^{\lambda_{\sigma_1},\lambda_{\sigma_2},\lambda_{\sigma_3}}\quad(\theta_{\omega}=1)\,,
\end{align}
while for $USp(N)$ the symmetry of the couplings is
\begin{align}
&C^{\lambda_1,\lambda_2,\lambda_3}=C^{\lambda_{\sigma_1},\lambda_{\sigma_2},\lambda_{\sigma_3}}\quad(\theta_{\omega}=1)\,,&
    &C^{\lambda_1,\lambda_2,\lambda_3}=-C^{\lambda_{\sigma_1},\lambda_{\sigma_2},\lambda_{\sigma_3}}\quad(\theta_{\omega}=-1)\,.
\end{align}

%%%%%%%%%%%%%%%%%%%%%%%%%%%%%%%%%%%%%%%%%%%%%%%%%%%%%%%%%%%%%
\section{SO(N) and USp(N) gauge groups}\label{AppendixD}
%%%%%%%%%%%%%%%%%%%%%%%%%%%%%%%%%%%%%%%%%%%%%%%%%%%%%%%%%%%%%
Without diving into explicit computation, we examine the $SO(N)$ and $USp(N)$ cases. For $SO(N)$, the Poisson bracket takes the form
\begin{equation}\label{SO(N)poisson_bracket}
[(\phi^{\lambda}_p)_{AB},(\phi^{\mu}_q)_{CD}]=\frac{\delta^{\lambda,-\mu}\delta^3(p+q)}{2p^+}\,(\delta_{AC}\delta_{BD}+\theta_{\lambda}\delta_{AD}\delta_{BC})\,,
\end{equation}
and also here, due to the antisymmetry of the Poisson bracket, we have $\theta_\lambda = \theta_{-\lambda}$.
Moreover, in this case, the phase can be related to the symmetry properties of the field under transposition
\begin{equation} 
[(\phi^{\lambda}_p)_{BA},(\phi^{\mu}_q)_{CD}]=\frac{\delta^{\lambda,-\mu}\delta^3(p+q)}{2p^+}\,(\delta_{BC}\delta_{AD}+\theta_{\lambda}\delta_{BD}\delta_{AC})= \theta_\lambda [(\phi^{\lambda}_p)_{AB},(\phi^{\mu}_q)_{CD}]\,,
\end{equation} 
where we used the identity $\theta_{\lambda}^2 = \theta_{\lambda} \theta_{-\lambda} = 1$ and we find the property $(\phi^{\lambda}_q)_{BA}=\theta_{\lambda}(\phi^{\lambda}_q)_{AB}$. This implies the following symmetry for the couplings:
\begin{align}\label{SO(n)_sym}
    &(\phi^{\lambda}_q)_{BA}=\theta_{\lambda}(\phi^{\lambda}_q)_{AB}&\Rightarrow&
    &C^{\lambda_1,\lambda_2,\lambda_3}=(-)^{\lambda_{123}}\theta_{\lambda_1}\theta_{\lambda_2}\theta_{\lambda_3}C^{\lambda_{\sigma_1},\lambda_{\sigma_2},\lambda_{\sigma_3}}\,,
\end{align}
where $\sigma\in\Sigma_3$ is an odd permutation. Given this, the Poisson bracket yields
\begin{equation}
    [\mathrm{Tr}(\phi^{\lambda_1}_{q_1}\phi^{\lambda_2}_{q_2}\phi^{\omega}_{q_{\omega}}),\mathrm{Tr}(\phi^{\lambda_3}_{q_3}\phi^{\lambda_4}_{q_4}\phi^{-\omega}_{-q_{\omega}})]=\frac{1}{2q^+_{\omega}}\Big(\theta_{\lambda_3}\theta_{\lambda_4}\mathrm{Tr}(\phi^{\lambda_1}_{q_1}\phi^{\lambda_2}_{q_2}\phi^{\lambda_4}_{q_4}\phi^{\lambda_3}_{q_3})+\theta_{\omega}\mathrm{Tr}(\phi^{\lambda_1}_{q_1}\phi^{\lambda_2}_{q_2}\phi^{\lambda_3}_{q_3}\phi^{\lambda_4}_{q_4})\Big)\,,
\end{equation}
where we contract only $\phi^{\omega}_{q_{\omega}}$, and the constraint becomes
\begin{align}\label{SOnconstraint}
\begin{split}
\sum_{\omega}(-)^{\omega}\text{Cycl}\Big[&\Big(\theta_{\omega}\mathcal{C}^{1234\omega}+(-)^{\lambda_{34}+\omega}\theta_{\lambda_3}\theta_{\lambda_4}\mathcal{C}^{1243\omega}\Big)\times\\
&\frac{(\lambda_1+\omega-\lambda_2)\beta_1-(\lambda_2+\omega-\lambda_1)\beta_2}{2(\beta_1+\beta_2)}\,
    \PPb_{12}^{\lambda_{12}+\omega-1}\PPb_{34}^{\lambda_{34}-\omega}\Big]=0\,.
\end{split}
\end{align}
Using \eqref{SO(n)_sym} we find the relation
\begin{equation}
    \mathcal{C}^{1243\omega}=(-)^{\lambda_{34}+\omega}\theta_{\lambda_3}\theta_{\lambda_4}\theta_{\lambda_\omega}\mathcal{C}^{1234\omega}\,.
\end{equation}
The constraint \eqref{SOnconstraint} can then be solved in the same way as for $U(N)$. If we additionally require that there be no relative signs among the various couplings, as in Metsaev's solution, we obtain the condition
\begin{align}
   &\theta_{\omega}(1+\theta_{\lambda_3}^2\theta_{\lambda_4}^2)=2(-)^{\omega}&
   &\Rightarrow&
   &\theta_{\omega}=(-)^{\omega}\,.
\end{align}
For $USp(N)$, the Poisson bracket is
\begin{equation}\label{USp(N)poisson_bracket}
[(\phi^{\lambda}_p)_{AB},(\phi^{\mu}_q)_{CD}]=\frac{\delta^{\lambda,-\mu}\delta^3(p+q)}{2p^+}\,(C_{AC}C_{BD}+\theta_{\lambda}C_{AD}C_{BC})\,,
\end{equation}
where $C_{AB}$ is the antisymmetric invariant tensor
\begin{align}
    &C_{AB}=-C_{BA}\,,&
    &C_{AB}C^{CB}=\delta_A^C\,.
\end{align}
The matrices $C$ raise and lower indices as follows: $V^A=C^{AB}V_B$ and $V^BC_{BA}=V_A$. As before, we can find the properties
\begin{align}\label{USp(n)_sym}
    &\theta_\lambda = \theta_{-\lambda}\,,&
    &(\phi^{\lambda}_q)_{BA}=\theta_{\lambda}(\phi^{\lambda}_q)_{AB}\,,&
    &C^{\lambda_1,\lambda_2,\lambda_3}=(-)^{\lambda_{123}+1}\theta_{\lambda_1}\theta_{\lambda_2}\theta_{\lambda_3}C^{\lambda_{\sigma_1},\lambda_{\sigma_2},\lambda_{\sigma_3}}\,,
\end{align}
where $\sigma\in\Sigma_3$ is an odd permutation. Given this, the Poisson bracket yields
\begin{equation}
    [\mathrm{Tr}(\phi^{\lambda_1}_{q_1}\phi^{\lambda_2}_{q_2}\phi^{\omega}_{q_{\omega}}),\mathrm{Tr}(\phi^{\lambda_3}_{q_3}\phi^{\lambda_4}_{q_4}\phi^{-\omega}_{-q_{\omega}})]=-\frac{1}{2q^+_{\omega}}\Big(\theta_{\lambda_3}\theta_{\lambda_4}\mathrm{Tr}(\phi^{\lambda_1}_{q_1}\phi^{\lambda_2}_{q_2}\phi^{\lambda_4}_{q_4}\phi^{\lambda_3}_{q_3})+\theta_{\omega}\mathrm{Tr}(\phi^{\lambda_1}_{q_1}\phi^{\lambda_2}_{q_2}\phi^{\lambda_3}_{q_3}\phi^{\lambda_4}_{q_4})\Big)\,,
\end{equation}
where we contract only $\phi^{\omega}_{q_{\omega}}$, and the constraint becomes
\begin{align}\label{USpnconstraint}
\begin{split}
\sum_{\omega}(-)^{\omega+1}\text{Cycl}\Big[&\Big(\theta_{\omega}\mathcal{C}^{1234\omega}-(-)^{\lambda_{34}+\omega}\theta_{\lambda_3}\theta_{\lambda_4}\mathcal{C}^{1243\omega}\Big)\times\\
&\frac{(\lambda_1+\omega-\lambda_2)\beta_1-(\lambda_2+\omega-\lambda_1)\beta_2}{2(\beta_1+\beta_2)}\, \PPb_{12}^{\lambda_{12}+\omega-1}\PPb_{34}^{\lambda_{34}-\omega}\Big]=0\,.
\end{split}
\end{align}
This constraint can then be solved as before. If we additionally require that there be no relative signs among the various couplings, as in Metsaev's solution, we obtain the condition
\begin{align}
   &\theta_{\omega}(1+\theta_{\lambda_3}^2\theta_{\lambda_4}^2)=2(-)^{\omega+1}&
   &\Rightarrow&
   \theta_{\omega}=(-)^{\omega+1}\,.
\end{align}
Using the values of $\theta_{\lambda}$ found for the two cases, we get the properties
\begin{align}\label{sym_gauge}
    SO(N):\quad
    (\phi^\lambda_q)_{AB}=&\;(-)^{\lambda}(\phi^\lambda_q)_{BA}\quad
    &&\Rightarrow\quad
    C^{\lambda_1,\lambda_2,\lambda_3}= C^{\lambda_{\sigma_1},\lambda_{\sigma_2},\lambda_{\sigma_3}}\,,\\
    USp(N):\quad
    (\phi^\lambda_q)_{AB}=&\;(-)^{\lambda+1}(\phi^\lambda_q)_{BA}\quad
    &&\Rightarrow\quad
    C^{\lambda_1,\lambda_2,\lambda_3}=C^{\lambda_{\sigma_1},\lambda_{\sigma_2},\lambda_{\sigma_3}}\,,
\end{align}
where $\sigma\in\Sigma_3$ is an odd permutation.
These results are consistent with the analysis in \cite{Skvortsov:2020wtf}, which was carried out under the assumption of Metsaev-type couplings only.
Notably, in all cases, fields with odd helicity always take values in the adjoint representation of the respective gauge group. This matches the Chan-Paton structure appearing in open string theory \cite{Marcus:1982fr}. 

Note also that the couplings are fully symmetric, which is consistent with the fact that, for example, Yang-Mills theory exists for all compact simple gauge groups, thus including both $SO(N)$ and $USp(N)$. This would not have been possible if the couplings turned out to be antisymmetric. 

Due to the symmetry of the couplings in \eqref{sym_gauge}, the solution to the constraint for both $SO(N)$ and $USp(N)$ --- once all colour-ordered constraints are taken into account --- takes the form
\begin{equation}
\boxed{
\begin{aligned}\label{symfinalsystem_SO(N)}
    &\mathcal{C}^{1234\omega}=\frac{k^{1234}_-(\Lambda-2)!}{2^{\Lambda-2}(\lambda_{12}+\omega-1)!(\lambda_{34}-\omega-1)!}\quad 
    \forall\;\omega\,,\quad\text{same for $(1324)$ and $(1423)$}\,,\\
    &k_-^{1234}=k_-^{1324}=k_-^{1423}\,,\qquad C^{\lambda_1,\lambda_1,0}C^{0,\lambda_2,\lambda_2}=\;\text{generic}\,.
\end{aligned}
}
\end{equation}
%%%%%%%%%%%%%%%%%%%%%%%%%%%%%%%%%%%%%%%%%%%%%%%%%%%%%%%%%%%%%
\section{All fields in the adjoint}\label{AppendixE}
%%%%%%%%%%%%%%%%%%%%%%%%%%%%%%%%%%%%%%%%%%%%%%%%%%%%%%%%%%%%%
Following \cite{Ponomarev:2016lrm}, we assume that all fields take values in the adjoint representation of some Lie algebra of internal symmetry with structure constant $f^{a_1a_2a_3}$. In this setup, the space-time part of the cubic vertices remains unchanged, while the coupling constant is multiplied by the structure constant
\begin{equation}\label{newcouplings}
    C^{\lambda_1,\lambda_2,\lambda_3}\quad\rightarrow\quad C^{\lambda_1,\lambda_2,\lambda_3}f^{a_1a_2a_3}\,.
\end{equation}
Since the structure constant is antisymmetric, even-derivative couplings become antisymmetric, and vice versa; the symmetry properties of the vertices are effectively reversed.
As a result, the new symmetry property for the coupling is
\begin{equation}\label{new_sym}
    C^{\lambda_1,\lambda_2,\lambda_3}=(-)^{\lambda_{123}+1}C^{\lambda_{\sigma_1},\lambda_{\sigma_2},\lambda_{\sigma_3}}\,.
\end{equation}
where $\sigma\in\Sigma_3$ represents an odd permutation.
This implies that odd-derivative vertices involving two identical fields can be nonzero. For example, the Yang-Mills vertex $C^{1,1,-1}\neq 0$ is allowed. We now rewrite Eq.~\eqref{LCholo} using the shorthand notation
\begin{equation}
    F_{1234}(A,B)+F_{1324}(B,C)+F_{1423}(C,A)=0\,,
\end{equation}
where each term represents a different line of Eq.~\eqref{LCholo}. Upon substituting \eqref{newcouplings} into this expression, we obtain%\etexco
\begin{equation}
    F_{1234}(A,B) f_{a_1a_2b}f^b_{\phantom{b}a_3a_4}+ F_{1324}(B,C) f_{a_1a_3b}f^b_{\phantom{b}a_2a_4}+ F_{1423}(C,A) f_{a_1a_4b}f^b_{\phantom{b}a_2a_3}=0\,,
\end{equation}
and using the Jacobi identity, we find two constraints:\\
$F_{1234}(A,B)=F_{1423}(C,A)$ :
\begin{align}\label{fabc1}
    \begin{split}
    &\sum_{\omega} (-)^{\omega}\big[((\lambda_{13}-\lambda_{24})A+(\lambda_{14}-\lambda_{23})B-2\omega C)\,\mathcal{C}^{1234\omega}(A-B)^{\lambda_{12}+\omega-1}(A+B)^{\lambda_{34}-\omega-1}\\
    &-((\lambda_{12}-\lambda_{34})C+(\lambda_{13}-\lambda_{24})A-2\omega B)\,\mathcal{C}^{1423\omega}(C-A)^{\lambda_{14}+\omega-1}(C+A)^{\lambda_{23}-\omega-1}\big]=0\,.
    \end{split}
\end{align}
$F_{1234}(A,B)=-F_{1324}(B,C)$ : 
\begin{align}\label{fabc2}
    \begin{split}
    &\sum_{\omega} \big[(-)^{\omega}((\lambda_{13}-\lambda_{24})A+(\lambda_{14}-\lambda_{23})B-2\omega C)\,\mathcal{C}^{1234\omega}(A-B)^{\lambda_{12}+\omega-1}(A+B)^{\lambda_{34}-\omega-1}\\
    &+(-)^{\lambda_{24}}((\lambda_{14}-\lambda_{23})B+(\lambda_{12}-\lambda_{34})C-2\omega A)\,\mathcal{C}^{1324\omega}(B-C)^{\lambda_{13}+\omega-1}(B+C)^{\lambda_{24}-\omega-1}\big]=0\,.
    \end{split}
\end{align}
To solve the constraints \eqref{fabc1} and \eqref{fabc2}, we can follow the same procedure applied previously for Eq.~\eqref{LCholocolour}.  Using \eqref{Unconstraint}, we rewrite here the colour-ordered constraints for $[1234]$ and $[1243]$, respectively:
\begin{align}
\nonumber
    &\sum_{\omega} \big[(-)^{\omega}\theta_{\omega}((\lambda_{13}-\lambda_{24})A+(\lambda_{14}-\lambda_{23})B-2\omega C)\,
    \mathcal{C}^{1234\omega}(A-B)^{\lambda_{12}+\omega-1}(A+B)^{\lambda_{34}-\omega-1}\\
    &+(-)^{\lambda_{14}}\theta_{\omega}((\lambda_{12}-\lambda_{34})C+(\lambda_{13}-\lambda_{24})A-2\omega B)\,
    \mathcal{C}^{4123\omega}(C-A)^{\lambda_{14}+\omega-1}(C+A)^{\lambda_{23}-\omega-1}\big]=0\,,\\
\nonumber
    &\sum_{\omega} \big[(-)^{\lambda_{34}}\theta_{\omega}((\lambda_{13}-\lambda_{24})A+(\lambda_{14}-\lambda_{23})B-2\omega C)\,
    \mathcal{C}^{1243\omega}(A-B)^{\lambda_{12}+\omega-1}(A+B)^{\lambda_{34}-\omega-1}\\
    &+(-)^{\Lambda+\omega}\theta_{\omega}((\lambda_{14}-\lambda_{23})B+(\lambda_{12}-\lambda_{34})C-2\omega A)\,
    \mathcal{C}^{3124\omega}(B-C)^{\lambda_{13}+\omega-1}(B+C)^{\lambda_{24}-\omega-1}\big]=0\,.
\end{align}
By setting $\theta_{\omega}=-1$ and applying the modified symmetry relation \eqref{nocolour_coupling_sym}, we match them to the constraints \eqref{fabc1} and \eqref{fabc2}, respectively.

From the previous analysis, we already know that a solution with $\theta_{\omega}=-1$ exists only for $\Lambda=2$, that is, in the presence of one-derivative interactions only. This corresponds to one of the special solutions found at the end of Appendix \ref{AppendixC}.
The solution is then given by
\begin{equation}
\boxed{
    C_1^{1234}=(-)^{\lambda_{24}} C^{4123}_1=(-)^{\lambda_{14}} C^{1324}_1\,.
    }
\end{equation}
If we further assume all helicities to be odd, we obtain
\begin{equation}
\boxed{
    C_1^{1234}=C^{4123}_1=C^{1324}_1\,.
    }
\end{equation}
In particular, the HS-SDYM theory truncated to only odd helicities constitutes a consistent solution. This agrees with the findings of \cite{Ponomarev:2016lrm}, although the specific truncations and possible extensions involving both odd and even helicities were not explicitly identified there. Nevertheless, the conclusion that consistent solutions involving fields all in the adjoint representation of the same internal symmetry algebra exist only for one-derivative theories is confirmed.

\newpage
\footnotesize
\providecommand{\href}[2]{#2}\begingroup\raggedright\endgroup


\begin{thebibliography}{100}

\bibitem{Fronsdal:1978rb}
C.~Fronsdal, ``Massless fields with integer spin,'' {\em Phys. Rev.} {\bf D18} (1978)
3624.
%%CITATION = PHRVA,D18,3624;%%.

\bibitem{Flato:1978qz}
M.~Flato and C.~Fronsdal, ``{One Massless Particle Equals Two Dirac Singletons: Elementary Particles in a Curved Space. 6.},'' {\em Lett.Math.Phys.} {\bf 2} (1978)
421--426.
%%CITATION = LMPHD,2,421;%%.

\bibitem{Flato:1980zk}
M.~Flato and C.~Fronsdal, ``{On {DIS} and Racs},'' {\em Phys. Lett. B} {\bf 97} (1980) 236--240.

\bibitem{Sezgin:2002rt}
E.~Sezgin and P.~Sundell, ``{Massless higher spins and holography},'' {\em Nucl.Phys.} {\bf B644} (2002) 303--370,
\href{http://arXiv.org/abs/hep-th/0205131}{{\tt hep-th/0205131}}.
%%CITATION = HEP-TH/0205131;%%.

\bibitem{Klebanov:2002ja}
I.~R. Klebanov and A.~M. Polyakov, ``{AdS dual of the critical $O(N)$ vector model},'' {\em Phys. Lett.} {\bf B550} (2002) 213--219,
\href{http://arXiv.org/abs/hep-th/0210114}{{\tt hep-th/0210114}}.
%%CITATION = HEP-TH/0210114;%%.

\bibitem{Sezgin:2003pt}
E.~Sezgin and P.~Sundell, ``{Holography in 4D (super) higher spin theories and a test via cubic scalar couplings},'' {\em JHEP} {\bf 0507} (2005) 044,
\href{http://arXiv.org/abs/hep-th/0305040}{{\tt hep-th/0305040}}.
%%CITATION = HEP-TH/0305040;%%.

\bibitem{Leigh:2003gk}
R.~G. Leigh and A.~C. Petkou, ``{Holography of the N=1 higher spin theory on AdS(4)},'' {\em JHEP} {\bf 0306} (2003) 011,
\href{http://arXiv.org/abs/hep-th/0304217}{{\tt hep-th/0304217}}.
%%CITATION = HEP-TH/0304217;%%.

\bibitem{Giombi:2011kc}
S.~Giombi, S.~Minwalla, S.~Prakash, S.~P. Trivedi, S.~R. Wadia, and X.~Yin, ``{Chern-Simons Theory with Vector Fermion Matter},'' {\em Eur. Phys. J.} {\bf C72} (2012) 2112,
\href{http://arXiv.org/abs/1110.4386}{{\tt 1110.4386}}.
%%CITATION = ARXIV:1110.4386;%%.

\bibitem{Dirac:1949cp}
P.~A.~M. Dirac, ``{Forms of Relativistic Dynamics},'' {\em Rev. Mod. Phys.} {\bf 21} (1949) 392--399.

\bibitem{Bengtsson:1983pg}
A.~K.~H. Bengtsson, I.~Bengtsson, and L.~Brink, ``{Cubic interaction terms for arbitrarily extended Supermultiplets},'' {\em Nucl. Phys.} {\bf B227} (1983)
41.
%%CITATION = NUPHA,B227,41;%%.

\bibitem{Bengtsson:1983pd}
A.~K.~H. Bengtsson, I.~Bengtsson, and L.~Brink, ``{Cubic interaction terms for arbitrary spin},'' {\em Nucl. Phys.} {\bf B227} (1983)
31.
%%CITATION = NUPHA,B227,31;%%.

\bibitem{Bengtsson:1986kh}
A.~K.~H. Bengtsson, I.~Bengtsson, and N.~Linden, ``{Interacting Higher Spin Gauge Fields on the Light Front},'' {\em Class. Quant. Grav.} {\bf 4} (1987)
1333.
%%CITATION = CQGRD,4,1333;%%.

\bibitem{Berends:1984wp}
F.~A. Berends, G.~J.~H. Burgers, and H.~Van~Dam, ``{On spin three selfinteractions},'' {\em Z. Phys.} {\bf C24} (1984)
247--254.
%%CITATION = ZEPYA,C24,247;%%.

\bibitem{Berends:1984rq}
F.~A. Berends, G.~J.~H. Burgers, and H.~van Dam, ``{On the theoretical problems in constructing interactions involving higher spin massless particles},'' {\em Nucl. Phys.} {\bf B260} (1985)
295.
%%CITATION = NUPHA,B260,295;%%.

\bibitem{Weinberg:1964ew}
S.~Weinberg, ``{Photons and Gravitons in S Matrix Theory: Derivation of Charge Conservation and Equality of Gravitational and Inertial Mass},'' {\em Phys. Rev.} {\bf 135} (1964)
B1049--B1056.
%%CITATION = PHRVA,135,B1049;%%.

\bibitem{Coleman:1967ad}
S.~R. Coleman and J.~Mandula, ``{All Possible Symmetries of the S Matrix},'' {\em Phys. Rev.} {\bf 159} (1967)
1251--1256.
%%CITATION = PHRVA,159,1251;%%.

\bibitem{Maldacena:2011jn}
J.~Maldacena and A.~Zhiboedov, ``{Constraining Conformal Field Theories with A Higher Spin Symmetry},''
\href{http://arXiv.org/abs/1112.1016}{{\tt 1112.1016}}.
%%CITATION = ARXIV:1112.1016;%%.

\bibitem{Fitzpatrick:2012cg}
A.~L. Fitzpatrick and J.~Kaplan, ``{AdS Field Theory from Conformal Field Theory},'' {\em JHEP} {\bf 02} (2013) 054, \href{http://arXiv.org/abs/1208.0337}{{\tt 1208.0337}}.

\bibitem{Boulanger:2013zza}
N.~Boulanger, D.~Ponomarev, E.~D. Skvortsov, and M.~Taronna, ``{On the uniqueness of higher-spin symmetries in AdS and CFT},'' {\em Int. J. Mod. Phys.} {\bf A28} (2013) 1350162,
\href{http://arXiv.org/abs/1305.5180}{{\tt 1305.5180}}.
%%CITATION = ARXIV:1305.5180;%%.

\bibitem{Alba:2013yda}
V.~Alba and K.~Diab, ``{Constraining conformal field theories with a higher spin symmetry in d=4},''
\href{http://arXiv.org/abs/1307.8092}{{\tt 1307.8092}}.
%%CITATION = ARXIV:1307.8092;%%.

\bibitem{Alba:2015upa}
V.~Alba and K.~Diab, ``{Constraining conformal field theories with a higher spin symmetry in $d> 3$ dimensions},''
\href{http://arXiv.org/abs/1510.02535}{{\tt 1510.02535}}.
%%CITATION = ARXIV:1510.02535;%%.

\bibitem{Sleight:2021iix}
C.~Sleight and M.~Taronna, ``{On the consistency of (partially-)massless matter couplings in de Sitter space},'' {\em JHEP} {\bf 10} (2021) 156, \href{http://arXiv.org/abs/2106.00366}{{\tt 2106.00366}}.

\bibitem{Fradkin:1986qy}
E.~S. Fradkin and M.~A. Vasiliev, ``{Cubic Interaction in Extended Theories of Massless Higher Spin Fields},'' {\em Nucl. Phys.} {\bf B291} (1987)
141.
%%CITATION = NUPHA,B291,141;%%.

\bibitem{Aragone:1979hx}
C.~Aragone and S.~Deser, ``{Consistency Problems of Hypergravity},'' {\em Phys. Lett.} {\bf B86} (1979)
161.
%%CITATION = PHLTA,B86,161;%%.

\bibitem{Krasnov:2021nsq}
K.~Krasnov, E.~Skvortsov, and T.~Tran, ``{Actions for self-dual Higher Spin Gravities},'' {\em JHEP} {\bf 08} (2021) 076, \href{http://arXiv.org/abs/2105.12782}{{\tt 2105.12782}}.

\bibitem{Bekaert:2010hp}
X.~Bekaert, N.~Boulanger, and S.~Leclercq, ``{Strong obstruction of the Berends-Burgers-van Dam spin-3 vertex},'' {\em J.Phys.} {\bf A43} (2010) 185401,
\href{http://arXiv.org/abs/1002.0289}{{\tt 1002.0289}}.
%%CITATION = ARXIV:1002.0289;%%.

\bibitem{Ponomarev:2017nrr}
D.~Ponomarev, ``{Chiral Higher Spin Theories and Self-Duality},'' {\em JHEP} {\bf 12} (2017) 141,
\href{http://arXiv.org/abs/1710.00270}{{\tt 1710.00270}}.
%%CITATION = ARXIV:1710.00270;%%.

\bibitem{Roiban:2017iqg}
R.~Roiban and A.~A. Tseytlin, ``{On four-point interactions in massless higher spin theory in flat space},'' {\em JHEP} {\bf 04} (2017) 139,
\href{http://arXiv.org/abs/1701.05773}{{\tt 1701.05773}}.
%%CITATION = ARXIV:1701.05773;%%.

\bibitem{Bekaert:2015tva}
X.~Bekaert, J.~Erdmenger, D.~Ponomarev, and C.~Sleight, ``{Quartic AdS Interactions in Higher-Spin Gravity from Conformal Field Theory},'' {\em JHEP} {\bf 11} (2015) 149,
\href{http://arXiv.org/abs/1508.04292}{{\tt 1508.04292}}.
%%CITATION = ARXIV:1508.04292;%%.

\bibitem{Maldacena:2015iua}
J.~Maldacena, D.~Simmons-Duffin, and A.~Zhiboedov, ``{Looking for a bulk point},'' {\em JHEP} {\bf 01} (2017) 013,
\href{http://arXiv.org/abs/1509.03612}{{\tt 1509.03612}}.
%%CITATION = ARXIV:1509.03612;%%.

\bibitem{Sleight:2017pcz}
C.~Sleight and M.~Taronna, ``{Higher-Spin Gauge Theories and Bulk Locality},'' {\em Phys. Rev. Lett.} {\bf 121} (2018), no.~17, 171604,
\href{http://arXiv.org/abs/1704.07859}{{\tt 1704.07859}}.
%%CITATION = ARXIV:1704.07859;%%.

\bibitem{Ponomarev:2017qab}
D.~Ponomarev, ``{A Note on (Non)-Locality in Holographic Higher Spin Theories},'' {\em Universe} {\bf 4} (2018), no.~1, 2,
\href{http://arXiv.org/abs/1710.00403}{{\tt 1710.00403}}.
%%CITATION = ARXIV:1710.00403;%%.

\bibitem{Neiman:2023orj}
Y.~Neiman, ``{Quartic locality of higher-spin gravity in de Sitter and Euclidean anti-de Sitter space},'' {\em Phys. Lett. B} {\bf 843} (2023) 138048, \href{http://arXiv.org/abs/2302.00852}{{\tt 2302.00852}}.

\bibitem{Blencowe:1988gj}
M.~Blencowe, ``{A Consistent Interacting Massless Higher Spin Field Theory in $D$ = (2+1)},'' {\em Class.Quant.Grav.} {\bf 6} (1989)
443.
%%CITATION = CQGRD,6,443;%%.

\bibitem{Bergshoeff:1989ns}
E.~Bergshoeff, M.~P. Blencowe, and K.~S. Stelle, ``{Area Preserving Diffeomorphisms and Higher Spin Algebra},'' {\em Commun. Math. Phys.} {\bf 128} (1990)
213.
%%CITATION = CMPHA,128,213;%%.

\bibitem{Campoleoni:2010zq}
A.~Campoleoni, S.~Fredenhagen, S.~Pfenninger, and S.~Theisen, ``{Asymptotic symmetries of three-dimensional gravity coupled to higher-spin fields},'' {\em JHEP} {\bf 1011} (2010) 007,
\href{http://arXiv.org/abs/1008.4744}{{\tt 1008.4744}}.
%%CITATION = ARXIV:1008.4744;%%.

\bibitem{Henneaux:2010xg}
M.~Henneaux and S.-J. Rey, ``{Nonlinear $W_{\infty}$ as Asymptotic Symmetry of Three-Dimensional Higher Spin Anti-de Sitter Gravity},'' {\em JHEP} {\bf 1012} (2010) 007,
\href{http://arXiv.org/abs/1008.4579}{{\tt 1008.4579}}.
%%CITATION = ARXIV:1008.4579;%%.

\bibitem{Grigoriev:2020lzu}
M.~Grigoriev, K.~Mkrtchyan, and E.~Skvortsov, ``{Matter-free higher spin gravities in 3D: Partially-massless fields and general structure},'' {\em Phys. Rev. D} {\bf 102} (2020), no.~6, 066003, \href{http://arXiv.org/abs/2005.05931}{{\tt 2005.05931}}.

\bibitem{Pope:1989vj}
C.~N. Pope and P.~K. Townsend, ``{Conformal Higher Spin in (2+1)-dimensions},'' {\em Phys. Lett.} {\bf B225} (1989)
245--250.
%%CITATION = PHLTA,B225,245;%%.

\bibitem{Fradkin:1989xt}
E.~S. Fradkin and V.~{\relax Ya}. Linetsky, ``{A Superconformal Theory of Massless Higher Spin Fields in $D$ = (2+1)},'' {\em Mod. Phys. Lett.} {\bf A4} (1989) 731.
[Annals Phys.198,293(1990)].
%%CITATION = MPLAE,A4,731;%%.

\bibitem{Grigoriev:2019xmp}
M.~Grigoriev, I.~Lovrekovic, and E.~Skvortsov, ``{New Conformal Higher Spin Gravities in $3d$},'' {\em JHEP} {\bf 01} (2020) 059,
\href{http://arXiv.org/abs/1909.13305}{{\tt 1909.13305}}.
%%CITATION = ARXIV:1909.13305;%%.

\bibitem{Sharapov:2024euk}
A.~Sharapov, E.~Skvortsov, and A.~Sukhanov, ``{Matter-coupled higher spin gravities in 3d: no- and yes-go results},'' {\em JHEP} {\bf 04} (2025) 155, \href{http://arXiv.org/abs/2409.12830}{{\tt 2409.12830}}.

\bibitem{Alkalaev:2020kut}
K.~Alkalaev and X.~Bekaert, ``{On BF-type higher-spin actions in two dimensions},'' {\em JHEP} {\bf 05} (2020) 158, \href{http://arXiv.org/abs/2002.02387}{{\tt 2002.02387}}.

\bibitem{Segal:2002gd}
A.~Y. Segal, ``{Conformal higher spin theory},'' {\em Nucl. Phys.} {\bf B664} (2003) 59--130,
\href{http://arXiv.org/abs/hep-th/0207212}{{\tt hep-th/0207212}}.
%%CITATION = HEP-TH/0207212;%%.

\bibitem{Tseytlin:2002gz}
A.~A. Tseytlin, ``{On limits of superstring in $AdS_5\times S^5$},'' {\em Theor. Math. Phys.} {\bf 133} (2002) 1376--1389, \href{http://arXiv.org/abs/hep-th/0201112}{{\tt hep-th/0201112}}.
[Teor. Mat. Fiz.133,69(2002)].
%%CITATION = HEP-TH/0201112;%%.

\bibitem{Bekaert:2010ky}
X.~Bekaert, E.~Joung, and J.~Mourad, ``{Effective action in a higher-spin background},'' {\em JHEP} {\bf 02} (2011) 048,
\href{http://arXiv.org/abs/1012.2103}{{\tt 1012.2103}}.
%%CITATION = ARXIV:1012.2103;%%.

\bibitem{Basile:2022nou}
T.~Basile, M.~Grigoriev, and E.~Skvortsov, ``{Covariant action for conformal higher spin gravity},'' {\em J. Phys. A} {\bf 56} (2023), no.~38, 385402, \href{http://arXiv.org/abs/2212.10336}{{\tt 2212.10336}}.

\bibitem{Metsaev:1991mt}
R.~R. Metsaev, ``{Poincare invariant dynamics of massless higher spins: Fourth order analysis on mass shell},'' {\em Mod. Phys. Lett.} {\bf A6} (1991)
359--367.
%%CITATION = MPLAE,A6,359;%%.

\bibitem{Metsaev:1991nb}
R.~R. Metsaev, ``{$S$ matrix approach to massless higher spins theory. 2: The Case of internal symmetry},'' {\em Mod. Phys. Lett.} {\bf A6} (1991)
2411--2421.
%%CITATION = MPLAE,A6,2411;%%.

\bibitem{Ponomarev:2016lrm}
D.~Ponomarev and E.~D. Skvortsov, ``{Light-Front Higher-Spin Theories in Flat Space},'' {\em J. Phys.} {\bf A50} (2017), no.~9, 095401,
\href{http://arXiv.org/abs/1609.04655}{{\tt 1609.04655}}.
%%CITATION = ARXIV:1609.04655;%%.

\bibitem{Ponomarev:2024jyg}
D.~Ponomarev, ``{Chiral higher-spin double copy},'' {\em JHEP} {\bf 01} (2025) 143, \href{http://arXiv.org/abs/2409.19449}{{\tt 2409.19449}}.

\bibitem{Adamo:2022lah}
T.~Adamo and T.~Tran, ``{Higher-spin Yang\textendash{}Mills, amplitudes and self-duality},'' {\em Lett. Math. Phys.} {\bf 113} (2023), no.~3, 50, \href{http://arXiv.org/abs/2210.07130}{{\tt 2210.07130}}.

\bibitem{Sperling:2017dts}
M.~Sperling and H.~C. Steinacker, ``{Covariant 4-dimensional fuzzy spheres, matrix models and higher spin},'' {\em J. Phys.} {\bf A50} (2017), no.~37, 375202,
\href{http://arXiv.org/abs/1704.02863}{{\tt 1704.02863}}.
%%CITATION = ARXIV:1704.02863;%%.

\bibitem{deMelloKoch:2018ivk}
R.~de~Mello~Koch, A.~Jevicki, K.~Suzuki, and J.~Yoon, ``{AdS Maps and Diagrams of Bi-local Holography},'' {\em JHEP} {\bf 03} (2019) 133,
\href{http://arXiv.org/abs/1810.02332}{{\tt 1810.02332}}.
%%CITATION = ARXIV:1810.02332;%%.

\bibitem{Metsaev:1993ap}
R.~R. Metsaev, ``{Generating function for cubic interaction vertices of higher spin fields in any dimension},'' {\em Mod. Phys. Lett.} {\bf A8} (1993)
2413--2426.
%%CITATION = MPLAE,A8,2413;%%.

\bibitem{Metsaev:2005ar}
R.~R. Metsaev, ``{Cubic interaction vertices for massive and massless higher spin fields},'' {\em Nucl. Phys.} {\bf B759} (2006) 147--201,
\href{http://arXiv.org/abs/hep-th/0512342}{{\tt hep-th/0512342}}.
%%CITATION = HEP-TH/0512342;%%.

\bibitem{Metsaev:2007rn}
R.~R. Metsaev, ``{Cubic interaction vertices for fermionic and bosonic arbitrary spin fields},'' {\em Nucl. Phys. B} {\bf 859} (2012) 13--69, \href{http://arXiv.org/abs/0712.3526}{{\tt 0712.3526}}.

\bibitem{Metsaev:2018xip}
R.~R. Metsaev, ``{Light-cone gauge cubic interaction vertices for massless fields in AdS(4)},'' {\em Nucl. Phys.} {\bf B936} (2018) 320--351,
\href{http://arXiv.org/abs/1807.07542}{{\tt 1807.07542}}.
%%CITATION = ARXIV:1807.07542;%%.

\bibitem{Boulanger:2008tg}
N.~Boulanger, S.~Leclercq, and P.~Sundell, ``{On The Uniqueness of Minimal Coupling in Higher-Spin Gauge Theory},'' {\em JHEP} {\bf 08} (2008) 056,
\href{http://arXiv.org/abs/0805.2764}{{\tt 0805.2764}}.
%%CITATION = ARXIV:0805.2764;%%.

\bibitem{Manvelyan:2010je}
R.~Manvelyan, K.~Mkrtchyan, and W.~Ruehl, ``{A Generating function for the cubic interactions of higher spin fields},'' {\em Phys.Lett.} {\bf B696} (2011) 410--415,
\href{http://arXiv.org/abs/1009.1054}{{\tt 1009.1054}}.
%%CITATION = ARXIV:1009.1054;%%.

\bibitem{Boulanger:2012dx}
N.~Boulanger, D.~Ponomarev, and E.~Skvortsov, ``{Non-abelian cubic vertices for higher-spin fields in anti-de Sitter space},'' {\em JHEP} {\bf 1305} (2013) 008,
\href{http://arXiv.org/abs/1211.6979}{{\tt 1211.6979}}.
%%CITATION = ARXIV:1211.6979;%%.

\bibitem{Francia:2016weg}
D.~Francia, G.~L. Monaco, and K.~Mkrtchyan, ``{Cubic interactions of Maxwell-like higher spins},'' {\em JHEP} {\bf 04} (2017) 068,
\href{http://arXiv.org/abs/1611.00292}{{\tt 1611.00292}}.
%%CITATION = ARXIV:1611.00292;%%.

\bibitem{Bekaert:2022poo}
X.~Bekaert, N.~Boulanger, A.~Campoleoni, M.~Chiodaroli, D.~Francia, M.~Grigoriev, E.~Sezgin, and E.~Skvortsov, ``{Snowmass White Paper: Higher Spin Gravity and Higher Spin symmetry},'' \href{http://arXiv.org/abs/2205.01567}{{\tt 2205.01567}}.

\bibitem{Bekaert:2002uh}
X.~Bekaert, N.~Boulanger, and M.~Henneaux, ``Consistent deformations of dual formulations of linearized gravity: A no-go result,'' {\em Phys. Rev.} {\bf D67} (2003) 044010,
\href{http://arXiv.org/abs/hep-th/0210278}{{\tt hep-th/0210278}}.
%%CITATION = HEP-TH/0210278;%%.

\bibitem{Sharapov:2022faa}
A.~Sharapov, A.~Sharapov, E.~Skvortsov, E.~Skvortsov, A.~Sukhanov, A.~Sukhanov, R.~Van~Dongen, and R.~Van~Dongen, ``{Minimal model of Chiral Higher Spin Gravity},'' {\em JHEP} {\bf 09} (2022) 134, \href{http://arXiv.org/abs/2205.07794}{{\tt 2205.07794}}. [Erratum: JHEP 02, 183 (2023)].

\bibitem{Sharapov:2022wpz}
A.~Sharapov, E.~Skvortsov, and R.~Van~Dongen, ``{Chiral higher spin gravity and convex geometry},'' {\em SciPost Phys.} {\bf 14} (2023), no.~6, 162, \href{http://arXiv.org/abs/2209.01796}{{\tt 2209.01796}}.

\bibitem{Sharapov:2022awp}
A.~Sharapov and E.~Skvortsov, ``{Chiral higher spin gravity in (A)dS4 and secrets of Chern{\textendash}Simons matter theories},'' {\em Nucl. Phys. B} {\bf 985} (2022) 115982, \href{http://arXiv.org/abs/2205.15293}{{\tt 2205.15293}}.

\bibitem{Sharapov:2022nps}
A.~Sharapov, E.~Skvortsov, A.~Sukhanov, and R.~Van~Dongen, ``{More on Chiral Higher Spin Gravity and convex geometry},'' {\em Nucl. Phys. B} {\bf 990} (2023) 116152, \href{http://arXiv.org/abs/2209.15441}{{\tt 2209.15441}}.

\bibitem{Sharapov:2023erv}
A.~Sharapov, E.~Skvortsov, and R.~Van~Dongen, ``{Strong homotopy algebras for chiral higher spin gravity via Stokes theorem},'' {\em JHEP} {\bf 06} (2024) 186, \href{http://arXiv.org/abs/2312.16573}{{\tt 2312.16573}}.

\bibitem{Skvortsov:2024rng}
E.~Skvortsov and Y.~Yin, ``{Low spin solutions of higher spin gravity: BPST instanton},'' {\em JHEP} {\bf 07} (2024) 032, \href{http://arXiv.org/abs/2403.17148}{{\tt 2403.17148}}.

\bibitem{Tran:2025yzd}
T.~Tran, ``{Self-dual pp-wave solutions in chiral higher-spin gravity},'' {\em JHEP} {\bf 03} (2025) 041, \href{http://arXiv.org/abs/2501.06445}{{\tt 2501.06445}}.

\bibitem{Skvortsov:2018jea}
E.~D. Skvortsov, T.~Tran, and M.~Tsulaia, ``{Quantum Chiral Higher Spin Gravity},'' {\em Phys. Rev. Lett.} {\bf 121} (2018), no.~3, 031601,
\href{http://arXiv.org/abs/1805.00048}{{\tt 1805.00048}}.
%%CITATION = ARXIV:1805.00048;%%.

\bibitem{Skvortsov:2020wtf}
E.~Skvortsov, T.~Tran, and M.~Tsulaia, ``{More on Quantum Chiral Higher Spin Gravity},'' {\em Phys. Rev.} {\bf D101} (2020), no.~10, 106001,
\href{http://arXiv.org/abs/2002.08487}{{\tt 2002.08487}}.
%%CITATION = ARXIV:2002.08487;%%.

\bibitem{Skvortsov:2020gpn}
E.~Skvortsov and T.~Tran, ``{One-loop Finiteness of Chiral Higher Spin Gravity},'' {\em JHEP} {\bf 07} (2020) 021, \href{http://arXiv.org/abs/2004.10797}{{\tt 2004.10797}}.

\bibitem{Tsulaia:2022csz}
M.~Tsulaia and D.~Weissman, ``{Supersymmetric quantum chiral higher spin gravity},'' {\em JHEP} {\bf 12} (2022) 002, \href{http://arXiv.org/abs/2209.13907}{{\tt 2209.13907}}.

\bibitem{Neiman:2024vit}
Y.~Neiman, ``{Higher-spin self-dual General Relativity: 6d and 4d pictures, covariant vs. lightcone},'' {\em JHEP} {\bf 07} (2024) 178, \href{http://arXiv.org/abs/2404.18589}{{\tt 2404.18589}}.

\bibitem{Skvortsov:2018uru}
E.~Skvortsov, ``{Light-Front Bootstrap for Chern-Simons Matter Theories},'' {\em JHEP} {\bf 06} (2019) 058,
\href{http://arXiv.org/abs/1811.12333}{{\tt 1811.12333}}.
%%CITATION = ARXIV:1811.12333;%%.

\bibitem{Jain:2024bza}
S.~Jain, D.~K. S, and E.~Skvortsov, ``{Hidden sectors of Chern-Simons matter theories and exact holography},'' {\em Phys. Rev. D} {\bf 111} (2025), no.~10, 106017, \href{http://arXiv.org/abs/2405.00773}{{\tt 2405.00773}}.

\bibitem{Aharony:2024nqs}
O.~Aharony, R.~R. Kalloor, and T.~Kukolj, ``{A chiral limit for Chern-Simons-matter theories},'' {\em JHEP} {\bf 10} (2024) 051, \href{http://arXiv.org/abs/2405.01647}{{\tt 2405.01647}}.

\bibitem{Ponomarev:2016cwi}
D.~Ponomarev, ``{Off-Shell Spinor-Helicity Amplitudes from Light-Cone Deformation Procedure},'' {\em JHEP} {\bf 12} (2016) 117,
\href{http://arXiv.org/abs/1611.00361}{{\tt 1611.00361}}.
%%CITATION = ARXIV:1611.00361;%%.

\bibitem{Monteiro:2022xwq}
R.~Monteiro, ``{From Moyal deformations to chiral higher-spin theories and to celestial algebras},'' {\em JHEP} {\bf 03} (2023) 062, \href{http://arXiv.org/abs/2212.11266}{{\tt 2212.11266}}.

\bibitem{Boulanger:2000rq}
N.~Boulanger, T.~Damour, L.~Gualtieri, and M.~Henneaux, ``{Inconsistency of interacting, multigraviton theories},'' {\em Nucl. Phys.} {\bf B597} (2001) 127--171,
\href{http://arXiv.org/abs/hep-th/0007220}{{\tt hep-th/0007220}}.
%%CITATION = HEP-TH/0007220;%%.

\bibitem{Fradkin:1986ka}
E.~S. Fradkin and M.~A. Vasiliev, ``Candidate to the role of higher spin symmetry,'' {\em Ann. Phys.} {\bf 177} (1987)
63.
%%CITATION = APNYA,177,63;%%.

\bibitem{Arkani-Hamed:2017jhn}
N.~Arkani-Hamed, T.-C. Huang, and Y.-t. Huang, ``{Scattering amplitudes for all masses and spins},'' {\em JHEP} {\bf 11} (2021) 070, \href{http://arXiv.org/abs/1709.04891}{{\tt 1709.04891}}.

\bibitem{Monteiro:2022lwm}
R.~Monteiro, ``{Celestial chiral algebras, colour-kinematics duality and integrability},'' {\em JHEP} {\bf 01} (2023) 092, \href{http://arXiv.org/abs/2208.11179}{{\tt 2208.11179}}.

\bibitem{Weinberg:1964ev}
S.~Weinberg, ``{Feynman Rules for Any Spin. 2. Massless Particles},'' {\em Phys. Rev.} {\bf 134} (1964)
B882--B896.
%%CITATION = PHRVA,134,B882;%%.

\bibitem{Tran:2021ukl}
T.~Tran, ``{Twistor constructions for higher-spin extensions of (self-dual) Yang-Mills},'' {\em JHEP} {\bf 11} (2021) 117, \href{http://arXiv.org/abs/2107.04500}{{\tt 2107.04500}}.

\bibitem{Herfray:2022prf}
Y.~Herfray, K.~Krasnov, and E.~Skvortsov, ``{Higher-spin self-dual Yang-Mills and gravity from the twistor space},'' {\em JHEP} {\bf 01} (2023) 158, \href{http://arXiv.org/abs/2210.06209}{{\tt 2210.06209}}.

\bibitem{Tran:2022tft}
T.~Tran, ``{Toward a twistor action for chiral higher-spin gravity},'' {\em Phys. Rev. D} {\bf 107} (2023), no.~4, 046015, \href{http://arXiv.org/abs/2209.00925}{{\tt 2209.00925}}.

\bibitem{Adamo:2016ple}
T.~Adamo, P.~Hähnel, and T.~McLoughlin, ``{Conformal higher spin scattering amplitudes from twistor space},'' {\em JHEP} {\bf 04} (2017) 021,
\href{http://arXiv.org/abs/1611.06200}{{\tt 1611.06200}}.
%%CITATION = ARXIV:1611.06200;%%.

\bibitem{Mason:2025pbz}
L.~Mason and A.~Sharma, ``{Chiral higher-spin theories from twistor space},'' \href{http://arXiv.org/abs/2505.09419}{{\tt 2505.09419}}.

\bibitem{Neiman:2023bkq}
Y.~Neiman, ``{Self-dual gravity in de Sitter space: Light-cone ansatz and static-patch scattering},'' {\em Phys. Rev. D} {\bf 109} (2024), no.~2, 024039, \href{http://arXiv.org/abs/2303.17866}{{\tt 2303.17866}}.

\bibitem{Lipstein:2023pih}
A.~Lipstein and S.~Nagy, ``{Self-Dual Gravity and Color-Kinematics Duality in AdS4},'' {\em Phys. Rev. Lett.} {\bf 131} (2023), no.~8, 081501, \href{http://arXiv.org/abs/2304.07141}{{\tt 2304.07141}}.

\bibitem{Chowdhury:2024dcy}
C.~Chowdhury, G.~Doran, A.~Lipstein, R.~Monteiro, S.~Nagy, and K.~Singh, ``{Light-cone actions and correlators of self-dual theories in AdS$_{4}$},'' {\em JHEP} {\bf 01} (2025) 172, \href{http://arXiv.org/abs/2411.04172}{{\tt 2411.04172}}.

\bibitem{Tran:2022amg}
T.~Tran, ``{Constraining higher-spin S-matrices},'' {\em JHEP} {\bf 02} (2023) 001, \href{http://arXiv.org/abs/2212.02540}{{\tt 2212.02540}}.

\bibitem{Tran:2025uad}
T.~Tran, ``{Anomaly-free twistorial higher-spin theories},'' \href{http://arXiv.org/abs/2505.13785}{{\tt 2505.13785}}.

\bibitem{Ren:2022sws}
L.~Ren, M.~Spradlin, A.~Yelleshpur~Srikant, and A.~Volovich, ``{On effective field theories with celestial duals},'' {\em JHEP} {\bf 08} (2022) 251, \href{http://arXiv.org/abs/2206.08322}{{\tt 2206.08322}}.

\bibitem{Ponomarev:2022atv}
D.~Ponomarev, ``{Invariant traces of the flat space chiral higher-spin algebra as scattering amplitudes},'' {\em JHEP} {\bf 09} (2022) 086, \href{http://arXiv.org/abs/2205.09654}{{\tt 2205.09654}}.

\bibitem{Ponomarev:2022ryp}
D.~Ponomarev, ``{Towards higher-spin holography in flat space},'' {\em JHEP} {\bf 01} (2023) 084, \href{http://arXiv.org/abs/2210.04035}{{\tt 2210.04035}}.

\bibitem{Ponomarev:2022qkx}
D.~Ponomarev, ``{Chiral higher-spin holography in flat space: the Flato-Fronsdal theorem and lower-point functions},'' {\em JHEP} {\bf 01} (2023) 048, \href{http://arXiv.org/abs/2210.04036}{{\tt 2210.04036}}.

\bibitem{Bengtsson:2014qza}
A.~K.~H. Bengtsson, ``{A Riccati type PDE for light-front higher helicity vertices},'' {\em JHEP} {\bf 09} (2014) 105,
\href{http://arXiv.org/abs/1403.7345}{{\tt 1403.7345}}.
%%CITATION = ARXIV:1403.7345;%%.

\bibitem{Basile:2024raj}
T.~Basile, ``{Massless chiral fields in six dimensions},'' \href{http://arXiv.org/abs/2409.12800}{{\tt 2409.12800}}.

\bibitem{Neville:1971zk}
R.~A. Neville and F.~Rohrlich, ``{Quantum field theory off null planes},'' {\em Nuovo Cim. A} {\bf 1} (1971) 625--644.

\bibitem{Heinzl:2000ht}
T.~Heinzl, ``{Light cone quantization: Foundations and applications},'' {\em Lect. Notes Phys.} {\bf 572} (2001) 55--142, \href{http://arXiv.org/abs/hep-th/0008096}{{\tt hep-th/0008096}}.

\bibitem{Marcus:1982fr}
N.~Marcus and A.~Sagnotti, ``{Tree Level Constraints on Gauge Groups for Type I Superstrings},'' {\em Phys. Lett. B} {\bf 119} (1982) 97--99.

\end{thebibliography}
\end{document}